\documentclass[fleqn,usenatbib]{mnras}

\usepackage{newtxtext,newtxmath}

\usepackage[T1]{fontenc}
\usepackage{ae,aecompl}

\usepackage{graphicx}	
\usepackage{amsmath}	
\usepackage{import}
\usepackage[dvipsnames]{xcolor}
\usepackage{bm}         

\def \kms {{\rm km\,s$^{-1}$}}

\def \sol {{\rm M$_\odot$}}

\def \micron{\hbox{$\upmu$m}}

\def \Kcmcb {{\rm K\,cm$^{-3}$}}

\makeatletter
\DeclareRobustCommand{\ion}[2]{%
  \text{#1\,\check@mathfonts\fontsize\sf@size\z@\selectfont #2}%
}
\makeatother

\newcommand{\HI}{\ion{H}{I}}
\newcommand{\HII}{\ion{H}{II}}

\def \Ha {\textup{H\ensuremath{\upalpha}}}
\def \Hb {\textup{H\ensuremath{\upbeta}}}
\newcommand{\SIIa}{[\ion{S}{II}]\ensuremath{\uplambda6716}}
\newcommand{\SIIb}{[\ion{S}{II}]\ensuremath{\uplambda6731}}
\newcommand{\SII}{[\ion{S}{II}]\ensuremath{\uplambda6716{+}\uplambda6731}}
\newcommand{\nii}{[\ion{N}{II}]}
\newcommand{\sii}{[\ion{S}{II}]}
\newcommand{\Rsii}{\ensuremath{R_{\sii}}}
\newcommand{\NII}{[\ion{N}{II}]\ensuremath{\uplambda6583}}
\newcommand{\OIII}{[\ion{O}{III}]\ensuremath{\uplambda5007}}
\newcommand{\OI}{[\ion{O}{I}]\ensuremath{\uplambda6300}}

\def \Te {\ensuremath{T_\mathrm{e}}}

\def \ne {\ensuremath{n_\mathrm{e}}}
\def \nemin {\ensuremath{n_\mathrm{e,min}}}
\def \nemax {\ensuremath{n_\mathrm{e,max}}}

\def \reff {\ensuremath{r_\mathrm{eff}}}
\def \rmin {\ensuremath{r_\mathrm{min}}}

\def \rgal {\ensuremath{R_\mathrm{gal}}}

\def \ptot {\ensuremath{P_\mathrm{tot}}}
\def \pdir {\ensuremath{P_\mathrm{rad}}}
\def \ptherm {\ensuremath{P_\mathrm{therm}}}
\def \pwind {\ensuremath{P_\mathrm{wind}}}
\def \pde {\ensuremath{P_\mathrm{de}}}
\def \pdess {\ensuremath{\langle P_\mathrm{de,\theta pc} \rangle_\mathrm{1kpc}}}
\def \pmax {\ensuremath{P_\mathrm{max}}}
\def \pmin {\ensuremath{P_\mathrm{min}}}

\def \Lbol {\ensuremath{L_\mathrm{bol}}}

\newcommand{\aifa}{Argelander-Institut f\"{u}r Astronomie, Universit\"{a}t Bonn, Auf dem H\"{u}gel 71, 53121, Bonn, Germany}
\newcommand{\ita}{Institut f\"{u}r Theoretische Astrophysik, Zentrum f\"{u}r Astronomie der Universit\"{a}t Heidelberg, Albert-Ueberle-Str 2, D-69120 Heidelberg, Germany}
\newcommand{\ari}{Astronomisches Rechen-Institut, Zentrum f\"{u}r Astronomie der Universit\"{a}t Heidelberg, M\"{o}nchhofstra\ss e 12-14, 69120 Heidelberg, Germany}
\newcommand{\princeton}{Department of Astrophysical Sciences, Princeton University, Princeton, NJ 08544 USA}
\newcommand{\arcetri}{INAF — Osservatorio Astrofisico di Arcetri, Largo E. Fermi 5, I-50125, Florence, Italy}
\newcommand{\mpe}{Max-Planck-Institut f{\"u}r extraterrestrische Physik, Giessenbachstra{\ss}e~1, D-85748 Garching, Germany}

\newcommand{\eso}{European Southern Observatory, Karl-Schwarzschild-Stra{\ss}e 2, 85748 Garching, Germany}
\newcommand{\ljmu}{Astrophysics Research Institute, Liverpool John Moores University, 146 Brownlow Hill, Liverpool L3 5RF, UK}
\newcommand{\stern}{Sternberg Astronomical Institute, Lomonosov Moscow State University, Universitetsky pr. 13, 119234 Moscow, Russia}
\newcommand{\alberta}{Department of Physics, University of Alberta, Edmonton, AB T6G 2E1, Canada}
\newcommand{\ohio}{Department of Astronomy, The Ohio State University, 140 West 18th Avenue, Columbus, OH 43210, USA}
\newcommand{\wyoming}{Department of Physics \& Astronomy, University of Wyoming, Laramie, WY 82071}
\newcommand{\mpia}{Max-Planck-Institute for Astronomy, K\"onigstuhl 17, D-69117 Heidelberg, Germany}
\newcommand{\iwr}{Universit\"{a}t Heidelberg, Interdisziplin\"{a}res Zentrum f\"{u}r Wissenschaftliches Rechnen, 69120 Heidelberg, Germany}
\newcommand{\anu}{Research School of Astronomy and Astrophysics, Australian National University, Weston Creek, ACT 2611, Australia}
 
\newcommand{\UGent}{Sterrenkundig Observatorium, Universiteit Gent, Krijgslaan 281 S9, B-9000 Gent, Belgium}
\newcommand{\carnegie}{The Observatories of the Carnegie Institution for Science, 813 Santa Barbara Street, Pasadena, CA 91101, USA} 
\newcommand{\Uchile}{Departamento de Astronomía, Universidad de Chile, Casilla 36-D, Santiago, Chile}
\newcommand{\nrao}{National Radio Astronomy Observatory, 520 Edgemont Road, Charlottesville, VA 22903, USA}
\newcommand{\UWaus}{International Centre for Radio Astronomy Research, University of Western Australia, 7 Fairway, Crawley, 6009, WA, Australia}

\title[Feedback mechanisms across nearby galaxies]
{Comparing the pre-SNe feedback and environmental pressures for 6000 \HII\ regions across 19 nearby spiral galaxies}

\author[A.~T.~Barnes et al.]
{A.~T.~Barnes,$^{1}$\thanks{E-mail: ashleybarnes.astro@gmail.com}
S.~C.~O.~Glover,$^{2}$
K.~Kreckel,$^{3}$
E.~C.~Ostriker,$^{4}$
F.~Bigiel,$^{1}$ 
F.~Belfiore,$^{5}$ \and
I.~Be\v{s}li\'c,$^{1}$
G.~A.~Blanc,$^{6,7}$
M.~Chevance,$^{3}$
D.~A.~Dale,$^{8}$
O.~Egorov,$^{3,9}$
C.~Eibensteiner,$^{1}$ \and
E.~Emsellem,$^{10}$ 
K.~Grasha,$^{11}$
B.~A.~Groves,$^{12}$ 
R.~S.~Klessen,$^{2,13}$ 
J.~M.~D.~Kruijssen,$^{3}$ \and
A.~K.~Leroy,$^{14}$ 
S.~N.~Longmore,$^{15}$
L.~Lopez,$^{14}$ 
R.~McElroy,$^{16}$
S.~E.~Meidt,$^{17}$ \and
E.~J.~Murphy,$^{18}$ 
E.~Rosolowsky,$^{19}$ 
T.~Saito,$^{16}$
F.~Santoro,$^{16}$ 
E.~Schinnerer,$^{16}$
A.~Schruba,$^{20}$ \and
J.~Sun,$^{14}$ 
E.~J.~Watkins$^{3}$ and
T.~G.~Williams$^{16}$
\\
Affiliations are listed after references. 
}

\date{Accepted 2021 October 11. Received 2021 September 29; in original form 2021 August 5}
\pubyear{2021}

\begin{document}
\label{firstpage}
\pagerange{\pageref{firstpage}--\pageref{lastpage}}
\maketitle

\begin{abstract}
The feedback from young stars (i.e. pre-supernova) is thought to play a crucial role in molecular cloud destruction. In this paper, we assess the feedback mechanisms acting within a sample of 5810 \HII\ regions identified from the PHANGS-MUSE survey of 19 nearby ($<$\,20\,Mpc)  star-forming, main sequence spiral galaxies (log($M_\star$/M$_\odot$)=  9.4 -- 11). These optical spectroscopic maps are essential to constrain the physical properties of the \HII\ regions, which we use to investigate their internal pressure terms. We estimate the photoionised gas (\ptherm), direct radiation (\pdir), and mechanical wind pressure (\pwind), which we compare to the confining pressure of their host environment (\pde). The \HII\ regions remain unresolved within our ${\sim}50{-}100$\,pc resolution observations, so we place upper (\pmax) and lower (\pmin) limits on each of the pressures by using a minimum (i.e. clumpy structure) and maximum (i.e. smooth structure) size, respectively. We find that the \pmax\ measurements are broadly similar, and for \pmin\ the \ptherm\ is mildly dominant. We find that the majority of \HII\ regions are over-pressured, $\ptot/\pde = (\ptherm+\pwind+\pdir)/\pde > 1$, and expanding, yet there is a small sample of compact \HII\ regions with $P_\mathrm{tot,max}/\pde < 1$ ($\sim$\,1\% of the sample). These mostly reside in galaxy centres ($R_\mathrm{gal}<1$\,kpc), or, specifically, environments of high gas surface density;  log($\Sigma_\mathrm{gas}$/$\mathrm{M_\odot} \mathrm{pc}^{-2}$)\,$\sim$2.5 (measured on kpc-scales). Lastly, we compare to a sample of literature measurements for \ptherm\ and \pdir\, to investigate how dominant pressure term transitions over around 5\,dex in spatial dynamic range and 10\,dex in pressure. 
\end{abstract}
\begin{keywords}
Galaxies: ISM --
Galaxies: star formation --
HII regions --
ISM: structure --
ISM: general
\end{keywords}



\section{Introduction}

\begin{figure*}
    \centering
	\includegraphics[width=\textwidth]{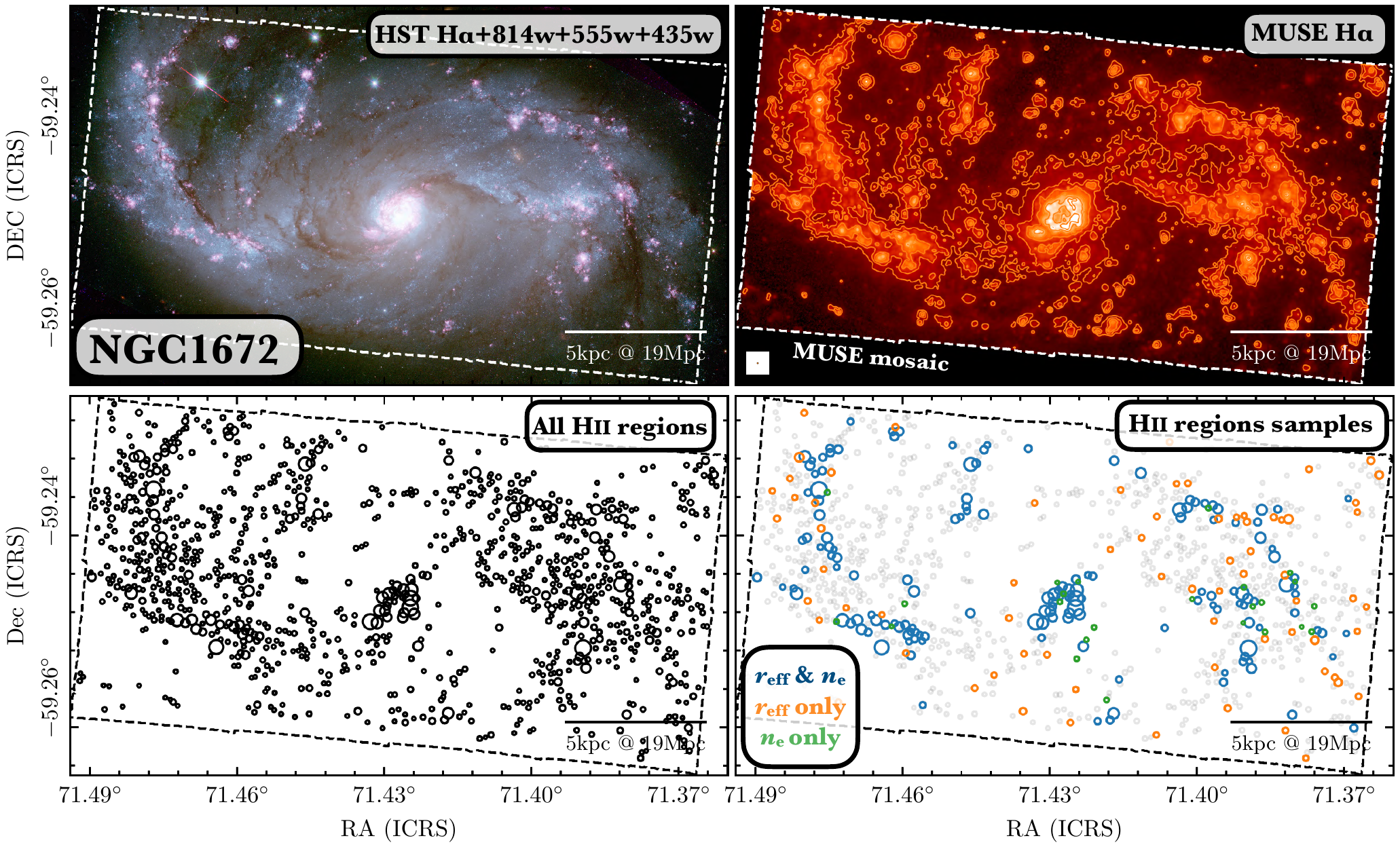}
    \caption{Overview of the \HII\ region samples for two of galaxies studied in this work (NGC\,1672 and NGC\,1300). {\em Upper left}: Three colour image composed of 814\,nm (red), 555\,nm (green) and 435\,nm (blue) wideband emission from the PHANGS-HST survey \citep{Lee2021}, and the \textit{HST} continuum-subtracted 658\,nm or \Ha\ (pink) narrowband emission. {\em Upper right}: MUSE \Ha\ emission map obtained as part of the PHANGS-MUSE survey \citep{Emsellem2021}. Also shown are the beam size of the MUSE \Ha\ observations in the lower left corner, and a scale bar of 5\,kpc in the lower right corner. {\em Lower left}: Full \HII\ region sample identified within each galaxy (\citealt{Santoro2021}; see also Section\,\ref{sec:sample}). Circle sizes represent the measured sizes of the \HII\ regions (\reff; see Section\,\ref{sec:physprops}). {\em Lower right}: Sample of \HII\ regions that have measurements of both the electron density ($n_\mathrm{e}$) and effective radius ($r_\mathrm{eff}$) shown by blue circles. Also shown are the samples that are resolved and below the low-density limit as orange circles, and that are not resolved and above the low-density limit as green circles (see Section\,\ref{sec:lowerlimits}).}
    \label{fig:rgb_main}
\end{figure*}

\begin{figure*}
    \centering
	\includegraphics[width=\textwidth]{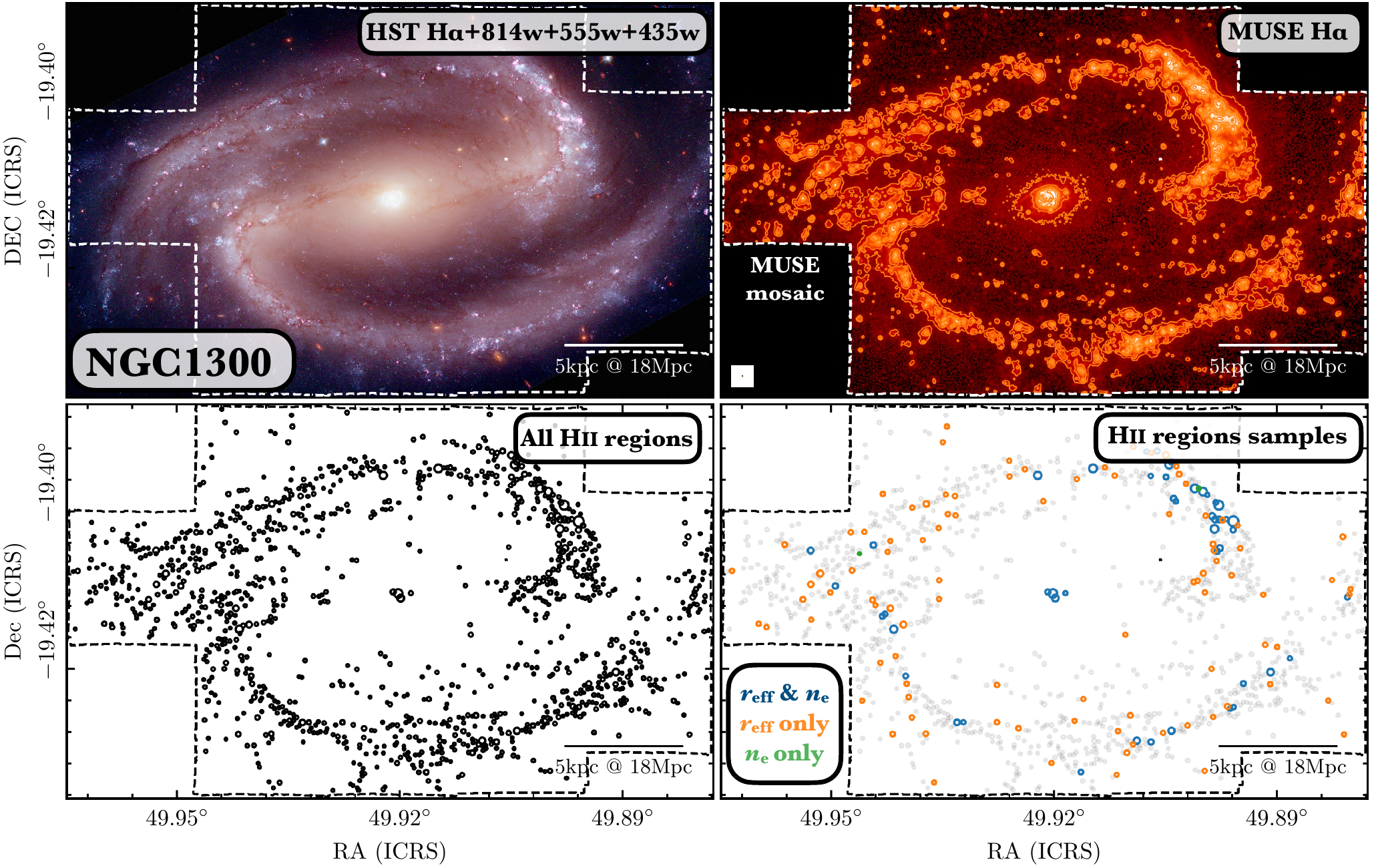}
   \contcaption{}
\end{figure*}

High-mass stars (${>}8$\,\sol) are fundamental for driving the evolution of galaxies across cosmic time, due to the large amount of energy and momentum -- stellar feedback -- that they inject into the interstellar medium (ISM) during their relatively short lifetimes \citep[e.g.][]{krumholz_2014}. This is crucial, as in the absence of any stellar feedback, the ISM would rapidly cool and form stars at a high efficiency, consuming most available gas in the galaxy on a short timescale incompatible with observations \citep[e.g.][]{White1978}. Recent simulations \citep[e.g.][]{dale_2012, dale_2013, Raskutti2016, Gatto2017, Rahner2017, Rahner2019, Kim2018, Kim_JG2021, Kannan2020,Jeffreson2021} and observational evidence \citep[e.g.][]{Grasha2018, Grasha2019, kruijssen19a, chevance20b, chevance20, Kim2021,Barrera-Ballesteros2021a,Barrera-Ballesteros2021b} suggest that feedback in the early (pre-supernova) stages of high-mass stars plays a critical role in destroying molecular clouds, and hence producing the low star formation efficiencies inferred for giant molecular clouds (GMCs) in the Milky Way \citep[ e.g.][]{Zuckerman1974, Krumholz2007, Murray2011, Evans2009, Evans2014, longmore_2013, Lee2016, barnes_2017} and in many other nearby galaxies \citep[e.g.][]{Leroy2008, leroy_2017, Utomo2018, schruba19,Sanchez2020,Sanchez2021}.


Stellar feedback from young stars and stellar clusters is heavily associated with \HII\ regions. In the idealised picture, \citet{stromgren_1939} described \HII\ regions as static, uniform density, spherical regions of ionized gas with a radius set by the balance of ionisation and recombination rates. However, over the following decades our understanding of several physical effects has led to departures from this simple static model: the dynamical expansion of an \HII\ region, if the pressure in the surrounding neutral medium cannot confine its ionized gas, deviates from sphericity due to nonuniform density, injection of energy and momentum by a stellar wind, absorption of hydrogen ionizing photons by dust grains and radiation pressure acting on gas and dust \citep[see e.g.][]{Kahn1954, Savedoff1955, Mathews1967, Mathews1969, Gail1979}. More recently, many works have focused on observationally quantifying the impact of the various feedback mechanisms on driving the expansion of feedback-driven bubbles by detailed studies of their feedback mechanisms, ionization structures, morphologies, dynamics and the stellar content across the Milky Way (e.g.\ \citealp{Rugel2019,Watkins2019,barnes20b, olivier2020}), the Small and Large Magellanic Clouds \citep{oey_1996a, oey_1996b,lopez_2011, lopez_2014, Pellegrini2010,Pellegrini2012, chevance16, mcleod_2019}, and in nearby galaxies (e.g.\ \citealp{mcleod_20,McLeod2021}). 

The dynamics and expansion of \HII\ regions may be driven by several possible sources of internal energy and momentum injection. By definition, \HII\ regions are filled with warm (${\sim}10^{4}$\,K) ionised hydrogen, which imparts an outward gas pressure \citep[e.g.][]{spitzer_1978}. Yet, in addition, several other forms of stellar feedback can drive the dynamics of \HII\ regions and deposit energy and momentum in the surrounding ISM: the direct radiation of stars \citep[e.g.][]{Dopita2005,Dopita2006,Krumholz2009, Peters2010, Fall2010, Murray2010b, Hopkins2011, Commercon2011, Rathjen2021}, the dust-processed infrared radiation \citep[e.g.][]{Thompson2005, Murray2010b, Andrews2011, Skinner2015,Tsang2018, Reissl2018}, stellar winds and supernovae \citep[SNe; e.g.][]{Yorke1989, Harper-Clark2009, Rogers2013}, and protostellar outflows/\linebreak[0]{}jets \citep[e.g.][]{Quillen2005, Cunningham2006, Li2006, Nakamura2008, Wang2010, Rosen2020}. While we have a good understanding of how individual stars or massive stellar populations produce each of these effects, the field still lacks a substantial number of quantitative observations for a diverse sample of \HII\ regions and their environments.

In this work, we investigate the role of early stellar feedback within a sample of \HII\ regions identified across the discs of $19$~nearby spiral galaxies (see Table\,\ref{tab:galprops}). To do so, this work exploits optical integral field unit spectroscopy (IFU; see \citealp{Sanchez2020} for a recent review of nearby galaxy IFU studies) from the Multi Unit Spectroscopic Explorer (MUSE; \citealp{bacon10}) instrument mounted on the Very Large Telescope (VLT) obtained as part of a VLT large programme (PI: Schinnerer). The spatial sampling and large field of view of MUSE allow us to analyse the properties of the ionized gas at high resolution (${\sim}50{-}100$\,pc) in systems as far as $19$\,Mpc (see Figure\,\ref{fig:rgb_main}; \citealt{Emsellem2021}). We use these observations to place limits on the sizes, luminosities and ultimately the feedback-related pressure terms (i.e.\ the direct radiation pressure, the pressure from stellar winds, and the ionized gas pressure) for each of the \HII\ regions. Contrasting these with the local environmental pressure, we can capture a snapshot of the physical and dynamical state of the ionized gas at the later evolutionary stages of \HII\ regions (around a few Myr). 

This paper is organised as follows. In Section\,\ref{sec:sampleselection}, we introduce the sample of $19$~nearby galaxies and the MUSE observations that are used to identify and study their \HII\ region populations. In Section\,\ref{sec:physprops}, we outline assumptions for the unresolved density distribution within each of the identified \HII\ regions and use these to estimate their physical properties. In Section\,\ref{sec:prescalc}, we place limits on the internal pressures within each \HII\ region. We compare how the pressure components vary across the galaxies, how the total internal pressure compares to the external pressure, and how these results compare to samples within the literature in Section\,\ref{sec:prescomp}. Finally, the main results of this paper are summarised in Section\,\ref{sec:summary}.
\section{Sample selection}\label{sec:sampleselection}

To study the young stellar feedback mechanisms, we require a large sample of \HII\ regions that have accurate measurements of ionising photon flux, electron density and/or their size (see Section\,\ref{sec:prescalc}). In this section, we outline the sample of galaxies studied in this work, introduce the MUSE/VLT observations taken as part of the PHANGS (Physics at High Angular Resolution in Nearby GalaxieS) survey (see \citealp{Emsellem2021, Lee2021, Leroy2021a}), and outline how these observations are used to identify a catalogue of around $23{,}699$ \HII\ regions (see Figure\,\ref{fig:rgb_main}). 

\subsection{Galaxy sample of PHANGS-MUSE}

\begin{table*}
    \centering
    \caption{Properties of the galaxy sample. We show in columns from left to right the galaxy name, central right ascension~(RA) and declination~(Dec), inclination~($i$), position angle~(PA), morphological type (Morph.), distance (Dist.), effective radius ($R_\mathrm{eff}$), globally averaged metallicity ($12+\log(\mathrm{O/H})$), total mass of atomic gas ($M_{\HI}$), molecular gas ($M_\mathrm{H_{2}}$) and stars ($M_\star$), and global star formation rate~(SFR).}
    \label{tab:galprops}
    
    \begin{tabular}{ccccccccccccc}
    \hline\hline
   Galaxy & RA & Dec & $i$ & PA & Morph. & Dist. & $R_\mathrm{eff}$ & Metal. & $M_{\HI}$ & $M_\mathrm{H_{2}}$ & $M_\star$ & SFR \\
 & $\mathrm{{}^{\circ}}$ & $\mathrm{{}^{\circ}}$ & $\mathrm{{}^{\circ}}$ & $\mathrm{{}^{\circ}}$ & & $\mathrm{Mpc}$ & $\mathrm{kpc}$ & 12+log(O/H) & log($\mathrm{M_{\odot}}$) & log($\mathrm{M_{\odot}}$) & log($\mathrm{M_{\odot}}$) & log($\mathrm{M_{\odot}\,yr^{-1}}$) \\
  & $(a)$ & $(a)$ & $(b)$ & $(b)$ & $(c)$ & $(d)$ & $(e)$ & $(f)$ & $(g)$ & $(h)$ & $(i)$  & $(i)$ \\
 \hline 

IC5332 & 353.615 & -36.101 & 26.9 & 74.4 & SABc & 9.0 & 3.6 & 8.39 & 9.3 & nan & 9.7 & -0.4 \\
NGC0628 & 24.174 & 15.784 & 8.9 & 20.7 & Sc & 9.8 & 3.9 & 8.51 & 9.7 & 9.4 & 10.3 & 0.2 \\
NGC1087 & 41.605 & -0.499 & 42.9 & 359.1 & Sc & 15.9 & 3.2 & 8.43 & 9.1 & 9.2 & 9.9 & 0.1 \\
NGC1300 & 49.921 & -19.411 & 31.8 & 278.0 & Sbc & 19.0 & 6.5 & 8.52 & 9.4 & 9.4 & 10.6 & 0.1 \\
NGC1365 & 53.402 & -36.140 & 55.4 & 201.1 & Sb & 19.6 & 2.8 & 8.54 & 9.9 & 10.3 & 11.0 & 1.2 \\
NGC1385 & 54.369 & -24.501 & 44.0 & 181.3 & Sc & 17.2 & 3.4 & 8.43 & 9.2 & 9.2 & 10.0 & 0.3 \\
NGC1433 & 55.506 & -47.222 & 28.6 & 199.7 & SBa & 18.6 & 4.3 & 8.57 & 9.4 & 9.3 & 10.9 & 0.1 \\
NGC1512 & 60.976 & -43.349 & 42.5 & 261.9 & Sa & 18.8 & 4.8 & 8.57 & 9.9 & 9.1 & 10.7 & 0.1 \\
NGC1566 & 65.002 & -54.938 & 29.5 & 214.7 & SABb & 17.7 & 3.2 & 8.57 & 9.8 & 9.7 & 10.8 & 0.7 \\
NGC1672 & 71.427 & -59.247 & 42.6 & 134.3 & Sb & 19.4 & 3.4 & 8.56 & 10.2 & 9.9 & 10.7 & 0.9 \\
NGC2835 & 139.470 & -22.355 & 41.3 & 1.0 & Sc & 12.2 & 3.3 & 8.41 & 9.5 & 8.8 & 10.0 & 0.1 \\
NGC3351 & 160.991 & 11.704 & 45.1 & 193.2 & Sb & 10.0 & 3.0 & 8.61 & 8.9 & 9.1 & 10.4 & 0.1 \\
NGC3627 & 170.063 & 12.991 & 57.3 & 173.1 & Sb & 11.3 & 3.6 & 8.55 & 9.1 & 9.8 & 10.8 & 0.6 \\
NGC4254 & 184.707 & 14.416 & 34.4 & 68.1 & Sc & 13.1 & 2.4 & 8.55 & 9.5 & 9.9 & 10.4 & 0.5 \\
NGC4303 & 185.479 & 4.474 & 23.5 & 312.4 & Sbc & 17.0 & 3.4 & 8.58 & 9.7 & 9.9 & 10.5 & 0.7 \\
NGC4321 & 185.729 & 15.822 & 38.5 & 156.2 & SABb & 15.2 & 5.5 & 8.57 & 9.4 & 9.9 & 10.7 & 0.6 \\
NGC4535 & 188.585 & 8.198 & 44.7 & 179.7 & Sc & 15.8 & 6.3 & 8.55 & 9.6 & 9.6 & 10.5 & 0.3 \\
NGC5068 & 199.728 & -21.039 & 35.7 & 342.4 & Sc & 5.2 & 2.0 & 8.34 & 8.8 & 8.4 & 9.4 & -0.6 \\
NGC7496 & 347.447 & -43.428 & 35.9 & 193.7 & Sb & 18.7 & 3.8 & 8.51 & 9.1 & 9.3 & 10.0 & 0.4 \\

    \hline\hline
    \end{tabular}

    \begin{minipage}{\textwidth}
    \vspace{1mm} References: 
    $(a)$ From \cite{Salo2015}.
    $(b)$ From \cite{Lang2020}, based on PHANGS \mbox{CO(2--1)} kinematics. For IC\,5332, we use values from \cite{Salo2015}.
    $(c)$ Morphological classification taken from HyperLEDA \citep{Makarov2014}.
    $(d)$ Source distances are taken from the compilation of \citet{Anand2021}. 
    $(e)$ $R_\mathrm{eff}$ that contains half of the stellar mass of the galaxy \citep{Leroy2021a}. 
    $(f)$ Averaged metallicity within the area mapped by MUSE, computed using the Scal method of \citet{Pilyugin2016}; see \citet{kreckel19} for more details. 
    $(g)$ Total atomic gas mass taken from HYPERLEDA \citep{Makarov2014}.
    $(h)$ Total molecular gas mass determined from PHANGS \mbox{CO(2--1)} observations (see \citealp{Leroy2021a}). CO was not detected at high enough significance in IC\,5332 to allow a molecular gas mass to be determined.
    $(i)$ Derived by \citet{Leroy2021a}, using \textit{GALEX}~UV and \textit{WISE}~IR photometry, following a similar methodology to \cite{Leroy2019}.
    \end{minipage}
      
\end{table*}

The parent galaxy sample of the overall PHANGS program was constructed according to the criteria outlined in \citet{Leroy2021a} and \citet{Emsellem2021}. Briefly, the PHANGS galaxies were selected to be observable by both ALMA (\citealp{Leroy2021a, Leroy2021b}) and MUSE (\citealp{Emsellem2021}; $-75^{\circ}\leq\delta\leq+25^{\circ}$), nearby ($5~\textrm{Mpc} \leq D \leq 17~\textrm{Mpc}$)\footnote{As part of the PHANGS-HST campaign, more accurate distances based on the tip of the red giant branch were determined \citep{Anand2021}, moving some galaxies slightly outside the original selection criteria.}, to allow star-forming regions and molecular clouds to be resolved at high spatial resolution (${\sim}100$\,pc), at low to moderate inclination to limit the effects of extinction and line-of-sight confusion ($i < 65^{\circ}$), and to be massive star-forming galaxies with $\log(M_{\star}/M_{\odot}) \gtrsim 9.75$, and $\log(\textrm{sSFR}/\textrm{yr}^{-1}) \gtrsim -11$. In this work, we use a subset of $19$~PHANGS galaxies that have been observed with the MUSE spectrograph on the VLT \citep[see][for full details]{Emsellem2021}. The sample of PHANGS-MUSE galaxies is given in Table\,\ref{tab:galprops} along with their key properties. For each galaxy, we tabulate the central right ascension~(RA) and declination~(Dec) from \citet{Salo2015}, and the inclination~($i$) and position angle~(PA) based on PHANGS \mbox{CO(2--1)} kinematics from \citet{Lang2020}. Insufficient CO emission was detected in IC\,5332 to allow the kinematics to be constrained, and so for this galaxy we use values from $i$~and PA from \citet{Querejeta2015}. We also show the source distances that are taken from the compilation of \citet{Anand2021}, which along with their Tip of the Red Giant Branch method (TRGB) estimates also include distances taken from \citet{Freedman2001}, \citet{Nugent2006}, \citet{Jacobs2009}, \citet{Kourkchi2017}, \citet{Shaya2017} and \citet{Kourkchi2020}. The deprojected galactocentric radii (in parsec) quoted in this work use these central positions, orientations and distance estimates. In Table\,\ref{tab:galprops}, we list the average metallicity within the region of each galaxy mapped by MUSE, computed using the Scal method of \citet{Pilyugin2016}, as discussed in more detail in \citet{kreckel19}. We also show (in~logarithmic units) mass estimates of the atomic gas \citep[$M_{\HI}$;][]{Makarov2014}, molecular gas \citep[$M_\mathrm{H2}$;][]{Leroy2021a} and stars \citep[$M_\star$;][]{Leroy2021a}, and the average star formation rate \citep[SFR;][]{Leroy2021a}.

\subsection{MUSE observations}
\label{sec:muse}

\begin{table}
\centering
\caption{Properties of the \HII\ catalogue. We show in columns from left to right the galaxy name, the FWHM of the Gaussian PSF of the homogenised (\textit{copt}) mosaic used to identify the \HII\ region sample \citep[see][]{Emsellem2021}, the number of \HII\ regions within the whole sample (see Section\,\ref{sec:sample}), and number of \HII\ regions that are resolved and above the low density limit (i.e. have measurements of both $r_\mathrm{eff}\,\&\,n_\mathrm{e,max}$, such that $r_\mathrm{min}\,\&\,n_\mathrm{e,min}$ can also be calculated), resolved and below the low density limit (i.e. $r_\mathrm{eff}$ only, such that $n_\mathrm{e,min}$ is estimated), and unresolved and above the low density limit (i.e. $n_\mathrm{e,max}$ only such that $r_\mathrm{min}$ is estimated; see Section\,\ref{sec:lowerlimits}). In section\,\ref{sec:prescalc}, the (2238+141) \HII\ regions with $r_\mathrm{min}\,\&\,n_\mathrm{e,max}$ are used to determine maximum pressure terms ($P_\mathrm{max}$) under the assumption of a \emph{clumpy} density structure, and the (2238+3431) with $r_\mathrm{eff}\,\&\,n_\mathrm{e,min}$ are used to determine minimum pressure terms ($P_\mathrm{min}$) under the assumption of a \emph{smooth} density structure (see Figure\,\ref{fig:toyfig}).}
\label{tab:galobsprops}

\begin{tabular}{cccccc}
\hline\hline
Galaxy & $\mathrm{FWHM}_\mathrm{PSF}$ & \multicolumn{4}{c}{Samples of \HII\ regions}\\ 
& & All & $r_\mathrm{eff}$\,\&\,$r_\mathrm{min}$ & $r_\mathrm{eff}$ & $r_\mathrm{min}$ \\
& & & $n_\mathrm{e,max}$\,\&\,$n_\mathrm{e,min}$ & $n_\mathrm{e,min}$ & $n_\mathrm{e,max}$ \\
 & $\mathrm{{}^{\prime\prime}}$ & $\#$ & $\#$ & $\#$ & $\#$ \\
 \hline
IC5332 & 0.87 & 630 & 6 & 120 & 0 \\
NGC0628 & 0.92 & 2369 & 99 & 399 & 0 \\
NGC1087 & 0.92 & 895 & 84 & 108 & 1 \\
NGC1300 & 0.89 & 1178 & 47 & 94 & 2 \\
NGC1365 & 1.15 & 866 & 90 & 56 & 19 \\
NGC1385 & 0.67 & 919 & 156 & 244 & 3 \\
NGC1433 & 0.91 & 1285 & 37 & 94 & 0 \\
NGC1512 & 1.25 & 485 & 19 & 30 & 11 \\
NGC1566 & 0.80 & 1654 & 204 & 187 & 16 \\
NGC1672 & 0.96 & 1069 & 152 & 70 & 24 \\
NGC2835 & 1.15 & 818 & 87 & 72 & 5 \\
NGC3351 & 1.05 & 821 & 21 & 113 & 1 \\
NGC3627 & 1.05 & 1012 & 171 & 142 & 14 \\
NGC4254 & 0.89 & 2576 & 360 & 437 & 8 \\
NGC4303 & 0.78 & 2211 & 374 & 311 & 16 \\
NGC4321 & 1.16 & 1416 & 121 & 89 & 20 \\
NGC4535 & 0.56 & 1476 & 72 & 421 & 0 \\
NGC5068 & 1.04 & 1469 & 96 & 388 & 1 \\
NGC7496 & 0.89 & 550 & 42 & 56 & 0 \smallskip \\
All & - & 23699 & 2238 & 3431 & 141 \\
\hline\hline
\end{tabular}
\end{table}

The MUSE Integral Field Unit (IFU) provides a $1\arcmin \times 1\arcmin$ field of view, $0.2\arcsec$ pixels, and a typical spectral resolution (FWHM) of ${\sim}2.5$\,\AA\ (or ${\sim}100$\,\kms) covering the spectral range $4800{-}9300$\,\AA. 
Observations of the $19$~galaxies are reduced using the {\tt pymusepipe} package. {\tt pymusepipe} was developed specifically for these observations by the PHANGS team,\footnote{\url{https://github.com/emsellem/pymusepipe}} and is a {\tt python} wrapper around the main processing steps of the data reduction conducted by the MUSE pipeline ({\tt MUSE DRS}; \citealp{Weilbacher2020}) accessed via {\tt EsoRex} command-line recipes.
A~complete discussion of the processing and reduction of the MUSE observations is presented in \citet{Emsellem2021}. The final reduced cubes have an angular resolution ranging between ${\sim}0.6\arcsec$ (for the subset observed using ground-layer correction adaptive optics) and ${\sim}1.2\arcsec$ (see Table\,\ref{tab:galobsprops}), which at the distances of our sample corresponds to physical scales of $25{-}70$\,pc at the corresponding galaxy distance (see Table\,\ref{tab:galprops}).

To identify and determine the properties of the \HII\ regions across each galaxy using their optical spectroscopic features, \citet{Emsellem2021} first produce emission line maps covering the full field of view. To do so, all data cubes are processed using the penalised pixel fitting Python package \citep[\textsc{pPXF};][]{cappellari2004, Cappellari2017}. These fits include \mbox{E-MILES} simple stellar population models \citep{Vazdekis2016} and a set of emission lines that are treated as additional Gaussian templates. A~detailed description of the spectral fitting process is presented in \citet{Emsellem2021}.

\subsection{Sample of \HII\ regions}
\label{sec:sample}

We make use of an ionised nebulae catalogue derived from PSF-homogenised \Ha\ line maps (the \textit{copt} data products described in \citealp{Emsellem2021}), as described in detail in 
\citet{Santoro2021}. Briefly, \citet{Santoro2021}  first run an implementation of \textsc{HIIphot} \citep{Thilker2000}, which has been adapted for use with integral field data. The final catalogue contains not only \HII\ regions, but also planetary nebulae and supernova remnants. The ionised nebulae spatial masks are then applied to the original data cube to extract integrated spectra for each object. The emission lines within these spectra are subsequently fitted using the same procedure as the one described in Section\,\ref{sec:muse}. This fitting procedure includes both strong lines (e.g.\ \Hb, \OIII, \Ha, \NII, \SIIa, \SIIb) as well as fainter, tempera\-ture-sensi\-tive auroral lines (e.g.\ [\ion{N}{II}]$\lambda$5755, [\ion{S}{III}]$\lambda$6312). We then employ the following selection criteria to identify \HII\ regions within the nebular catalogue:
\begin{enumerate}
    \item We remove all nebulae that are flagged as being smaller than a single point spread function (PSF; see Table\,\ref{tab:galobsprops}) separated from the edge of the field of view in order to ensure that we do not include regions that have artificially small sizes because they lie partially outside of the field of view.
    \item We require all of the strong lines used to compute line ratio diagnostics (see below) to be detected with signal-to-noise greater than three.\footnote{In practice, this restriction is not strictly necessary, as regions where not all of the strong lines are detected are almost always too faint and too small for us to be able to derive an estimate of their internal pressures.}
    \item We apply three emission line ratio diagnostic diagrams \citep[BPT diagrams;][]{baldwin81} to separate the nebulae photoionised by high-mass stars from those ionised by other sources (e.g.\ Active Galactic Nuclei). Nebulae are classified as \HII\ regions if they fall below the \citet{Kauffmann2003} line in the $\OIII/\Hb$ vs.\ $\NII/\Ha$ diagram and below the \citet{Kewley2001} line in the $\OIII/\Hb$ vs.\ $\SII/\Ha$ and $\OIII/\Hb$ vs.\ $\OI/\Ha$ diagrams.\footnote{The source identification routine, which omits diffuse emission, and these high emission line thresholds should mitigate the contamination from other sources of ionisation in our \HII\ region sample (e.g.\ shocks; e.g. see \citealp{Espinosa-Ponce2020} and references therein). This will be, however, investigated further in a future version of the catalogue (Section\,\ref{subsec:future}).}
    \item We discard regions with \Ha\ velocity dispersions that exceed $100$~\kms, which are likely compact supernova remnants.\footnote{Note that this limit is more conservative than the one adopted in \citet{Santoro2021}.}
\end{enumerate}

We note that \citet{Santoro2021} conducted the source identification on the \Ha\ emission maps from MUSE that have not been extinction corrected. The extinction correction is then determined for the \Ha\ flux within the source masks using the Balmer decrement (see section\,\ref{subsec:ionphot}). Hence, there is no correction for the surface brightness dimming within extincted regions during the source identification stage of our analysis. Due to the high sensitivity of our MUSE observations, however, we can typically recover the \Ha\ emission out to the point where the \HII\ regions merge with the diffuse ionised gas (DIG). In addition, we do not directly account for the contribution of the DIG in our determination of the \HII\ region line fluxes, as this was minimised by the selection of the source identification parameters in the \textsc{HIIphot} package. That said, the contribution of the DIG to the \HII\ regions studied in this work is expected to be low, given that here we have explicitly chosen to study the brightest regions due to our high flux selection thresholds (signal-to-noise greater than three). Doing so is also not trivial, and can introduce large uncertainties into the remaining fluxes. Lastly, the physical interpretation of removing the DIG in our pressures analysis is not clear (see section\,\ref{sec:prescalc}). The leakage of ionised gas from the HII regions is the main contributor to the DIG \citep{Belfiore2021}, yet this gas could still provide a contribution to the pressure terms we measure.

The above selection criteria remove $7798$ sources from the initial catalogue of $31{,}497$ objects, hence leaving ${\sim}75$ per cent of the sources ($23{,}699$) as \HII\ regions. The number of \HII\ regions within each galaxy is summarised in Table\,\ref{tab:galobsprops}. To achieve a final sample of ${\sim}6000$ \HII\ regions, which is used throughout this work, we imposed further selection criteria on size and density measurements (outlined in Section\,\ref{sec:physprops}). The total number of identified \HII\ regions within each galaxy is presented in Table\,\ref{tab:galprops}, and ranges from a few hundred to a few thousand. In Figure\,\ref{fig:rgb_main}, we show the distribution of the \Ha\ emission line and \HII\ region catalogue compared to an optical \textit{HST} three colour composite \citep{Lee2021} for two galaxies in our sample (NGC\,1300 and NGC\,1672). 
\section{Physical properties}
\label{sec:physprops}

We use the wealth of information provided by the MUSE observations to estimate several fundamental physical properties for the \HII\ regions in our catalogue: the \HII\ region sizes~(\reff), electron densities~(\ne) and ionisation rates~($Q$), as well as the mass~($M_\mathrm{cl}$), bolometric luminosity~($L_\mathrm{bol}$), total mass loss rate~($\dot M$) and mechanical luminosity~($L_\mathrm{mech}$) of the cluster or association powering each region.

With regard to the \HII\ region sizes, we face the complication that the size that we ideally want to measure is the characteristic radius at which the majority of the mass of the ionised gas is located, since this is more relevant for understanding the dynamics of the \HII\ region and the interplay between the different pressure terms than the maximum physical extent of the ionised region. For an \HII\ region that is well-described by the classical Str\"omgren sphere solution \citep{stromgren_1939}, or one with a shell-like morphology, this is comparable to the extent of the \HII\ region, but for a partially embedded or blister-type \HII\ region, this is not necessarily the case. 

\begin{figure}
    \centering
	\includegraphics[width=1\columnwidth]{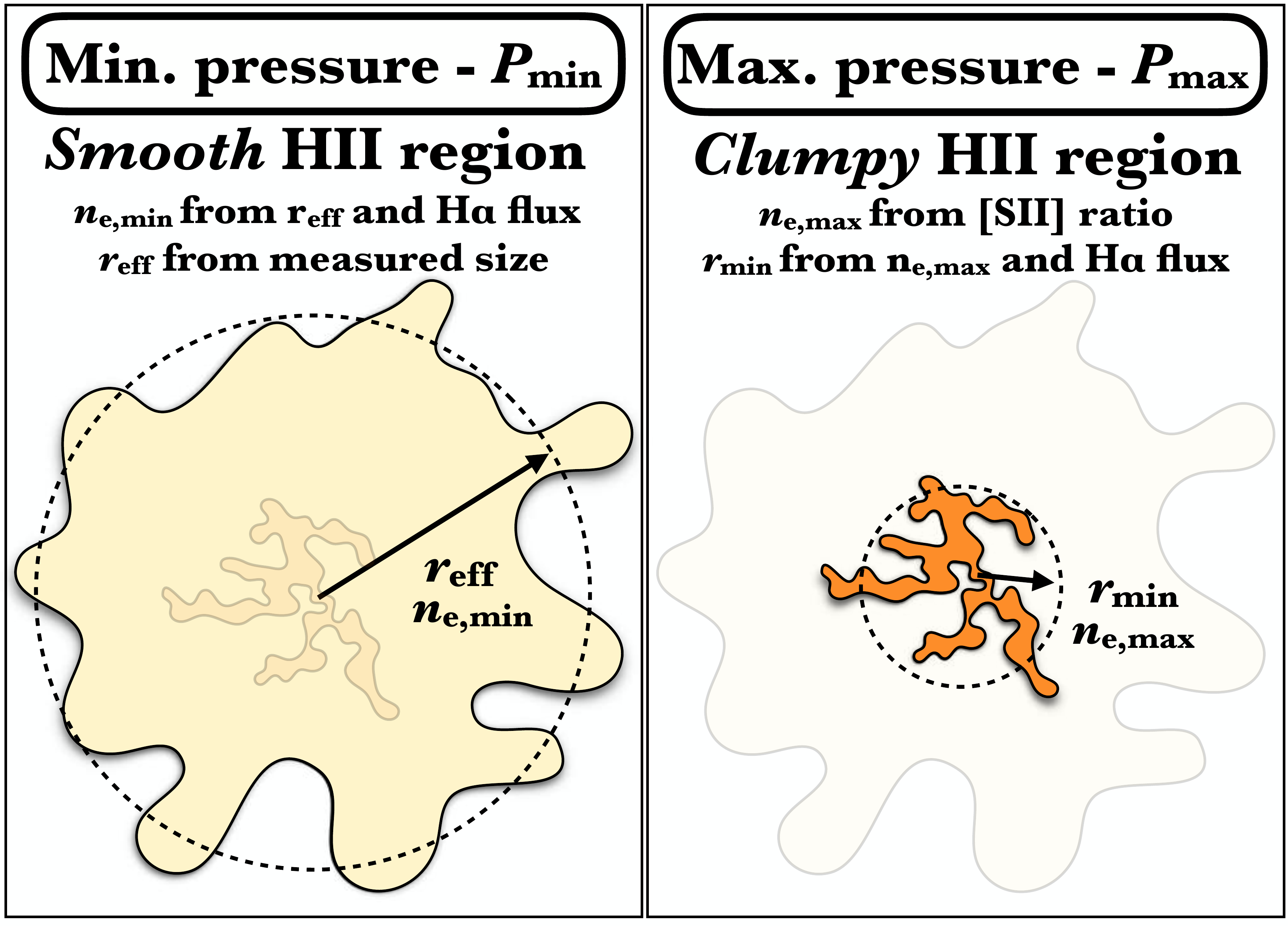}
    \caption{Schematic diagram representative of an \HII\ region in our sample. As the linear resolution of our observations is of the order a few tens of parsec at the distance of the galaxies in our sample, in this work we consider two limiting cases for the unresolved density distribution within an \HII\ region (see Section\,\ref{sec:physprops}). {\em Left panel:} On the one extreme, we posit that the density distribution is \emph{smooth}, and the properties of the \HII\ regions are hence determined for the effective spherical radius determined from the resolved \HII\ regions size (i.e. the measured~\reff). {\em Right panel:} On the other extreme, we posit that the \HII\ regions have a \emph{clumpy} density structure, with the clumps located close to the centre of the region, as might be expected, e.g., in an \HII\ region in the process of breaking out from a molecular cloud. In this case, the properties are determined using the minimum volume within which these dense clumps can be distributed while remaining consistent with the measured \Ha\ flux and the density inferred from the \sii\ doublet. The corresponding size scale in this case is \rmin. Our calculated pressure terms therefore each have two limits: a maximum ($P_\mathrm{max}$) at \rmin\ and a minimum ($P_\mathrm{min}$) at \reff\ (Section\,\ref{sec:prescalc}).}
    \label{fig:toyfig}
\end{figure}

At the resolution of our MUSE observations, we cannot easily distinguish these different \HII\ region morphologies, and so we instead consider two limiting cases for the distribution of ionised gas within the observed \HII\ regions. In one limit, we assume that the ionised gas is \emph{smoothly} distributed throughout the measured volume of the \HII\ region, and the properties of the \HII\ regions are hence determined for the maximum radius (i.e.\ the measured~\reff; Section\,\ref{subsubsec:reff}). In the other limit, we assume that the ionised gas is \emph{clumpy}, with most of the mass located in dense clumps that lie close to the centre of the \HII\ region. In this case, the properties are determined using the minimum volume within which these clumps can be accommodated while remaining
consistent with the measured electron density (Section \ref{subsubsec:electrondensity}) and \Ha\ flux. The radial length scale associated with this minimum volume is hereafter denoted as \rmin. These are, of course, not the only possibilities -- for instance, a shell-like \HII\ region may have most of its gas in dense clumps that are located far from the ionising source at $r \sim \reff$ -- but for our purposes we restrict our attention to these two limiting cases as they will later allow us to put upper and lower limits on the various pressure terms.  
These assumptions are illustrated in Figure\,\ref{fig:toyfig} and are discussed further in Section\,\ref{sec:lowerlimits}.

\subsection{Measured effective radii -- \reff}
\label{subsubsec:reff}

\begin{figure}
    \centering
	\includegraphics[width=\columnwidth]{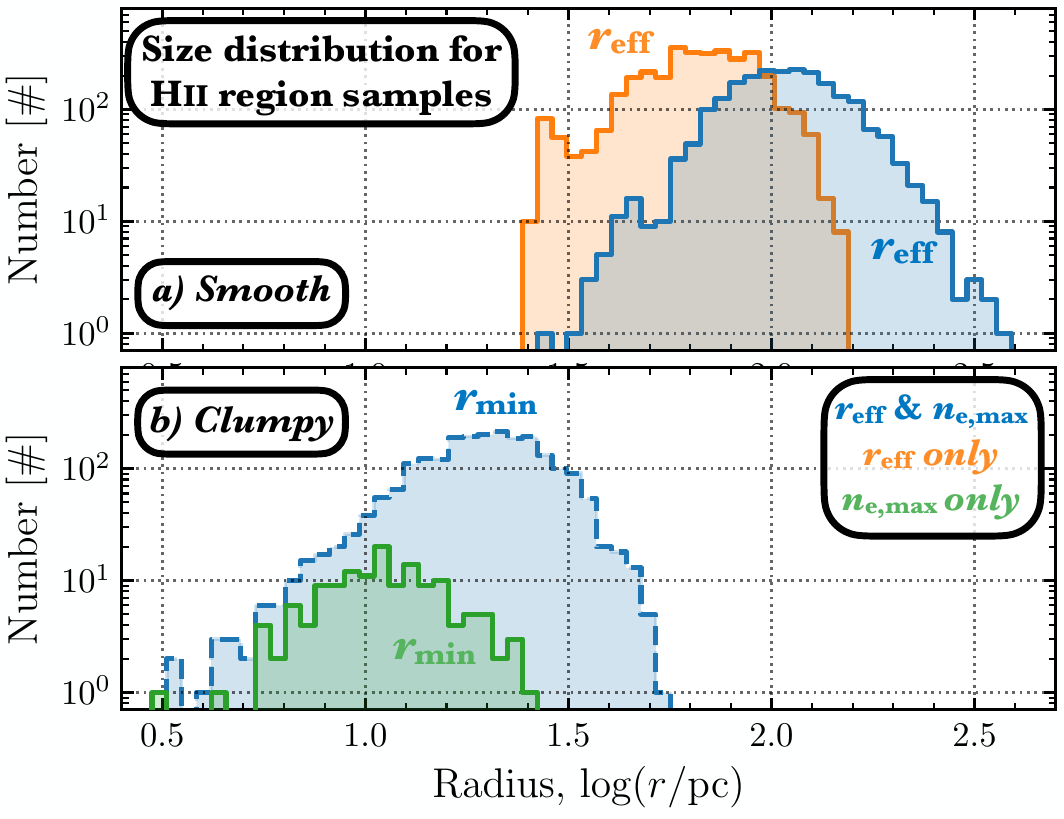}
    \caption{Histograms of the size distribution for the \HII\ region sample for assumptions of a \emph{a) smooth} and \emph{b) clumpy} unresolved density distribution (see Figure\,\ref{fig:toyfig}). {\em Upper panel:} We show \reff\ for the resolved ($\reff \ge \mathrm{FWHM}_\mathrm{PSF}$; see Section\,\ref{subsubsec:reff}) sources as both blue (solid outline) and orange histograms. The blue and orange histograms differentiate the samples with and without \nemax\ electron density measurements, respectively (see Section\,\ref{subsubsec:electrondensity}). {\em Lower panel:} Where possible, we also determine a size (\rmin) using the \nemax\ determined from \Rsii\ and the \Ha\ flux (see Section~\ref{sec:lowerlimits}). These \rmin\ distributions are shown in green and blue (dashed line). Note that the \HII\ regions that are both resolved and are below the low-density limit have both \rmin\ and \reff\ estimates, and, hence, the blue histograms have the same sample size.}
    \label{fig:hist_radius}
\end{figure}

We estimate the effective angular radii for the \HII\ regions in the catalogue by circularising the area contained within each \HII\ region; $\reff = \sqrt{a/\pi}$,\footnote{We do not correct $\reff$ for the additional broadening of the PSF for the following reasons. A comparison of the MUSE observations to the available higher resolution ($\sim$\,0.05\arcsec) HST \Ha\ images showed that many of the \HII\ regions we identify are large (partial) shells (Barnes et al. in prep). We found that a simple quadrature subtraction of the $\mathrm{FWHM}_\mathrm{PSF}$ (see Table\,\ref{tab:galobsprops}) did not accurately recover the shell radii. Moreover, the PSF broadening is minimal for our sample of \HII\ regions with reliable $\reff$. As these are significantly far away from the PSF size; our resolved threshold of $\reff < \mathrm{FWHM}_\mathrm{PSF}$ (i.e.\ the effective diameter must be at least two times the resolution limit). For example, the quadrature subtraction of PSF for \HII\ with $\reff$ within a factor of two of $\mathrm{FWHM}_\mathrm{PSF}$ gives $\sim$10\% reduction in sizes.} where $a$ is the area enclosed by the boundary identified using the \textsc{HIIphot} routine (i.e.\ above some intensity threshold, not a fitted ellipse).
An inherent problem with many automated source identification algorithms, such as with \textsc{HIIphot}, is their tendency to separate compact emission into distinct sources that have sizes comparable to the point spread function (or resolution) of the input observations. The result of this is the identification of a large sample of unresolved or only marginally resolved (point) sources, the sizes for which are either unconstrained or highly uncertain. As we require an accurate measure of \HII\ region sizes for our pressure analysis (see Section\,\ref{sec:dir_prescalc}), we consider regions to be resolved only if their effective radius satisfies the resolution criterion $\reff \ge \mathrm{FWHM}_\mathrm{PSF}$, i.e.\ the effective diameter must be at least two times the resolution limit of the observations for each galaxy (see Table\,\ref{tab:galobsprops}). This threshold was chosen to include the \HII\ regions that are significantly more extended than the observational limits, yet without substantially limiting our sample for the most distant galaxies. We consider all regions smaller than the $\reff < \mathrm{FWHM}_\mathrm{PSF}$ limit to be unresolved, and do not make use of the values of $\reff$ derived for these regions in our later analysis. For the subset of these unresolved regions for which we can determine the electron  density (see Section \ref{subsubsec:electrondensity}), we can place a lower limit on their sizes, as discussed later in Section\,\ref{sec:lowerlimits}. Unresolved regions without a well-determined electron density cannot be assigned meaningful values of either \reff\ or \rmin\ and are not considered further in our analysis.

The physical effective radius of each \HII\ region in units of parsec is determined using the source distance given in Table\,\ref{tab:galprops}. We find that the \reff\ size range across the whole sample of \HII\ regions (including both resolved and unresolved sources) is $16.6$\,pc to $388.4$\,pc (median: $54.6$\,pc), while for the resolved sub-sample of \HII\ regions it is $26.2$\,pc to $388.4$\,pc (median: $80.64$\,pc). In Figure\,\ref{fig:hist_radius}, we show two distributions of \reff: one for regions that are resolved and that have a measured electron density (blue histogram with solid outline) and one for regions that are resolved but that do not have a measured electron density (orange histogram). In the Figure, we also show the distribution of $\rmin$ for those regions in which it can be calculated (see Section\,\ref{sec:lowerlimits} below).

\subsection{Measured electron densities -- \nemax}
\label{subsubsec:electrondensity}

\begin{figure}
    \centering
	\includegraphics[width=1\columnwidth]{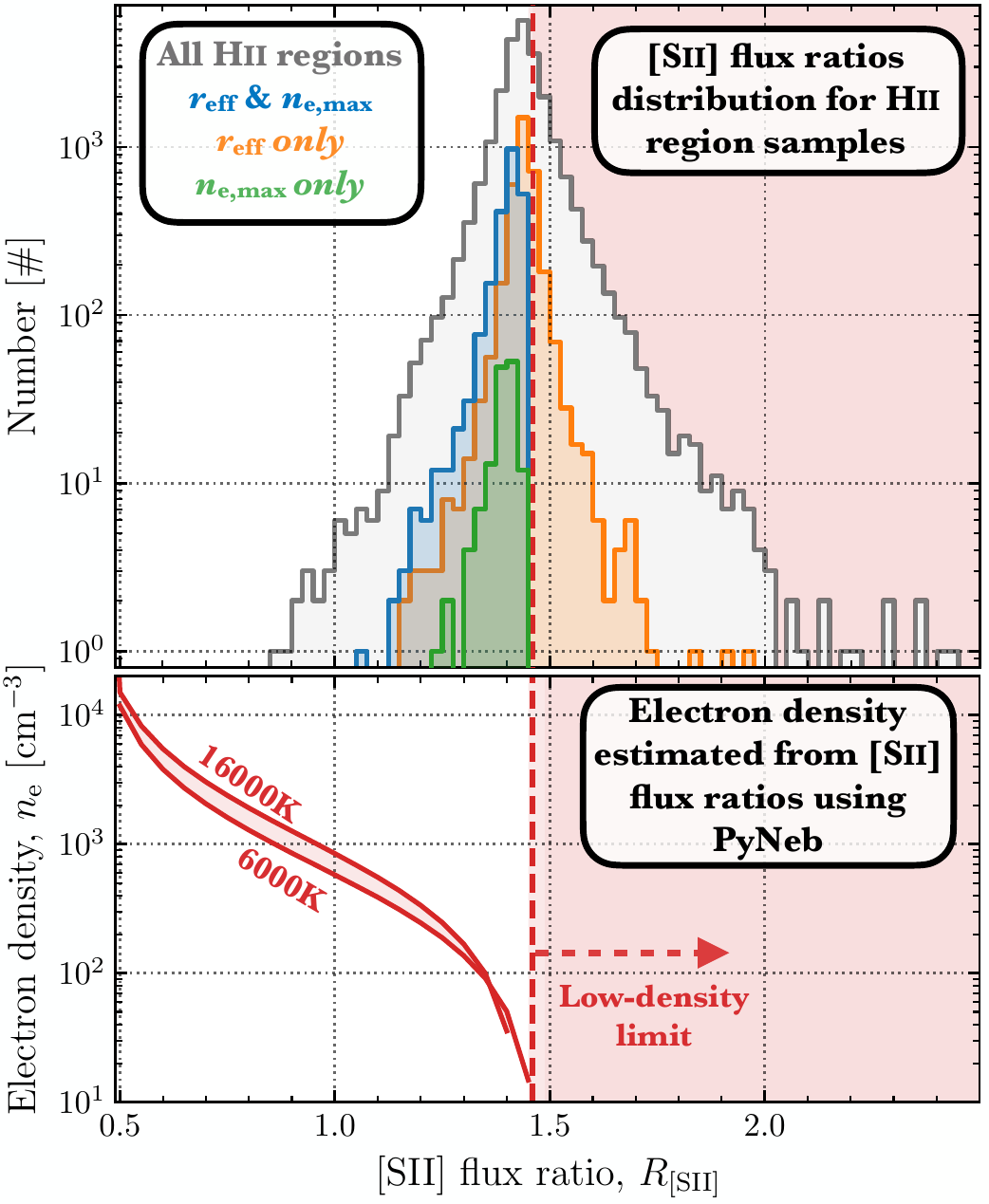}
    \caption{{\em Upper panel:} Histogram of the $\Rsii = F_{\SIIa} / F_{\SIIb}$ line ratio across the \HII\ region sample. We show the distribution of \Rsii\ for the full sample of \HII\ regions as a grey histogram. The vertical dashed red line and shaded region shows the low-density limit (also shown in the lower panel). The blue and green histograms show the distributions of \HII\ regions that are statistically distinguishable from the low-density limit (see Figure\,\ref{fig:hist_ne}), and have resolved and unresolved sizes (see Section\,\ref{subsubsec:reff}). The orange histogram shows the \HII\ regions that are indistinguishable from the low-density limit, yet are resolved (see Figure\,\ref{fig:hist_ne}). {\em Lower panel:} The dashed curve shows the conversion between \Rsii\ and electron density (\ne, see right $y$-axis) as determined from \textsc{PyNeb} assuming two electron temperatures that span the observed range of approximately $6000$\,K to $16{,}000$\,K.}
    \label{fig:hist_RSII}
\end{figure}

\begin{figure}
    \centering
	\includegraphics[width=1\columnwidth]{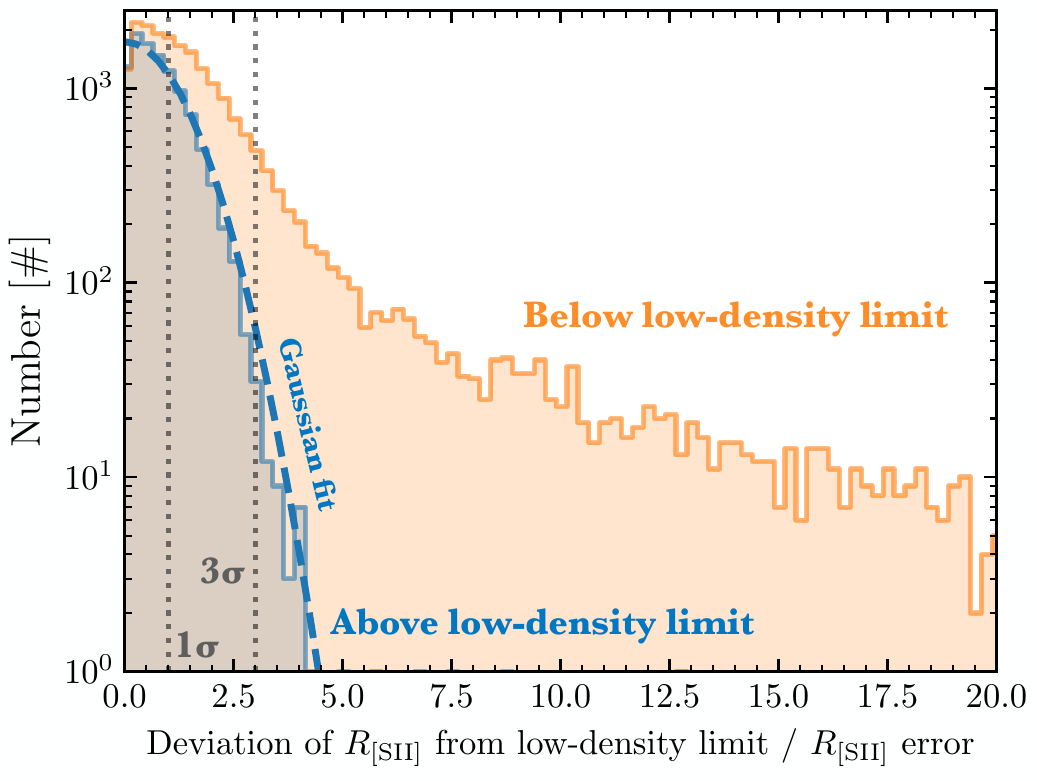}
    \caption{Histogram of the absolute difference of the $\Rsii = F_{\SIIa} / F_{\SIIb}$ line ratio from the low-density limit, normalized by the error in the line ratio for each \HII\ region. The orange and blue histograms show the distributions of \Rsii\ above and below an electron tempera\-ture-depen\-dent low-density limit (see Section\,\ref{subsubsec:electrondensity}). The vertical dotted lines show a difference of 1 and 3\,$\sigma_{\Rsii}$ from the low-density limit. The dashed curve shows the Gaussian function fit to the histogram distribution of \HII\ regions above the \Rsii\ limit, which has a standard deviation of $\sim$\,1\,$\sigma_{\Rsii}$. We highlight the non-Gaussian tail for those sources below the low-density limit.}
    \label{fig:hist_RSIIerr}
\end{figure}

To calculate the electron density of the \HII\ regions in our sample~(\nemax), we use the \textsc{PyNeb} package \citep{Luridiana2015}. \textsc{PyNeb} is a Python module for the analysis of emission lines. It solves the equilibrium equations and determines level populations for one or several user-selected model atoms and ions. We use \textsc{PyNeb} to solve for the electron density within each \HII\ region given the flux ratio $\Rsii = F_{\SIIa} / F_{\SIIb}$, and a value for the electron temperature (\Te; Belfiore et al. in prep).\footnote{The electron temperature is determined from the nitrogen auroral lines using \textsc{PyNeb}, and will be presented by Belfiore et al. (in prep). Briefly, the method uses the \nii\ ion-based auroral-to-nebular line ratio, $(\nii\uplambda6584 + \nii\uplambda6548) / \nii\uplambda5755$, the value of which is sensitive to the temperature of the ionised gas. 
A~density of $100$\,cm$^{-3}$ is used in \textsc{PyNeb} for the purposes of calculating $T_{\rm e}$ from this ratio, although the values obtained are insensitive to this choice.}
 In the lower panel of Figure\,\ref{fig:hist_RSII}, we show the \textsc{PyNeb} solutions for $n_\mathrm{e}$ as a function of \Rsii\ for two values of the electron temperature that are representative of the extremes of the temperature distribution for the \HII\ region sample ($6000$ to $16{,}000$\,K; Belfiore et al. in prep). \Rsii\ has two limiting values: a high density limit at $\Rsii \sim 0.4$ that is reached at number densities above a few $1000 \: {\rm cm^{-3}}$ and a low density limit at $\Rsii \sim 1.45$ that is reached at number densities below a few 100 to 10s~cm$^{-3}$. For \HII\ regions with densities between these two limits, measuring the value of \Rsii\ allows us to infer the \sii\ emission-weighted mean density of the ionised gas. In the case of a clumpy \HII\ region, this density will primarily reflect that of the gas in the clumps and may significantly exceed the mean density of the ionised gas in the \HII\ region as a whole. For that reason, we refer to this estimate loosely as the ``maximum'' electron density for the \HII\ region, which we will later compare with a ``minimum'' electron density estimate derived using a different technique. Note also that the inferred density depends on the electron temperature, but as Figure\,\ref{fig:hist_RSII} shows, this dependence is relatively weak.

In the upper panel of Figure\,\ref{fig:hist_RSII}, we show the distribution of values for \Rsii\ that we measure for the whole sample of \HII\ regions (grey). We see that the distribution peaks just above $\Rsii > 1.4$, i.e.\ at a value comparable to the one we expect to recover on the low density limit. However, we also see that many of the values we measure for \Rsii\ lie above this limiting value (indicated by the red shaded region in the Figure). These values are unphysical and so we assume that they are due to the statistical errors in our measurements of the fluxes of the \sii\ lines, which introduce an error into the calculated line ratio. To check this, we calculate for each region the absolute difference of $\Rsii$ from the low density limiting value, normalised by the uncertainty in the value of $\Rsii$ for that region ($\sigma_{\Rsii}$). This uncertainty is calculated using the formal errors in the fluxes of the two \sii\ lines, adjusted upwards by a factor of 1.43 to account for the fact that these formal errors are still somewhat under-estimated in the latest version of the MUSE data reduction, likely due to imperfect sky substraction.\footnote{See the detailed discussion of this issue in \citet{Emsellem2021}.} We use a tempera\-ture-depen\-dent low-density limit of $\Rsii = 1.49 - 3.94\times10^{-6}\,\Te$. For regions that do not have reliable estimates of the electron temperature, as can happen if the $\nii\uplambda5755$ auroral line is not detected, we adopt a representative electron temperature of $8000$\,K, which yields $\Rsii = 1.46$. Finally, we account for the uncertainty in $\Rsii$ that arises due to the statistical error in the $\Te$ measurement by combining this in quadrature with the line flux uncertainties when computing $\sigma_{\Rsii}$.

We show the distribution of the normalized absolute differences in Figure\,\ref{fig:hist_RSIIerr}. The blue histogram corresponds to \HII\ regions with values of $\Rsii$ above the low density limiting value, while the orange histogram shows the \HII\ regions with values of \Rsii\ below this limit. We see that the distribution of \Rsii\ values in the unphysical region above the low density limit is Gaussian, with a standard deviation of 1, consistent with what we would expect if all of these regions have a true value of \Rsii\ at or very close to the low density limiting value. For \HII\ regions with measured \Rsii\ below this limit, we recover a Gaussian distribution for low values of the normalized deviation and a clear non-Gaussian tail for higher deviations. In order to exclude regions which are consistent with Gaussian noise around the low-density limit, we select \HII\ regions that are at least $3\sigma_{\Rsii}$ away from the low-density limit, where the Gaussian distribution becomes sub-dominant (see dashed vertical grey line in Figure\,\ref{fig:hist_RSIIerr}). The \Rsii\ distributions for the samples of \HII\ regions, that are significantly below the low-density limit, are shown as blue and green histograms in Figure\,\ref{fig:hist_RSII}. Note that these fall below the histogram for the whole sample (shown in grey) as, even below the low-density limit, the uncertainties of \Rsii\ can be large, causing some values to be indistinguishable from the low-density limit. We also show the distribution of \HII\ regions that are indistinguishable from the low-density limit and have resolved sizes, for which we will calculate lower limits for the electron density (see Section\,\ref{sec:lowerlimits}).

\begin{figure}
    \centering
	\includegraphics[width=\columnwidth]{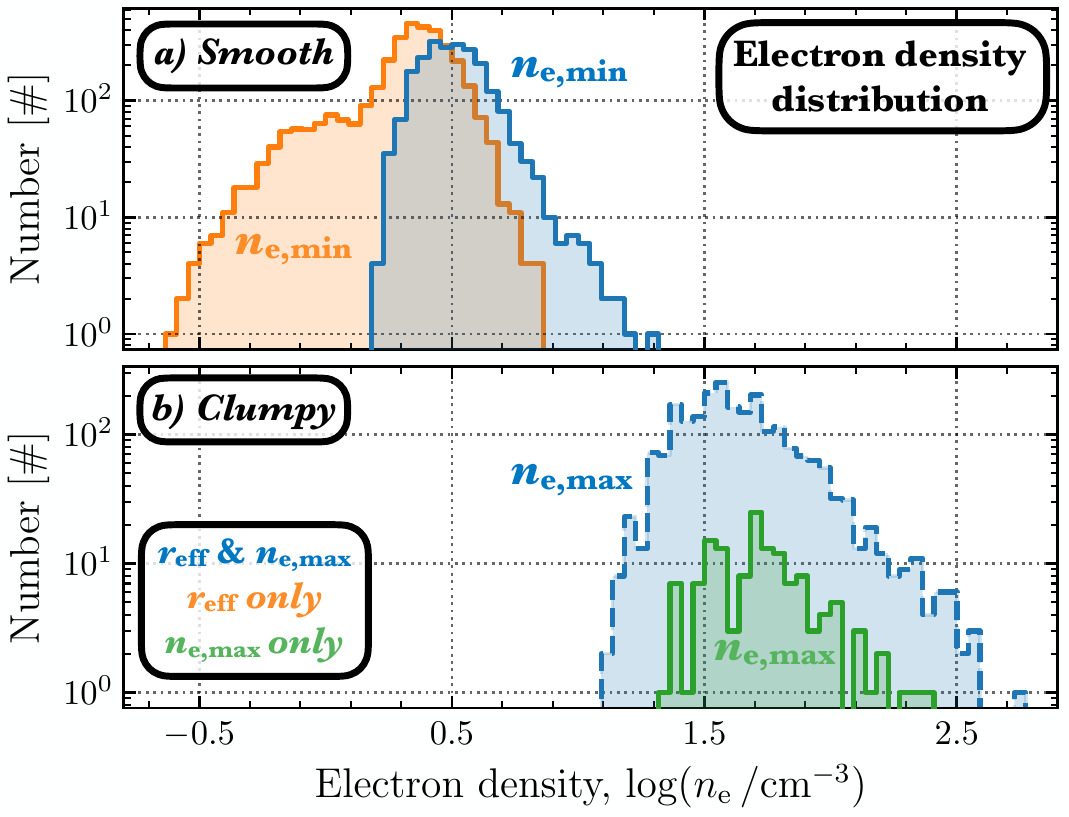}
    \caption{Histograms of the electron density ($n_\mathrm{e}$) distribution for the \HII\ region sample for assumptions of a \emph{a) smooth} and \emph{b) clumpy} unresolved density distribution (see Figure\,\ref{fig:toyfig}). {\em Upper panel:} The blue (solid line) and green histograms show the \nemax\ distributions for \HII\ regions that are statistically distinguishable from the low-density limit (see Figure\,\ref{fig:hist_ne}). The blue and green histograms differentiate the samples with resolved and unresolved sizes (see Section\,\ref{subsubsec:reff}). {\em Lower panel:} We derive an $n_\mathrm{e,min}$ for the resolved \HII\ regions using the measured effective radius ($r_\mathrm{eff}$) and extinction-corrected \Ha\ flux ($F_\mathrm{H\upalpha}$; see Section\,\ref{sec:lowerlimits}). We show the distribution of $n_\mathrm{e,min}$ as orange and blue (dashed line) histograms. Note that the \HII\ regions that are both resolved and are below the low-density limit have both \nemax\ and \nemin\ estimates, and, hence, the blue histograms have the same sample size.}
    \label{fig:hist_ne}
\end{figure}

The histogram distribution of allowed \nemax\ measurements below the low-density limit are shown in Figure\,\ref{fig:hist_ne}. Here we differentiate those that are resolved and have both \nemax\ and \reff\ measurements (in blue), and those that are unresolved and have only \nemax\ measurements (see Section\,\ref{subsubsec:reff}). We find that the \ne~range across these samples of \HII\ regions is $13$\,cm$^{-3}$ to $537$\,cm$^{-3}$ (median: $40.5$\,cm$^{-3}$), respectively. In Figure\,\ref{fig:hist_ne}, we also show the distribution of the (minimum) electron densities (\nemin), which are determined using a different method for the sample of resolved sources (see Section\,\ref{sec:lowerlimits}). These values of the electron density are similar to those determined from other IFU studies of resolve \HII\ regions within nearby galaxies (e.g.\ NGC 628: \citealp{Rousseau-Nepton2018}; NGC300: \citealp{mcleod_20}). We note, however, that they sit at the lower end of the values estimated from lower spectral and angular resolution studies (e.g. \citealp{Sanchez2012, Sanchez2015, Espinosa-Ponce2020}). The comparison between resolved and unresolved studies presents an interesting avenue for future investigations.

Overall, out of our initial sample of 23,699 \HII\ regions, a total of 5810 ($\sim 25$\%) have measurements of either \reff\ or \nemax. $5669$ ($98$\%) of these regions have sizes above our resolution limit and hence have valid \reff\ measurements, whilst $2379$ ($41$\%) have densities large enough to allow us to distinguish \Rsii\ from its low density limiting value, thereby allowing us to determine \nemax\ for these regions. Finally, we have both measurements for a total of $2238$ \HII\ regions (see Table\,\ref{tab:galobsprops} for a summary). We remind the reader that these measurements correspond to our two limiting cases (see Figure\,\ref{fig:toyfig}): the ionised gas is \emph{smoothly} distributed throughout the measured volume of the \HII\ region (i.e.\ \reff), or the ionised gas is \emph{clumpy} and fills only some fraction of the \HII\ region close to the source (i.e.\ \nemax). Note that the requirement of a resolved size or accurate electron density measurement, biases these samples to the brightest and largest \HII\ regions within each galaxy. In Section\,\ref{sec:lowerlimits}, we use the estimates of maximum sizes, \reff\ (or electron densities, \nemax), of the \HII\ regions to place lower limits on the electron densities, \nemin\ (or sizes, \rmin); or in other words, we also determine the electron density and sizes for both our assumptions of a \emph{smooth} and \emph{clumpy} ionised gas density distribution. To do so, however, we must first outline how the ionising photon rate within each \HII\ region is calculated. 

\subsection{Ionisation rate -- $Q$}
\label{subsec:ionphot}

To calculate the ionisation rate ($Q$), we first calculate the \Ha\ luminosity from $L_{\Ha} = 4 \pi D^2 F_{\Ha}$, where $F_{\Ha}$ is the extinction-corrected \Ha\ flux computed by \citet{Santoro2021} and $D$ is the distance to each galaxy given in Table\,\ref{tab:galprops}. The extinction correction is computed by measuring the reddening using the H$\alpha$/H$\beta$ ratio measured for each \HII\ region and applying a correction assuming the \cite{ODonnell1994} reddening law with $R_V = 3.1$ and a theoretical H$\alpha$/H$\beta = 2.86$. For optically thick nebulae \citep[case~B recombination;][]{Osterbrock2006} at $\Te = 10{,}000$\,K, the ionisation rate is given as $Q \approx L_{\Ha} (\upalpha_\mathrm{B}/\upalpha^\mathrm{eff}_{\Ha}) / (h\nu_{\Ha}) = L_{\Ha} / (0.45 h\nu_{\Ha})$, where the total recombination coefficient of hydrogen is $\upalpha_\mathrm{B} = 2.59\times10^{-13}$\,cm$^3$s$^{-1}$, the effective recombination coefficient (i.e. the rate coefficient for recombinations resulting in the emission of an \Ha\ photon)  is $\upalpha^\mathrm{eff}_{\Ha} \approx 1.17\times10^{-13}$\,cm$^3$s$^{-1}$ \citep{Osterbrock2006}, $\nu_{\Ha}$ is the frequency of the \Ha\ emission line and $h$~is the Planck constant. 

We find that $Q$~ranges from $10^{49.0 - 52.8}$\,s$^{-1}$ across the sample of \HII\ regions with both $\nemax$ and $\reff$ measurements. We find that $Q$~ranges from $10^{49.1 - 51.6}$\,s$^{-1}$ across the $\nemax$\,\emph{only} sample, and $10^{47.6 - 51.0}$\,s$^{-1}$ across the $\reff$\,\emph{only} sample. The fact that we recover systematically lower~$Q$ values for the 
$\reff$\,\emph{only} sample is easily understood: the \HII\ regions that are weaker in \Ha\ emission ($Q \propto F_{\Ha}$) are also weaker in the \SIIa\ and \SIIb\ emission lines, causing larger errors on the \Rsii\ ratio, making it harder to distinguish \Rsii\ in these regions from the low-density limit. 
Finally, note that the value of $Q$ we derive for each \HII\ region does not depend on the escape fraction of ionising photons from that \HII\ region ($f_{\rm esc}$), since $Q$ here refers only to the ionisation rate of gas within the \HII\ region.

\subsection{Minimum radii and electron densities -- \rmin\ and \nemin}
\label{sec:lowerlimits}

For the \HII\ regions with sizes greater than the resolution limit, the measured \reff\ represents an estimate of their maximum extent. However, as mentioned previously, in cases where the \HII\ region is clumpy and the clumps are close to the ionising source, the average distance of the clumps from the source is a more appropriate measure of the \HII\ region size from the point of view of understanding its dynamics. Our observations do not have sufficient resolution to allow us to measure this distance directly. However, for regions where we have a measure of the electron density from the \sii doublet, we can put a lower limit on this size, which we hereafter denote as \rmin. 
We can do this because the extinction-corrected \Ha\ luminosity of an \HII\ region is determined by three quantities: the electron temperature (measured as explained in the previous section), the root-mean-squared density of the gas and the volume of the \HII\ region. If we assume that the root-mean-squared density is the same as the density we measure from \sii, then we can straightforwardly solve for the volume, and hence the size of the region if we approximate it as a sphere. The rms density could of course be lower than our \sii-derived density if the \sii-bright clumps fill only a small fraction of the volume, but is unlikely to be larger than this value. The estimate of the \HII\ region size that we get from this argument is therefore a lower limit on the true size, complementing the upper limit we get from \reff. Note also that we can also derive a value of \rmin\ for unresolved \HII\ regions for which we cannot measure an accurate \reff, so long as we have a measure of \nemax\ for these regions. Finally, we can also apply the same logic to derive an estimate of the minimum density of our resolved \HII\ regions, \nemin, by fixing the volume and solving for the density.

Our expressions for \rmin\ and \nemin\ are therefore simply:
%
\begin{equation}
\begin{gathered}
 \rmin = \left(\frac{1}{\nemax^2} \frac{3Q}{4\pi\alpha_\mathrm{B}(\Te)}\right)^{1/3}, \\
 \nemin = \left(\frac{1}{\reff^3} \frac{3Q}{4\pi\alpha_\mathrm{B}(\Te)}\right)^{1/2},
 \end{gathered}
 \label{equ:recomb}
\end{equation}
where $Q$ is the previously determined ionisation rate and $\alpha_\mathrm{B}$ is the case~B recombination coefficient. For $\alpha_\mathrm{B}$, we use the following accurate fit from \citet{Hui97}, based on \citet{Ferland92}:
\begin{equation}
\alpha_\mathrm{B}(\Te) = \frac{2.753\times10^{-14} \, (315\,614 / \Te)^{1.5}} {\left[1.0 + (115\,188 / \Te)^{0.407}\right]^{2.242}}~,
\end{equation}
where \Te\ is the electron temperature (in units of Kelvin). We use estimates of \Te\ from the nitrogen auroral lines where available, and otherwise assume a representative value of $\Te = 8000$\,K (Belfiore et al. in prep). Varying this representative electron temperature between $5000$ and $15{,}000$\,K only causes a factor of ${\sim}\sqrt{2}$ difference in the estimated sizes and densities. From Equation\,\eqref{equ:recomb}, these limits on density and radius can be related by
\begin{equation}
 \left(\frac{\reff}{r_\mathrm{min}}\right)^{3} = \frac{V_\mathrm{max}}{V_\mathrm{min}} = \left(\frac{n_\mathrm{e,max}}{n_\mathrm{e,min}}\right)^{2}, 
 \label{equ:volume}
\end{equation}
where $V_\mathrm{max}$ and $V_\mathrm{min}$ are the minimum and maximum volumes.

The values of \rmin\ we derive from this approach range from a few parsecs to a few tens of parsecs, as illustrated in Figure\,\ref{fig:hist_radius}. For the resolved regions, they are typically around a factor of ten smaller than \reff. We also see that the values of \rmin\ that we derive for resolved \HII\ regions are generally larger than those we derive for unresolved regions. This is a consequence of the \HII\ region size-luminosity relationship: larger \HII\ regions tend to also be brighter, and hence the minimum volume of dense gas required to produce their observed \Ha\ luminosities is larger. 

In Table\,\ref{tab:galobsprops}, we list the number of \HII\ regions in each galaxy for which we can derive both \rmin\ and \nemin\ (2238 regions in total), only \nemin\ (3431 regions) or only \rmin\ (141 regions). 

\begin{figure*}
    \centering
	\includegraphics[width=\textwidth]{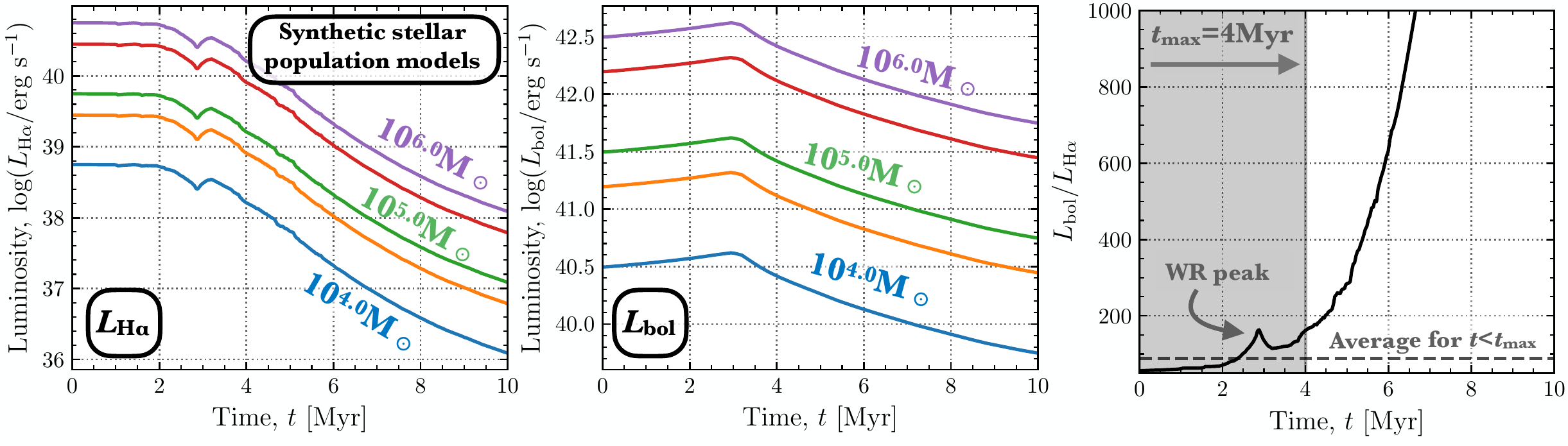}
    \caption{Early time evolution of the luminosity from the \textsc{starburst99} model \citep{Leitherer1999}. \emph{Left and centre panels:} We show the \Ha\ luminosity ($L_{\Ha}$) and the bolometric luminosity ($L_\mathrm{bol}$) for cluster with masses of $M_\star = 10^{4}$ to $10^{6}$\,\sol,in steps of 0.5\,dex. \emph{Right panel:} We show the ratio of $L_\mathrm{bol}/L_{\Ha}$, which is the same for all cluster masses. We label the average $L_\mathrm{bol}/L_{\Ha} \sim 88$ between $t=0$\,Myr and $t_\mathrm{max} = 4$\,Myr, where $t_\mathrm{max}$ is the time for $L_{\Ha}$ to drop by half an order of magnitude from the zero-age main sequence (see Section\,\ref{sec:sb99_lbol}). We also label the peak in the mass loss rate seen at ${\sim}3.5$\,Myr that corresponds to the time at which winds are the most effective \citep{Leitherer1999, Rahner2017, Rahner2019}, as the most massive O~stars are in their Wolf--Rayet phase but have not yet exploded as supernovae. The estimate of $L_\mathrm{bol}$ from $L_{\Ha}$ for each \HII\ region is used in the calculation of the direct radiation pressure (\pdir; see Section\,\ref{sec:dir_prescalc}).}
    \label{fig:sb99_lum}
\end{figure*}

\subsection{Stellar population models}
\label{sec:sb99}

Lastly, we require a final set of properties for our \HII\ region sample before determining their internal pressure terms, which we obtain from synthetic stellar population modelling. Namely, in this section we estimate their bolometric luminosity ($L_\mathrm{bol}$), cluster mass ($M_\mathrm{cl}$), mass loss rate ($\dot M$) and mechanical luminosity ($L_\mathrm{mech}$). We employ the \textsc{starburst99} model \citep{Leitherer1999},\footnote{\url{http://www.stsci.edu/science/starburst99/docs/default.html}} adopting the default parameter set and varying the cluster mass between $10^4$ to $10^6$\,\sol. Of note within the default parameter set, we use the Evolution wind model \citep{Leitherer1992} for the calculation of the wind power, the mass-loss rates from the Geneva models with no rotation, an instantaneous star formation burst populating a with Kroupa initial mass function (IMF; \citealp{kroupa_2001}). We assess the evolution of the ionisation, (wind) feedback power and mass-loss across a time range of ${\sim}0$ to $10$\,Myr (i.e.\ requiring the \textsc{quanta}, \textsc{snr}, \textsc{power}, \textsc{yield}, \textsc{spectrum}, \textsc{ewidth} outputs from \textsc{starburst99}).

\subsubsection{Bolometric luminosity -- $L_\mathrm{bol}$}

\label{sec:sb99_lbol}

We wish to estimate the bolometric luminosity of the cluster(s) responsible for the \HII\ regions in order to determine the radiation pressure (see Section\,\ref{sec:dir_prescalc}). In Figure\,\ref{fig:sb99_lum}, we show the bolometric luminosity and \Ha\ luminosity as a function of time for a range of cluster masses (also see \citealp{Agertz2013}), computed with the assumption that no ionising photons escape from the \HII\ region (i.e. that $f_{\rm esc} = 0$). We see that the bolometric luminosity remains relatively constant across the first $10$\,Myr (varying by only ${\sim}1$\,dex), whereas the \Ha\ luminosity drops significantly after ${\sim}2$\,Myr (${\sim}3$\,dex). Since our \HII\ region sample is constructed from observed \Ha\ emission, by definition they must have lifetimes less than the lifetime of ionising radiation. We then consider the maximum age ($t_\mathrm{max}$) as the time when $L_{\Ha}$ has dropped significantly from the zero-age value ($L_{\Ha}(t_0)$). We set $t_\mathrm{max}$ at $4$\,Myr, where $L_{\Ha}$ has decreased by half an order of magnitude (a~factor of around~3). This $t_\mathrm{max}$ includes the zero-age main sequence (ZAMS) and Wolf--Rayet phases of the high-mass stars (as labelled on Figure\,\ref{fig:sb99_lum}). We note that the choice of $t_\mathrm{max}$ is somewhat arbitrary; however, we do not see any appreciative change in our results by increasing it to, e.g., $5$\,Myr, where $L_{\Ha}$ has dropped by an order of magnitude, or decreasing it to, e.g., $0$\,Myr, to only include the ZAMS. 

In the right panel of Figure\,\ref{fig:sb99_lum}, we show the ratio of the bolometric to \Ha\ luminosity. We find an average value of $L_\mathrm{bol}/L_{\Ha} \approx 88$ between $t_0 = 0$\,Myr and $t_\mathrm{max} \sim 4$\,Myr. To check if this is reasonable, we compare to other estimates of $L_\mathrm{bol}/L_{\Ha}$. Firstly, we can derive a robust lower limit on $L_\mathrm{bol}$ by assuming that the only contribution to it is from the ionising radiation of the high-mass stars. In that case, $L_\mathrm{bol} = \langle h\nu \rangle Q$, where $\langle h\nu \rangle \sim 15$\,eV is a reasonable estimate for the mean energy of an ionising photon, and $Q$~is the ionisation rate derived in Section~\ref{subsec:ionphot} above. This gives a lower limit to the conversion factor of $L_\mathrm{bol}/L_{\Ha} \approx 18$. Secondly, there is the conversion presented by \citet{kennicutt_2012}, which accounts for a stellar population that fully samples the IMF and the stellar age distribution. This conversion is given as $L_\mathrm{bol}/L_{\Ha} \approx 138$. As the \HII\ regions in this work are assumed to be relatively young (given their bright \Ha\ emission), we expect the correct conversion to be somewhere between these two estimates. Hence, $L_\mathrm{bol}/L_{\Ha} \approx 88$ is reasonable for our sample, and is used throughout this work to estimate the $L_\mathrm{bol}$ for each \HII\ region.

Finally, we note that although we assume here that $f_{\rm esc} = 0$ for simplicity, we know that in reality some ionising photons will escape from the \HII\ regions into the diffuse ISM. Estimates of the average value of $f_{\rm esc}$ for a population of \HII\ regions vary \citep[see e.g.\ the discussion in][]{chevance20c} but are typically in the range $f_{\rm esc} = 0.3$--0.6, and so accounting for this would increase our estimates of $L_\mathrm{bol}$ by around a factor of 2. In practice, the impact on the radiation pressure will be smaller than this, since the photons that escape from the \HII\ region obviously do not contribute to the radiation pressure, and so we feel justified in neglecting this complication in our current study.

\subsubsection{Cluster mass, mass loss rate and mechanical luminosity -- $M_\mathrm{cl}, \dot M \text{ and } L_\mathrm{mech}$}
\label{sec:sb99_wind}

\begin{figure*}
    \centering
	\includegraphics[width=\textwidth]{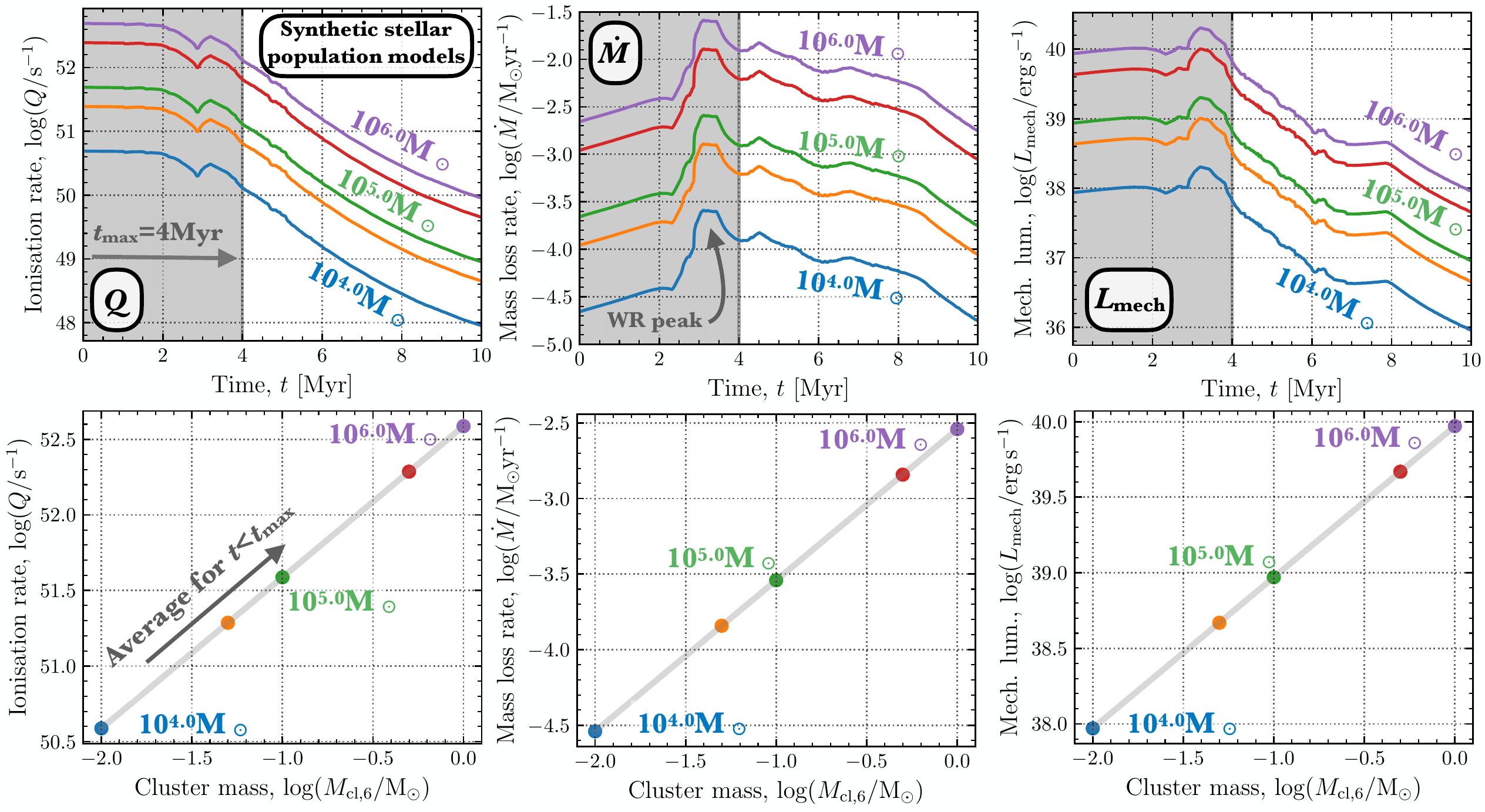}
    \caption{\emph{Upper panels (left to right):} Early time evolution of the ionisation rate ($Q$), mass loss rate ($\dot M$), and mechanical luminosity ($L_\mathrm{mech}$) from the \textsc{starburst99} model \citep{Leitherer1999} for cluster with masses of $M_\mathrm{cl} = 10^{4}$ to $10^{6}$\,\sol. The shaded region represents $t < t_\mathrm{max}$, where $t_\mathrm{max}$ is the time for $L_{\Ha}$ to drop by half an order of magnitude from the zero-age main sequence (see Section\,\ref{sec:sb99_lbol}). We label the peak in the mass loss rate seen at ${\sim}3.5$\,Myr that corresponds to the time at which winds are the most effective \citep{Leitherer1999, Rahner2017, Rahner2019}, as the most massive O~stars are in their Wolf--Rayet phase but have not yet exploded as supernovae. \emph{Lower panels (left to right):} The average $Q$, $\dot M$ and $L_\mathrm{mech}$ between $t=0$\,Myr and $t_\mathrm{max} = 4$\,Myr plotted as a function of the cluster mass ($M_\mathrm{cl,6} = M_\mathrm{cl}/10^6$\,\sol). We use the measured $Q$ for each \HII\ region to infer $M_\mathrm{cl}$. We then use this $M_\mathrm{cl}$ to estimate $L_\mathrm{mech}$ and $\dot M$, which are used in the calculation of the wind ram pressure (\pwind; see Section\,\ref{sec:wind_prescalc}).}
    \label{fig:sb99_mass}
\end{figure*}

We investigate how changing the cluster mass ($M_\mathrm{cl}$) of the \textsc{starburst99} model varies the ionisation rate~($Q$). We can then use the measured ionisation rate of each \HII\ region in the catalogue to estimate its cluster mass, which we then also use to estimate its mass loss rate~($\dot M$) and mechanical luminosity~($L_\mathrm{mech}$; also see e.g. \citealp{Dopita2005, Dopita2006}). 
The upper left panel of Figure\,\ref{fig:sb99_mass} shows $Q$ as a function of time~($t$) for a range of $M_\mathrm{cl}$. Similarly to $L_{\Ha}$, we see that $Q$ is higher for higher $M_\mathrm{cl}$, and suffers a strong decrease after ${\sim}4$\,Myr. As before, we then average $Q$ within a time of~$0$ to~$4$\,Myr. This time averaged~$Q$ is then plotted as a function of the $M_\mathrm{cl,6} = M_\mathrm{cl} / 10^6$\,\sol\ in the lower left panel of Figure\,\ref{fig:sb99_mass}. Plotted in log--log space, we see the relation is linear, with a constant of $Q / M_\mathrm{cl,6} = 10^{52.5}$~(s$^{-1}/$\sol). We use this conversion factor with the estimate of $Q$ (Section\,\ref{subsec:ionphot}) to determine $M_\mathrm{cl}$ for each of the \HII\ regions in our sample (see Figure\,\ref{fig:histprops}). As $Q$ is directly estimated from the observed $L_\mathrm{H\alpha}$ emission, we also outline that $L_\mathrm{H\alpha} / M_\mathrm{cl,6} = 10^{40.5}$~(erg~s$^{-1}/$\sol) and, for completeness, $L_\mathrm{bol} / M_\mathrm{cl,6} = 10^{42.5}$~(erg~s$^{-1}/$\sol). 

In the upper central and right panels of Figure\,\ref{fig:sb99_mass}, we show the time evolution of the mass loss rate ($\dot M$) and mechanical luminosity ($L_\mathrm{mech}$), respectively. We see that $\dot M$ has an overall increase relative to its zero-age main sequence when averaged over the shown timescale of $10$\,Myr, which is in contrast to the sharp declines seen in the $L_{\Ha}$, $M_\mathrm{cl}$ and $L_\mathrm{mech}$. The peak seen at ${\sim}3.5$\,Myr corresponds to the time at which winds are the most energetic \citep{Leitherer1999, Rahner2017, Rahner2019}, as the most massive O~stars are in their Wolf--Rayet phase but have not yet exploded as supernovae. The time-averaged $\dot M$ and $L_\mathrm{mech}$ are shown as a function of $M_\mathrm{cl}$ in the lower centre and right panels, respectively. We see that $\dot M/M_\mathrm{cl,6} = 10^{-2.3}$\,(\sol\,yr$^{-1}$/\sol) and $L_\mathrm{mech}/M_\mathrm{cl,6} = 10^{40.0}$\,(erg\,s$^{-1}$/\sol). We use these conversion factors to estimate $L_\mathrm{mech}$ and $\dot M$ for each \HII\ region within the sample, which are used in the following section to estimate the wind ram pressure (see Section\,\ref{sec:wind_prescalc}).

\section{Pressure calculation}\label{sec:prescalc}

\subsection{Internal pressure components}\label{sec:int_prescalc}

In this section, we will place quantitative observational constraints on the main feedback mechanisms driving the expansion of our large sample of \HII\ regions. We will use these constraints to then examine if the feedback mechanisms differ with evolutionary timescale. In this section, we will also identify the local environmental conditions surrounding the \HII\ regions. By contrasting the internal and external properties of the \HII\ region, we will investigate different dependencies on initial and current environmental conditions. To do so, we first determine the components of the internal pressure within an \HII\ region (see also \citealp{lopez_2011, lopez_2014, Pellegrini2011, mcleod_2019, barnes20b,olivier2020}). 

In this work, we consider three pressure terms that can be determined from our catalogue of \HII\ regions:   
\begin{enumerate}
\item thermal gas pressure (\ptherm),
\item direct radiation pressure (\pdir),\footnote{The direct radiation pressure studied in this work does not account for trapping, as we do not have access to high enough resolution infrared observations to probe dust reprocessed emission.}
\item wind ram pressure (\pwind).\footnote{Here we do not consider the hot X-ray emitting gas pressure produced via shocks from strong winds, as we do not have access to adequate $\sim$\,0.1 to 1\,keV X-ray observations for all our galaxies. Nonetheless, it is worth noting that this was found to be sub-dominant on larger scales \citep{lopez_2011,lopez_2014}.}
\end{enumerate}
We then assume that the total internal pressure of an \HII\ region is equal to the sum of these three components $\ptot = \ptherm+\pdir+\pwind$; i.e.\ assuming all components act independently and combine constructively to create a net positive internal pressure. The calculation of these various internal pressure components is outlined in this section. Note that throughout this work we will refer to the pressure terms in units of \Kcmcb\ or e.g.\ $P/k_\mathrm{B}$ (where $P/k_\mathrm{B}\,[\text{\Kcmcb}] = P/1.38 \times 10^{-16}$\,[dyn\,cm$^{-2}$]). 

The following calculations are simplistic in the sense that they do not account for the leaking of radiation or material into the diffuse ionised gas \citep[][Belfiore et al. in prep]{Kim_JG2019}, and the cancellation of radiation forces from distributed sources \citep[e.g.][]{Kim2018}, which may act to reduce our calculated pressures \citep{chevance20c}. In addition, we cannot constrain the unresolved density distribution within the ionised gas \citep[e.g.][]{Kado-Fong2020}, yet, in the previous section, we have placed limits on the various physical properties for the \HII\ regions assuming that they have a \emph{smooth} or \emph{clumpy} density profile (see Figure\,\ref{fig:toyfig}). Throughout this next section, we continue to use these two simple assumptions when calculating the various pressure terms. For each \HII\ region, we define a maximum pressure (\pmax) calculated for the smallest volume (i.e.\ using \nemax\ and \rmin), and a minimum pressure (\pmin) calculated for the largest volume (i.e.\ using \nemin\ and \reff).

\subsubsection{Direct radiation pressure -- \pdir}
\label{sec:dir_prescalc}

The intense radiation field produced by the young stellar populations within \HII\ regions can exert large pressure on the surrounding material. This direct radiation pressure is related to the change in momentum of the photons produced by the stellar population. Hence, it is directly proportional to their total bolometric luminosity (\Lbol), assuming that all of the luminosity is absorbed once (see e.g.\ \citealp{Krumholz2009,draine_2011} for discussion of radiative trapping effects, and see \citealp{Reissl2018} for a multifrequency radiative transfer calculation of the spectral shifting as stellar radiation travels through the gas). The volume-averaged direct radiation pressure (\pdir) is then given as \citep[e.g.][]{lopez_2011},
\begin{equation}
\pdir/k_\mathrm{B} = \frac{3 \Lbol}{4 \pi r^{2} c k_\mathrm{B}}~,
\label{eq:Pdir}
\end{equation}
where \Lbol\ is the bolometric luminosity (see Section\,\ref{sec:sb99_lbol}). In Equation\,\eqref{eq:Pdir}, we use $r=\reff$ (Section\,\ref{subsubsec:reff}) for a measure of the minimum direct radiation pressure ($P_\mathrm{rad,min}$) and the minimum radius $r=\rmin$ (Section\,\ref{sec:lowerlimits}) for a measure of the maximum ($P_\mathrm{rad,max}$; i.e., due to the $\pdir \propto \reff^{-2}$ dependence). Equation\,\eqref{eq:Pdir} refers to the volume-averaged pressure, which is appropriate here as this work aims at understanding the large-scale dynamics of the \HII\ regions (e.g.\ the total energy and pressure budget for each source; see e.g.\ \citealp{barnes20b}), as opposed to the force balance at the surface of an empty shell \citep[see][]{mcleod_2019}. 

\subsubsection{Wind (ram) pressure -- \pwind}
\label{sec:wind_prescalc}

In their early evolutionary stages, high-mass stars can produce strong stellar winds that can result in mechanical pressure within \HII\ regions. The pressure from these winds has been inferred directly \citep[e.g.][]{mcleod_2019, mcleod_20} or indirectly (e.g.\ from shock heated gas; \citealp{lopez_2011, lopez_2014, olivier2020}) for several \HII\ regions within the literature. Here, we determine the ram pressure of winds for our \HII\ region sample, i.e.\ the pressure exerted on the shell due to momentum transfer from the wind. While the classical energy-conserving solution of \citet{weaver_1977} would produce much higher pressure, recent theory and numerical simulations show that mixing at the interface between hot and cool gas leads to strong cooling \citep{Lancaster2021a, Lancaster2021b}, though the effect could be diminished in the presence of magnetic fields \citep{Rosen2021}. As a consequence, the pressure is within a factor of a few of the input ram pressure of the wind. The wind ram pressure is thus calculated as, 
\begin{equation}
\pwind/k_\mathrm{B} = \frac{3 \dot M v_\mathrm{wind}}{4\pi r^{2}k_\mathrm{B}}~,
\label{eq:Pwind}
\end{equation}
where $\dot M$ is the mass loss rate (Section\,\ref{sec:sb99_wind}) and $v_\mathrm{wind}$ is the wind velocity. The wind velocity is calculated as,
\begin{equation}
v_\mathrm{wind} = \left( \frac{2 L_\mathrm{mech}}{\dot M} \right)^{0.5}\sim2500 \,\mathrm{km~s^{-1}}~.
\label{eq:vwind}
\end{equation}
where $L_\mathrm{mech}$ is the mechanical luminosity (Section\,\ref{sec:sb99_wind}). Again, we use $r=\reff$ (Section\,\ref{subsubsec:reff}) for the minimum wind ram pressure ($P_\mathrm{wind,min}$) and the minimum radius $r=\rmin$ (Section\,\ref{sec:lowerlimits}) for the maximum wind ram pressure ($P_\mathrm{wind,max}$; i.e., due to the $\pwind \propto \reff^{-2}$ dependence).

\subsubsection{Thermal gas pressure -- \ptherm}
\label{sec:therm_prescalc}

The young high-mass stars (${>}8$\,\sol) produce a large flux of hydrogen ionising Lyman continuum photons, which maintain the high ionisation fraction observed within \HII\ regions. The photoionised gas is heated by the stellar population to temperatures typically within the range of $5000$ to $15{,}000$\,K. The thermal pressure of this ionised gas is set by the ideal gas law, 
\begin{equation}
\ptherm/k_\mathrm{B} = (\ne+n_\mathrm{H}+n_\mathrm{He}) \approx 2\ne \Te~, 
\label{eq:ptherm}
\end{equation}
where the factor of~2 comes from the assumption that all He is singly ionised. We determine \ptherm\ using values of the electron temperature (\Te) determined from the nitrogen auroral lines or, where not available, we adopt a representative value of $\Te = 8000$\,K. Here, we use the maximum electron density (\nemax) determined using the sulphur line ratio (i.e.\ at $\rmin$; Section\,\ref{subsubsec:electrondensity}) for the maximum thermal pressure ($P_\mathrm{therm,max}$) and the minimum \nemin\  (i.e.\ at \reff; Section\,\ref{sec:lowerlimits}) for the minimum thermal pressure ($P_\mathrm{therm,min}$; i.e.\ due to the $\ptherm \propto \ne$ dependence). 

\subsection{External (dynamical) pressure components}
\label{sec:ext_prescalc}

In this section, we outline the method used to calculate the external pressure components acting against the internal pressures (outlined above), to confine the \HII\ regions and limit their expansion. To do so, we use the dynamical equilibrium pressure (\pde), which is an indirect measurement of the ambient pressure consisting of the sum of thermal, turbulent, magnetic pressure, and the ambient radiation and cosmic rays \citep[see e.g.][]{Kim2013,Kim2015b}. The most simplistic `classic' form of \pde\ includes the gas self-gravity and the weight of the gas in the potential well of the stars, and is commonly adopted within the literature \citep[e.g.][]{spitzer42,elmegreen89,elmegreen94,gallagher18a,schruba19,Barrera-Ballesteros2021a}. 

In this work, we use the dynamical equilibrium pressure calculated in a set of kpc-sized hexagonal apertures covering each galaxy's sky footprint. We take these values of \pde\ directly from \citet{sun20}, and provide a short summary of how these measurements are calculated below. These authors estimate \pde\ as,
\begin{equation}
    \pde = \dfrac{\pi G}{2} \Sigma_\mathrm{gas,1kpc}^2 + \Sigma_\mathrm{gas,1kpc} \sqrt{2G\rho_\mathrm{*,1kpc}} \sigma_{\mathrm{gas,z}},
    \label{equation:PDE}
\end{equation}
where the first term is the weight due to the self-gravity of the ISM disk and the second term is the weight of the ISM due to stellar gravity in the limit that the gas layer's scale height is smaller than that of the stellar disk \citep[e.g.][]{spitzer42, elmegreen89, Wong2002,Blitz2004, ostriker10}.  
In this equation, $\Sigma_\mathrm{gas,1kpc}=\Sigma_\mathrm{mol,1kpc} +\Sigma_\mathrm{atom,1kpc}$ is the total gas surface density, $\rho_\mathrm{*,1kpc}$ is the stellar mass volume density near the disk midplane and $\sigma_{\text{gas,z}}$ is the velocity dispersion of the gas perpendicular to the disk. 

The kpc-scale molecular gas surface density, $\Sigma_\mathrm{mol,1kpc}$, is calculated from the CO\,(2–1) intensity $I_\mathrm{CO, 1kpc}$ from the PHANGS-ALMA survey \citep{Leroy2021a}, assuming a constant CO\,(2–1)/(1-0) ratio of 0.7 \citep{denBrok2021}, and the metallicity=dependent CO-to-H$_2$ conversion factor ($\alpha_\mathrm{CO}$) described in \citet{sun20}. Radial metallicity measurements were estimated using the galaxy mass-metallicity relation reported by \citet{Sanchez2019}, and a universal radial metallicity gradient \citep{Sanchez2014}. The kpc-scale atomic gas surface density, $\Sigma_\mathrm{atom,1kpc}$, is calculated from the HI 21 cm line intensity $I_\mathrm{HI, 1kpc}$ using data from the PHANGS-VLA project (PI: D.~Utomo), the EveryTHINGS project (PI: K.~Sandstrom), as well as existing data from VIVA \citep{Chung2009}, THINGS \citep{Walter2008}, and VLA observations associated with HERACLES \citep{leroy_2013}. The kpc-scale stellar mass surface density, $\Sigma_\mathrm{*,1kpc}$, is calculated from the (dust-corrected) 3.6\micron\ specific surface brightness $I_\mathrm{3.6, 1kpc}$ from {\em Spitzer}, assuming a mass-to-light ratio of 0.47 \citep{McGaugh2014}. All surface densities were corrected for the projection effect from the galaxy inclination.

The stellar mass volume density is given as (e.g. \citealp{blitz06,Leroy2008,ostriker10}),
\begin{equation}
    \rho_\mathrm{*,1kpc} = \frac{\Sigma_\mathrm{*,1kpc}}{4H_\mathrm{*}} = \frac{\Sigma_\mathrm{*,1kpc}}{0.54R_\mathrm{*}},
    \label{}
\end{equation}
where $H_\mathrm{*}$ is the stellar disk scale height, and $R_\mathrm{*}$ is the radial scale length of the stellar disk from the S4G photometric decompositions of the {\em Spitzer} 3.6\,\micron\ images \citep{Salo2015}. The first part of this equation assumes an isothermal density profile along the vertical direction; $\rho_*(z) \propto \mathrm{sech}^2 [z/(2H_*)]$ \citep{vanderKruit1988}, and the second part assumes a fixed stellar disk flattening ratio $R_\mathrm{*}/H_\mathrm{*} = 7.3$ (\citealp{Kregel2002, sun20}).

The velocity dispersion of the gas perpendicular to the disk is given as the mass-weighted average velocity dispersion of molecular and atomic phases,
\begin{equation}
    \sigma_{\mathrm{gas,z}} = f_\mathrm{mol} \left< \sigma_{\mathrm{mol,\theta\mathrm{pc}}} \right>_\mathrm{1kpc} + (1-f_\mathrm{mol})\sigma_{\mathrm{atom}}
    \label{}
\end{equation}
where $f_\mathrm{mol}$ is the fraction of gas mass in the molecular phase. \citet{sun20} adopt a fixed atomic gas velocity dispersion $\sigma_{\mathrm{atom}} = 10$\kms\ (e.g. \citealp{Leroy2008}), which we also use here.


\citet{sun20} provide the values of \pde\ averaged within kpc-sized apertures for 12 of the 19 galaxies studied in this work. Of the 7 galaxies without \pde\ measurements, NGC1365, NGC1433, NGC1512, NGC166, NGC1672, and NGC7496 have no available high-resolution \HI\ observations, and IC5332 lacks any significant CO\,(2-1) emission in the PHANGS-ALMA data \citep{Leroy2021a}.

In this work, we want to compare the estimated internal pressure in each \HII\ region to this kpc-scale estimate of \pde. We note, however, that multiple \HII\ regions could be located in the same kpc-sized aperture, in which case the \pde\ values used for such comparison are identical. Moreover, we note that these kpc-scale estimates do not account for the smaller scale density fluctuations on the scales of the \HII\ regions. \citet{sun20} did introduced a modified, cloud-scale dynamical equilibrium pressure, \pdess\ (their equation\,15), which treats the clumpy molecular ISM and diffuse atomic ISM separately, allowing them to have a different geometry \citep[also see e.g.][]{ostriker10, schruba19}. 
\citet{sun20} find that the \pdess\ range between factors of 2 to 10 higher than \pde. Either  \pde\ or \pdess\ could be an appropriate estimate of the ambient pressure of an \HII\ region depending on its location; if embedded inside a cloud, then \pdess\ may be relevant (i.e.\ akin to the initial conditions of the \HII\ region), yet if outside a cloud (i.e.\ a more evolved state), \pde\ would be more appropriate. As the \HII\ regions studied here have been identified from \Ha\ emission, they are not highly obscured, and therefore are most likely not embedded within molecular clouds. Whilst this is true for the majority of cases, there is a known cross-over between the \Ha\ emitting phase and the embedded phase, which typically corresponds to around a third of the total \Ha\ emitting lifetime \citep[e.g.][]{Kim2021}. The effect of the local environment and the initial conditions of the \HII\ regions will be assessed in detail in future work, and here we adopt \pde\ for the external dynamical pressure.

\section{Pressures Comparison}\label{sec:prescomp}

\subsection{Global variations in the pressure components}

\begin{figure*}
    \centering
	\includegraphics[width=1.8\columnwidth]{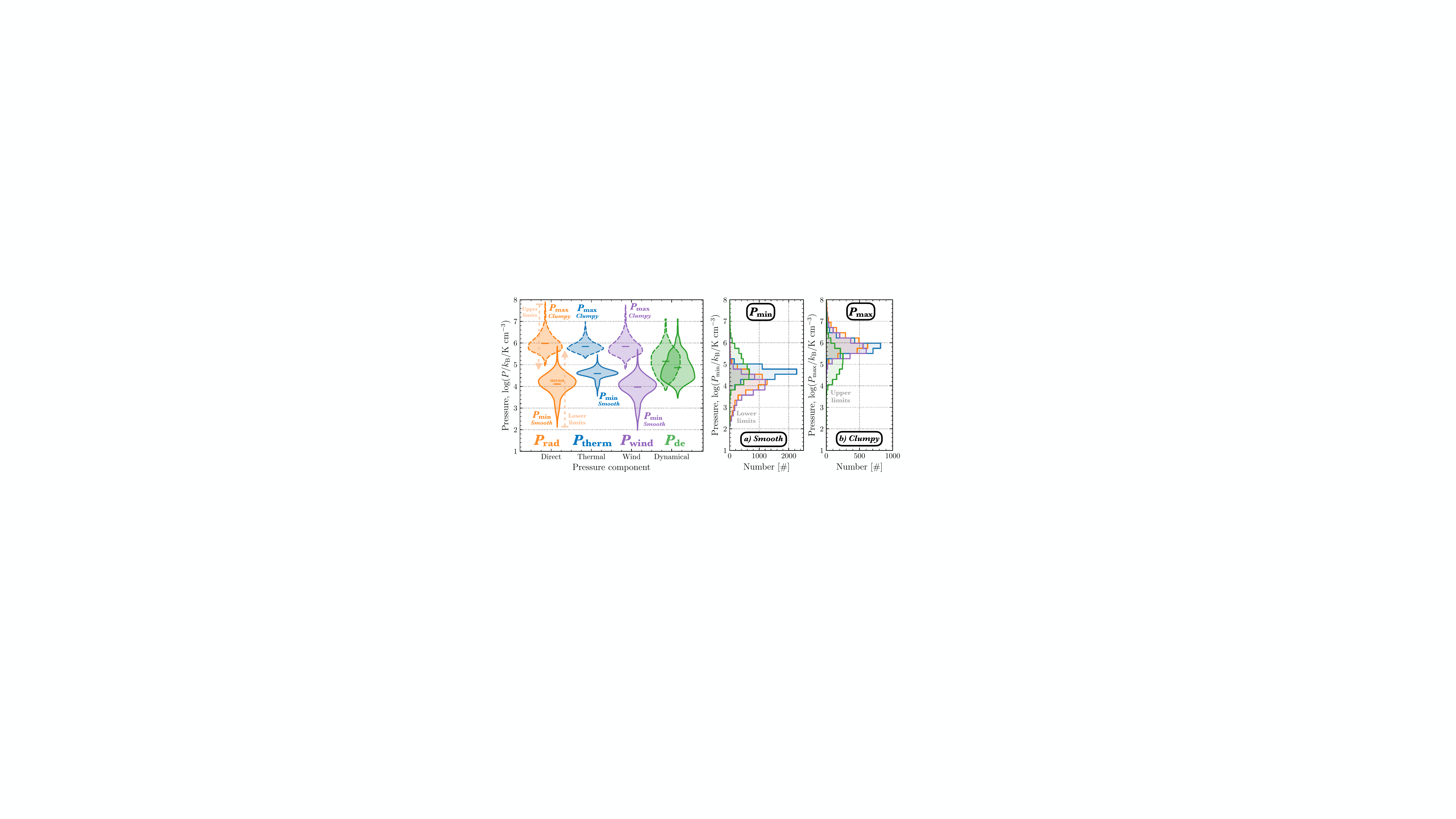}
    \caption{Distribution of the various pressure components determined across the \HII\ region sample for all galaxies for assumptions of a \emph{a) smooth} and \emph{b) clumpy} unresolved density distribution (see Figure\,\ref{fig:toyfig}). We show the internal pressure components of the direct radiation pressure (\pdir) in orange, thermal pressure from the ionised gas (\ptherm) in blue, and ram pressure from the wind (\pwind) in purple. We show the distribution of the external confining pressure or dynamical pressure (\pde) in green. {\em Left panel:} The KDE distributions are represented as violin plots, where the mean value of each distribution is highlighted by a horizontal bar and the width of each distribution corresponds to the logarithmic scale histogram distribution. The distributions for both the maximum pressure limit (\pmax) calculated for the smallest volume, and minimum pressure limit (\pmin) calculated for the largest volume are shown with dashed and solid outlines, respectively. The dashed, faded arrows highlight that \pmax\ and \pmin\ represent upper and lower limits on the internal pressure. {\em Right panels:} We show the histogram distributions for \pmin\ and \pmax\ separately.}
    \label{fig:pres_hist}
\end{figure*}

In this section, we compare the global variations of the pressure components across and between the galaxies in our sample. In Figure\,\ref{fig:pres_hist}, we show the total distribution of each pressure component for all galaxies. In the violin plot (left panel), we use a kernel density estimation (KDE) to compute the smoothed distribution for both the minimum (\pmin) and maximum (\pmax) pressure limits.\footnote{All the kernel density distributions used in this work are based on $200$ points to evaluate each of the Gaussian kernel density estimations, and Scott's Rule is used to calculate the estimator bandwidth \citep{scott15}.} In the right two panels of Figure\,\ref{fig:pres_hist}, we compare the histogram distributions for \pmin\ and \pmax\ separately. We list the mean and standard deviation of each \pmin\ and \pmax\ pressure component across the whole galaxy sample in Table\,\ref{tab:meanpressures}.

The first thing to note in Figure\,\ref{fig:pres_hist} is the one to two orders of magnitude difference between the mean values of \pmin\ and \pmax. To understand this difference, we outline here how the $\pmax/\pmin$ are related. For radiation and winds, $\pmax/\pmin = (\reff/\rmin)^2 = (\nemax/\nemin)^{4/3}$, and from Figure\,\ref{fig:hist_ne} we see that there is around $1.5$\,dex difference between the centres of the \nemin\ and \nemax\ distributions. This would then translate to the two orders of magnitude ratio of $\pmax/\pmin$ for direct radiation and winds seen in Figure\,\ref{fig:pres_hist} ($1.8$\,dex given in Table\,\ref{tab:meanpressures}). For the thermal pressure, $P_\mathrm{therm,max}/P_\mathrm{therm,min} = \nemax/\nemin$, which then would also be consistent with the $1.2$\,dex difference observed in Figure\,\ref{fig:pres_hist} (also see Table\,\ref{tab:meanpressures}).

\begin{table*}

    \centering
    \caption{Mean maximum (\pmax) and minimum (\pmin) pressures of the \HII\ regions across the galaxy sample (Section\,\ref{sec:prescalc}). The \pmax\ and \pmin\ are determined by assuming a \emph{clumpy} or \emph{smooth} unresolved density distribution for each \HII\ region, respectively (see Figure\,\ref{fig:toyfig}). We show in columns from left to right the galaxy name, the direct radiation pressure (\pdir), the thermal pressure (\ptherm), the wind pressure (\pwind), the dynamical pressure (\pde). The over-pressure, or fraction of total internal pressure divided by the external pressure ($\ptot/\pde$), is given in the final column. Several galaxies were not included in the sample from \citet{sun20} due to no available \HI\ observations or due to the lack of significant CO emission (e.g. IC\,5332), and hence have no \pde\ measurement. We determine the mean and standard deviation of each pressure component after taking the logarithm of the values. Note that pressures are in units of $\log(P/k_\mathrm{B}) = \log(\mathrm{K\,cm^{-3}})$.}
    \label{tab:meanpressures}

\begin{tabular}{ccccccccccc}
\hline\hline
 & \multicolumn{5}{c}{Maximum pressure (\pmax) -- {\em clumpy} -- log($\mathrm{K\,cm^{-3}}$)} & \multicolumn{5}{c}{Minimum pressure (\pmin) -- {\em smooth} -- log($\mathrm{K\,cm^{-3}}$)} \\ 
Galaxy & $P_\mathrm{rad}$ & $P_\mathrm{therm}$ & $P_\mathrm{wind}$ & $P_\mathrm{de}$ & $P_\mathrm{tot}/P_\mathrm{de}$ & $P_\mathrm{rad}$ & $P_\mathrm{therm}$ & $P_\mathrm{wind}$ & $P_\mathrm{de}$ & $P_\mathrm{tot}/P_\mathrm{de}$ \\
 \hline
All galaxies & 6.0$\pm$0.4 & 5.9$\pm$0.2 & 5.8$\pm$0.4 & 5.2$\pm$0.6 & 1.2$\pm$0.5 & 4.1$\pm$0.5 & 4.6$\pm$0.2 & 4.0$\pm$0.5 & 4.9$\pm$0.6 & 0.0$\pm$0.5 \\
IC5332 & 5.7$\pm$0.3 & 5.8$\pm$0.2 & 5.6$\pm$0.3 & -- & -- & 3.5$\pm$0.4 & 4.4$\pm$0.2 & 3.3$\pm$0.4 & -- & -- \\
NGC0628 & 5.8$\pm$0.3 & 5.8$\pm$0.2 & 5.7$\pm$0.3 & 4.6$\pm$0.3 & 1.7$\pm$0.4 & 3.9$\pm$0.4 & 4.6$\pm$0.2 & 3.8$\pm$0.4 & 4.5$\pm$0.3 & 0.2$\pm$0.3 \\
NGC1087 & 5.9$\pm$0.4 & 5.7$\pm$0.2 & 5.7$\pm$0.4 & 5.1$\pm$0.5 & 1.2$\pm$0.4 & 4.2$\pm$0.4 & 4.6$\pm$0.2 & 4.1$\pm$0.4 & 5.0$\pm$0.4 & -0.1$\pm$0.3 \\
NGC1300 & 5.9$\pm$0.3 & 5.8$\pm$0.2 & 5.8$\pm$0.3 & 4.8$\pm$0.8 & 1.5$\pm$0.6 & 3.9$\pm$0.6 & 4.4$\pm$0.3 & 3.7$\pm$0.6 & 4.6$\pm$0.7 & 0.3$\pm$0.6 \\
NGC1365 & 6.1$\pm$0.5 & 5.9$\pm$0.3 & 6.0$\pm$0.5 & -- & -- & 4.3$\pm$0.7 & 4.6$\pm$0.3 & 4.1$\pm$0.7 & -- & -- \\
NGC1385 & 5.9$\pm$0.3 & 5.8$\pm$0.2 & 5.8$\pm$0.3 & 5.2$\pm$0.4 & 1.1$\pm$0.4 & 4.1$\pm$0.5 & 4.6$\pm$0.2 & 4.0$\pm$0.5 & 5.0$\pm$0.4 & -0.1$\pm$0.3 \\
NGC1433 & 5.9$\pm$0.3 & 5.8$\pm$0.2 & 5.7$\pm$0.3 & -- & -- & 3.8$\pm$0.5 & 4.4$\pm$0.3 & 3.6$\pm$0.5 & -- & -- \\
NGC1512 & 6.0$\pm$0.4 & 5.9$\pm$0.3 & 5.9$\pm$0.4 & -- & -- & 3.8$\pm$0.6 & 4.3$\pm$0.3 & 3.6$\pm$0.6 & -- & -- \\
NGC1566 & 6.0$\pm$0.3 & 5.9$\pm$0.2 & 5.9$\pm$0.3 & -- & -- & 4.2$\pm$0.4 & 4.6$\pm$0.2 & 4.1$\pm$0.4 & -- & -- \\
NGC1672 & 6.1$\pm$0.6 & 5.9$\pm$0.3 & 6.0$\pm$0.6 & -- & -- & 4.4$\pm$0.5 & 4.6$\pm$0.2 & 4.2$\pm$0.5 & -- & -- \\
NGC2835 & 5.7$\pm$0.3 & 5.8$\pm$0.2 & 5.5$\pm$0.3 & 4.4$\pm$0.1 & 1.8$\pm$0.3 & 3.9$\pm$0.4 & 4.5$\pm$0.2 & 3.8$\pm$0.4 & 4.3$\pm$0.1 & 0.4$\pm$0.2 \\
NGC3351 & 6.4$\pm$0.7 & 6.2$\pm$0.4 & 6.2$\pm$0.7 & 5.4$\pm$1.3 & 1.5$\pm$0.9 & 3.9$\pm$0.5 & 4.5$\pm$0.2 & 3.7$\pm$0.5 & 4.3$\pm$0.8 & 0.4$\pm$0.6 \\
NGC3627 & 6.1$\pm$0.4 & 5.9$\pm$0.2 & 6.0$\pm$0.4 & 5.6$\pm$0.5 & 0.9$\pm$0.5 & 4.5$\pm$0.4 & 4.8$\pm$0.2 & 4.4$\pm$0.4 & 5.3$\pm$0.6 & -0.2$\pm$0.5 \\
NGC4254 & 6.0$\pm$0.3 & 5.9$\pm$0.2 & 5.8$\pm$0.3 & 5.4$\pm$0.4 & 1.0$\pm$0.4 & 4.3$\pm$0.4 & 4.7$\pm$0.2 & 4.1$\pm$0.4 & 5.3$\pm$0.5 & -0.3$\pm$0.4 \\
NGC4303 & 6.0$\pm$0.3 & 5.8$\pm$0.2 & 5.9$\pm$0.3 & 5.2$\pm$0.4 & 1.2$\pm$0.4 & 4.3$\pm$0.3 & 4.7$\pm$0.1 & 4.2$\pm$0.3 & 5.1$\pm$0.4 & -0.2$\pm$0.4 \\
NGC4321 & 6.2$\pm$0.4 & 6.0$\pm$0.3 & 6.0$\pm$0.4 & 5.1$\pm$0.7 & 1.5$\pm$0.6 & 4.3$\pm$0.5 & 4.6$\pm$0.2 & 4.1$\pm$0.5 & 4.9$\pm$0.7 & -0.0$\pm$0.5 \\
NGC4535 & 6.1$\pm$0.4 & 5.9$\pm$0.3 & 5.9$\pm$0.4 & 4.7$\pm$0.6 & 1.8$\pm$0.6 & 3.9$\pm$0.5 & 4.6$\pm$0.2 & 3.8$\pm$0.5 & 4.4$\pm$0.4 & 0.3$\pm$0.4 \\
NGC5068 & 5.4$\pm$0.4 & 5.7$\pm$0.3 & 5.3$\pm$0.4 & 4.4$\pm$0.2 & 1.6$\pm$0.4 & 3.8$\pm$0.4 & 4.6$\pm$0.2 & 3.6$\pm$0.4 & 4.4$\pm$0.2 & 0.3$\pm$0.3 \\
NGC7496 & 5.9$\pm$0.3 & 5.8$\pm$0.2 & 5.8$\pm$0.3 & -- & -- & 4.0$\pm$0.5 & 4.5$\pm$0.2 & 3.9$\pm$0.5 & -- & -- \\
\hline\hline
\end{tabular}
\end{table*}

We now compare the relative difference between the various pressure components considering either their maximal (\pmax) or minimal (\pmin) values. In Figure\,\ref{fig:pres_hist}, we see that the maximum internal pressures are all relatively similar, with mean values of around $\pmax/k_\mathrm{B} \sim 10^{6}$\,\Kcmcb. On the other hand, the minimum values appear relatively different, with the direct radiation and wind pressures having values around $\pmin/k_\mathrm{B} \sim 10^{4}$\,\Kcmcb\ and the thermal pressures being around a factor of~$4$ higher ($P_\mathrm{therm,min}/k_\mathrm{B} \sim 10^{4.6}$\,\Kcmcb). Comparing to the external dynamical pressure, we see that that typically $\pmin < \pde < \pmax$. Interestingly, we see that values of \pde\ determined towards those \HII\ regions with \pmax\ measurements are slightly ($0.3$\,dex) higher than towards those with \pmin\ measurements.
 
Lastly, in Figure\,\ref{fig:pres_hist}, we compare the external pressure (\pde) associated with each \HII\ region. Note that as each \HII\ region may not have both \pmin\ and \pmax\ measurements (see Table\,\ref{tab:galobsprops}), the differences in the \pde\ distributions are caused by these different samples rather than a difference in the \pde\ measurement method (section\,\ref{sec:ext_prescalc}). In addition, several galaxies were not included in the sample from \citet{sun20} due to no available \HI\ observations or due to the lack of CO significant emission (e.g. IC\,5332), and hence have no \pde\ measurement. We see that the mean of our \pde\ distribution is higher by ${\sim}0.5$\,dex than that shown in Fig.\,1 of \citet{sun20}. This is due to the fact that the majority of \HII\ regions within our samples are identified towards the spiral arms and centres of the galaxies, which have systematically higher \pde\ values than the galaxy averages. This can be seen in the upper right panel of Figure\,\ref{fig:maps_pres1}, which shows the \pde\ apertures taken from \citet{sun20} overlaid with the \HII\ region sample within NGC\,4321. 
 
 \begin{figure*}
    \centering
	\includegraphics[width=\textwidth]{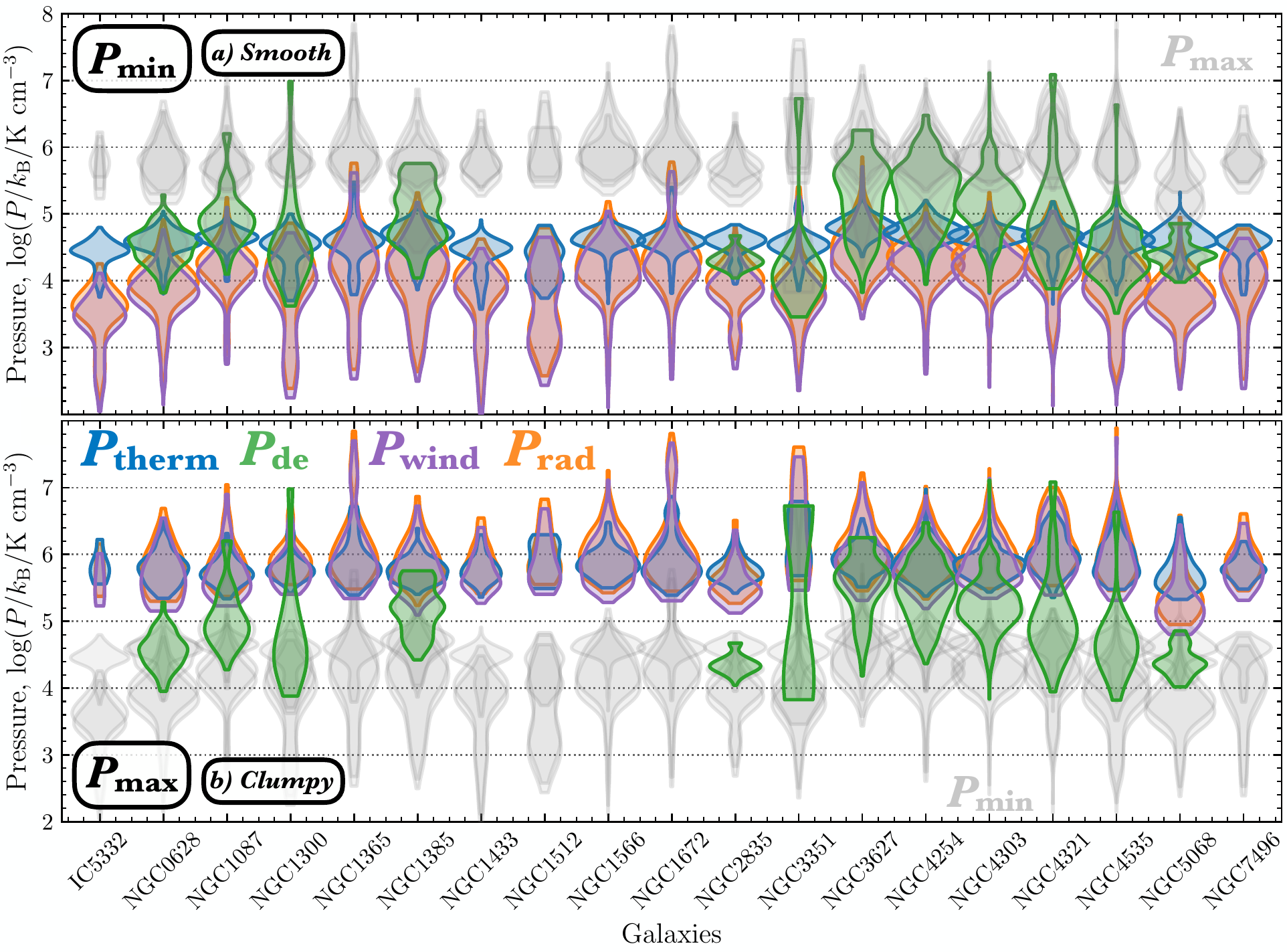}
    \caption{Distribution of the various pressure components determined across the \HII\ region sample within individual galaxies for assumptions of a \emph{a) smooth} and \emph{b) clumpy} unresolved density distribution (see Figure\,\ref{fig:toyfig}). We show the internal pressure components of the direct radiation pressure in orange (\pdir), thermal pressure from the ionised gas in blue (\ptherm), and ram pressure from the wind in purple (\pwind). Note that several galaxies were not included in the sample from \citet{sun20} due to no available \HI\ observations or due to the lack of significant CO emission (e.g. IC\,5332), and hence have no \pde\ measurement. We show the distribution of the external confining pressure, or dynamical pressure (\pde), in green. The KDE distributions are represented by violin plots, where the width of each distribution corresponds to the logarithmic scale histogram distribution. {\em Upper panel:} The distributions for the maximum pressure limit (\pmax) are shown in colour, and minimum pressure limit (\pmin) in faded grey. {\em Lower panel:} The distributions for the \pmin\ are shown in colour, and \pmax\ in faded grey.}
    \label{fig:pres_histindv}
\end{figure*}
 
In Figure\,\ref{fig:pres_histindv}, we show the pressure components determined within the \HII\ regions for each galaxy. The upper and lower panels show the separate distributions for \pmax\ and \pmin, respectively. Here, we again see that all the pressure terms are similar for \pmax, yet for \pmin, \ptherm\ are consistently larger than \pdir\ and \pwind. Moreover, we see that in general $\pmin < \pde < \pmax$ for individual galaxies. We do not see any significant deviations from these trends within the individual galaxies, and are careful to compare the distributions between galaxies given the systematic biases of our \HII\ sample. We list the mean and standard deviation of each pressure component for individual galaxies in Table\,\ref{tab:meanpressures}.

\subsection{Pressure components as a function of size and position}

\begin{figure*}
    \centering
	\includegraphics[width=0.85\textwidth]{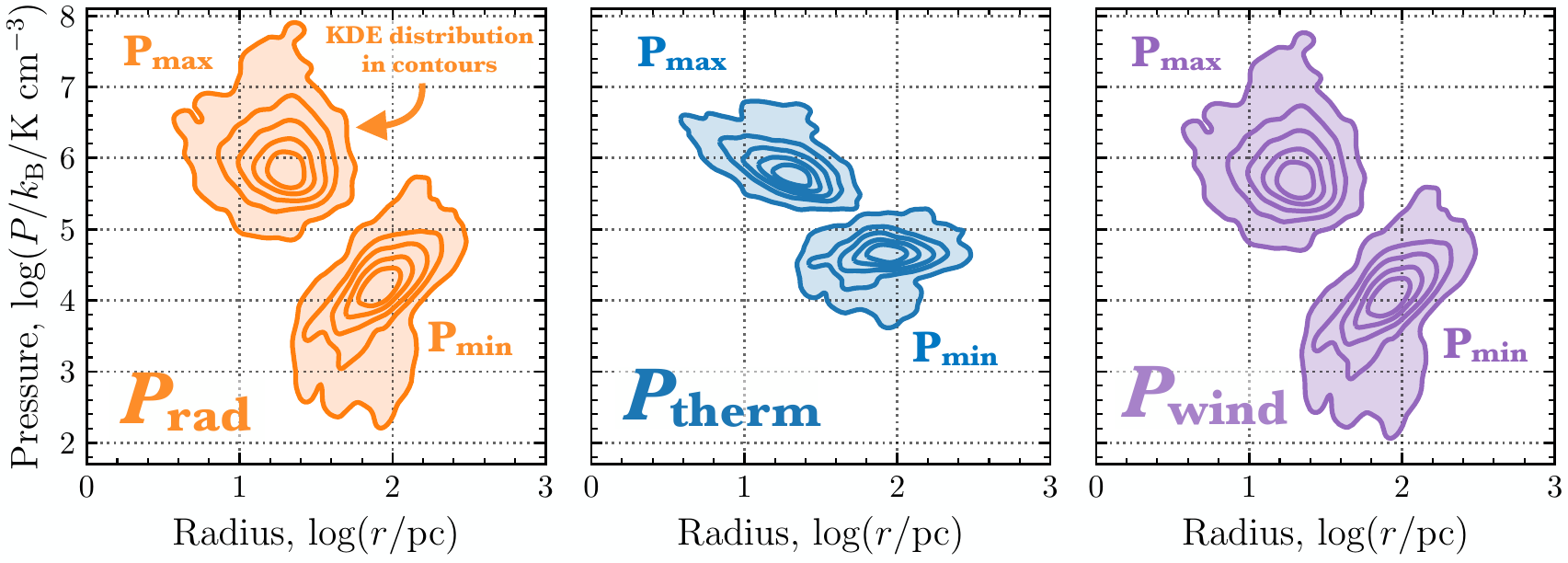}
    \caption{Pressure components as a function of the radius (log--log axes). We show the internal pressure components of the thermal pressure from the direct radiation pressure (\pdir), the ionised gas (\ptherm), and wind ram pressure (\pwind) as orange, blue and green contours, respectively ({\em left to right panels}). The contours show the Gaussian KDE distribution in levels that include 99, 90, 75, 50, and 25 per cent of the points. We show the distributions for both the maximum pressure limit (\pmax) calculated for the smallest volume, and minimum pressure limit (\pmin) calculated for the largest volume (as labelled).}
    \label{fig:scat_Pcompreff_all}
\end{figure*}

\begin{figure*}
    \centering
	\includegraphics[width=1\textwidth]{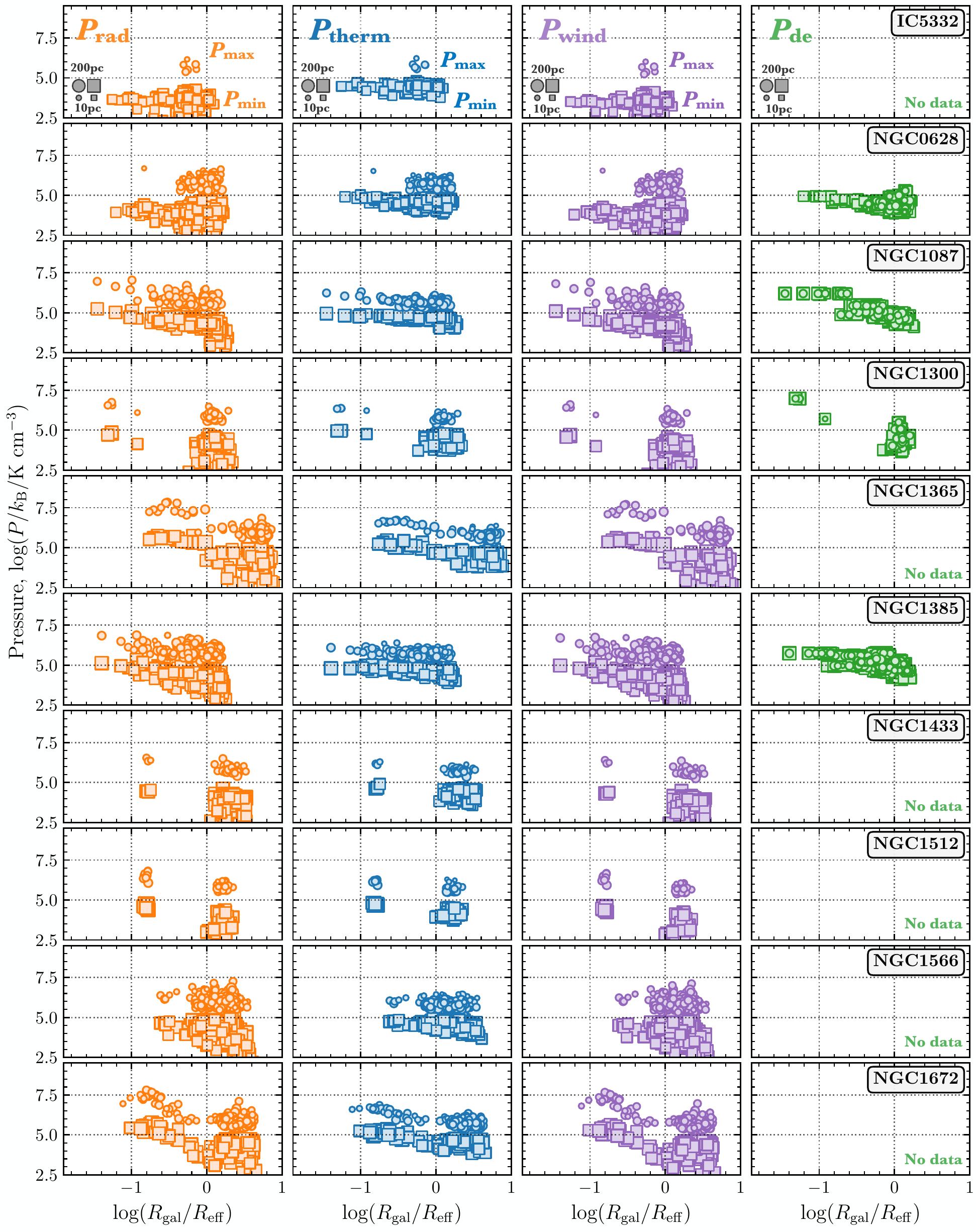} \vspace{-0.5cm}
    \caption{Pressure components as a function of the galactocentric radius (normalised to the $R_\mathrm{eff}$ radius; see Table\,\ref{tab:galprops}) for each galaxy in the sample. We show the internal pressure components of the direct radiation pressure (\pdir), the thermal pressure from the ionised gas (\ptherm), and wind ram pressure (\pwind) as orange, blue and green points, respectively ({\em left to right panels}). We show the external confining pressure, or dynamical pressure (\pde), as the green points. The size of each point has been scaled to the radius of the \HII\ region, and corresponds to the scale shown in the lower left of each panel. We highlight the points that correspond to the maximum pressure limit (\pmax) and minimum pressure limit (\pmin; Section\,\ref{sec:int_prescalc}) as circles and squares, respectively.}
    \label{fig:scat_PcompRgal_all}
\end{figure*}

\begin{figure*}
    \centering
	\includegraphics[width=1\textwidth]{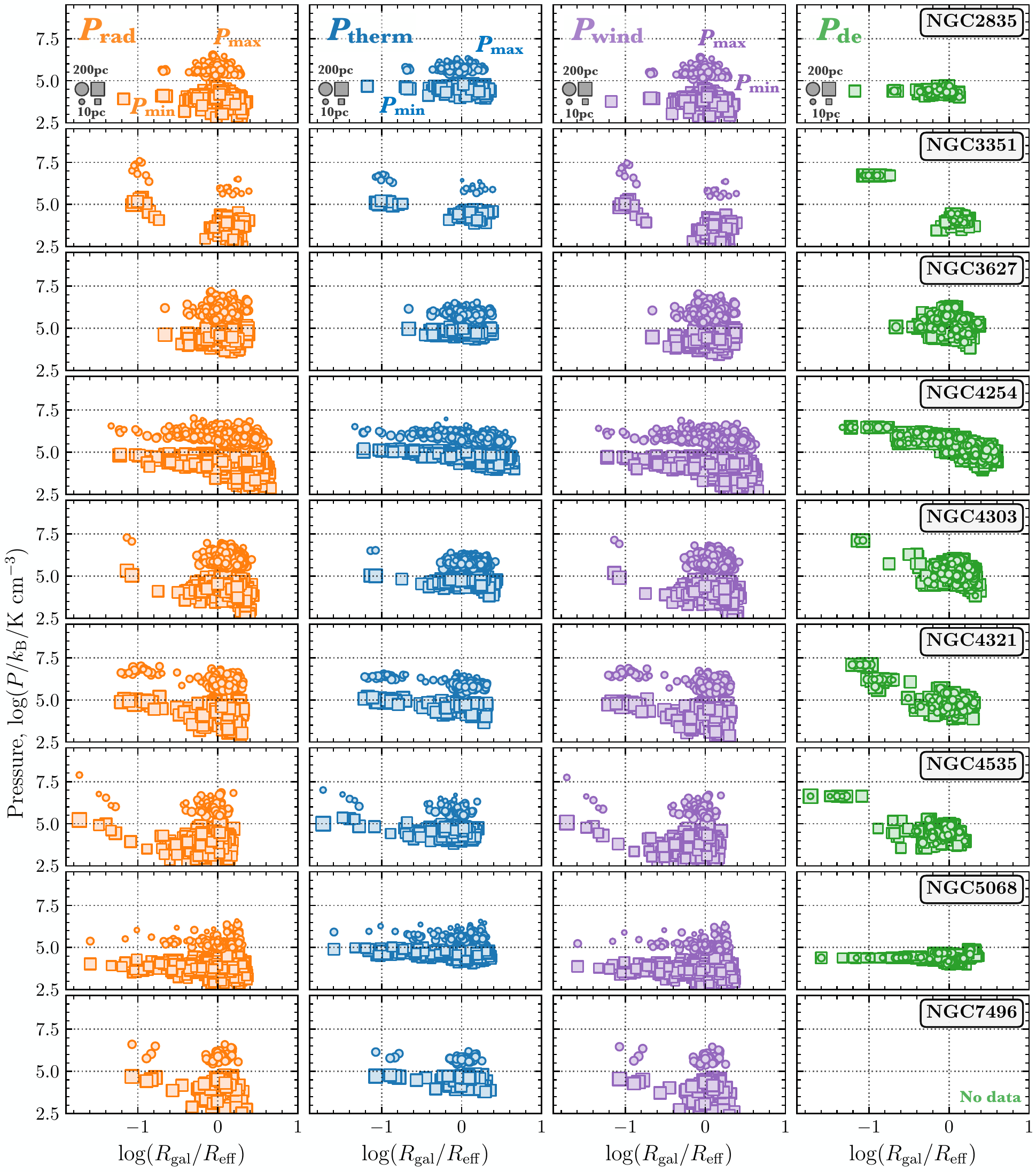}
    \contcaption{}
\end{figure*}

In this section, we assess how the various pressure components vary as a function of the sizes of the \HII\ regions and their position within the host galaxies. In Figure\,\ref{fig:scat_Pcompreff_all}, we show the minimum and maximum pressure limits for \pdir, \ptherm\ and \pwind\ as a function of the radius of the \HII\ regions, where \pmax\ is plotted at $\rmin$ and \pmin\ is at $\reff$. Due to the high density of individual measurements on this plot, we show the KDE distribution as contours that increase to include 99, 90, 75, 50, and 25 per cent of the data points of each \pmax\ or $\pmin$. 

The first thing to note in Figure\ \ref{fig:scat_Pcompreff_all} is that the distributions for \pdir\ and \pwind\ are very similar; albeit $\pdir \gtrsim \pwind$. This is due to the fact that they are both calculated using the \Ha\ emission and have the same radial dependence of $r^{-2}$ (Section\,\ref{sec:prescalc}). On the contrary, \ptherm\ uses the \ne\ calculated from the \sii\ line ratio, and the pressure calculation has no radial dependence, hence it is independent of \pdir\ and \pwind. The distribution of \ptherm\ in Figure\,\ref{fig:scat_Pcompreff_all} is therefore different to both \pdir\ and \pwind. 

It is worth quickly reviewing the biases within our \HII\ region samples before continuing the discussion of Figure\,\ref{fig:scat_Pcompreff_all} further. Firstly, we identified our initial sample of \HII\ regions using an automated algorithm on \Ha\ emission observations, which have a finite resolution and sensitivity. Hence, this means we could be missing detections of unresolved and weak \HII\ regions, or multiple compact and clustered \HII\ regions within complex environments (e.g. galaxy centres). Moreover, we may be missing lower surface brightness, larger \HII\ regions. In other words, our samples of \HII\ regions are biased to the brightest \HII\ regions across the galaxies (see section\,\ref{sec:sample}). Secondly, in determining \pmax\ or \pmin, we have split this initial sample into two sub-samples: those with \rmin\ and/or \reff\ measurements, respectively (see Table\,\ref{tab:galobsprops} for these sample sizes within each galaxy). In the case of \pmax, for example, this requires determination of \nemax, which is calculated from the \Rsii\ ratio (Section\,\ref{subsubsec:electrondensity}). Therefore, the distribution of \pmax\ does not include sources for which this ratio is statistically indistinguishable from the low density limit, and hence measurements of \pmax\ for larger, lower-density \HII\ regions will be missing. We are then cautious in drawing conclusions for the radial size dependence of the pressure terms due to the number of biases affecting the size distribution of our \HII\ region samples.

The above discussion may then explain the different dependencies we observe between \pmax\ or \pmin\ for each pressure component. We see that \pmax\ suffers a moderate decline with increasing \HII\ region radius, which one may expect as e.g. $\pdir \propto r^{-2}$ by definition (Section\,\ref{sec:dir_prescalc}). \pmin, on the other hand, shows an increase with increasing \HII\ region radius. The reason for this increase is not clear, but we speculate that this could be a result of the lack of either large, diffuse \HII\ regions or compact, clustered \HII\ regions in our sample.

In Figure\,\ref{fig:scat_PcompRgal_all}, we show how the pressure components for each galaxy vary as a function of the galactocentric radius normalised to $R_\mathrm{eff}$ (see Table\,\ref{tab:galprops}). In this figure, the size of each point is proportional to the \reff\ or \rmin\ of each \HII\ region. We see that our \HII\ region sample spans $\log(R_\mathrm{gal}/R_\mathrm{eff})$ of $-2\text{ to }1$, and hence covers a large range of galactic environments; from central molecular zones to outer edges of discs (see also Figure\,\ref{fig:rgb_main}). 

In the majority of galaxies, we see that the external pressure (\pde) shows a systematic increase by several orders of magnitude towards the centres. A~systematic increase in \pde\ is expected as the gas and stellar surface densities increase towards galaxy centres. However, notable exceptions are NGC\,2835 and NGC\,5068 that appear to have a relatively constant \pde\ across the whole disc (within a $1$\,dex scatter), which could be a result of these having lower than average atomic, molecular and stellar masses for the sample. Moreover, we see that NGC\,3627 has a large scatter \pde\ (around $4$\,dex) within the disc, which could be a result of the strong bar \citep[e.g.][]{Beslic2021}, or the strong ongoing tidal interaction in the Leo triplet (e.g. \citealp{Zhang1993}). Again, however, we caution any interpretation of the galaxy-to-galaxy variations seen here, given the systematic biases affecting the \HII\ region sample (section\,\ref{sec:sample} and \ref{sec:sampleselection}).  

Interestingly, we also see that both the internal \pmax\ or \pmin\ pressures generally show systematic increases towards galaxy centres (e.g. see NGC\,1365 and NGC\,4535), albeit with some significant scatter within discs (e.g. NGC\,1566). The increase in \pdir, \ptherm\ and \pwind\ towards centres is however smaller than the relative increase in \pde. For example, in the case of NGC\,4321, within the disc at $R_\mathrm{gal}/R_\mathrm{eff} = 1$, $\pde \sim 10^{4.5}$\,\Kcmcb\ and $P_\mathrm{rad,max} \sim 10^{6}$\,\Kcmcb, while near the centre at $R_\mathrm{gal}/R_\mathrm{eff} = 0.1$, $\pde \sim 10^{7.0}$\,\Kcmcb\ and $P_\mathrm{rad,max} \sim 10^{6.5}$\,\Kcmcb\ (see also NGC\,3351 and NGC\,4254). This is then a factor of ${\sim}300$ increase in \pde\ towards the centre, yet only a factor of ${\sim}3$ increase in $P_\mathrm{rad,max}$ (similarly minor relative increases are observed for \ptherm\ and \pwind, and the measurements for \pmin). 

It is not entirely clear why \HII\ regions should be more highly internally pressured within galaxy centres. For example, galaxy centres typically have higher metallicities, and hence cooling is more efficient within \HII\ regions and electron temperatures are lower. On the other hand, the more highly pressured environment causes higher local gas densities (see for electron density measurements e.g. \citealp{Herrera-Camus2016}). The latter could plausibly be an order of magnitude or more (i.e. effecting both \ptherm\ and \pdir), whilst the former is at most a factor of two ($\ptherm \propto \Te$). From an observational side, we have the most trouble getting consistent boundaries (and hence sizes) for the HII regions within the centres, as they're often quite clustered and sitting on a high diffuse ionized gas background \citep{Santoro2021}. If larger objects are preferentially identified in the centres (i.e. smaller regions merged into one larger region), then this would act to increase the relative difference in the pressure compared to the discs. One thing to note is that \rmin\ never gets below $\sim$\,10pc, but in the Galactic Centre \HII\ region sizes are at most a $\sim$\, few 1pc in size (e.g. \citealp{barnes20b}). In section\,\ref{subsec:future} we return how this can be addressed in future.

\subsection{Total internal pressure as a function of external pressure}

\begin{figure}
    \centering
	\includegraphics[width=0.9\columnwidth]{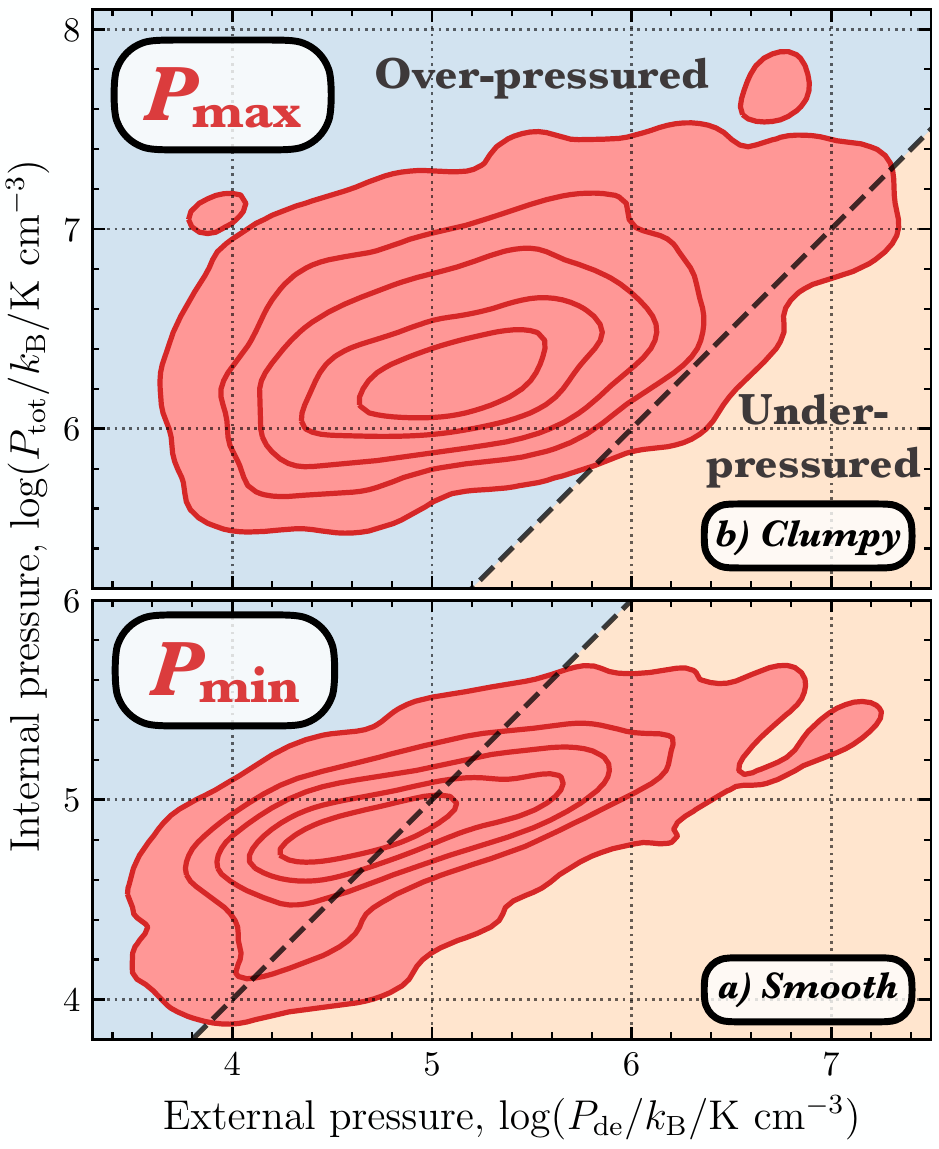}
    \caption{Total internal pressure ($\ptot = \ptherm+\pwind+\pdir$) of each \HII\ region as a function of its external or dynamical pressure (\pde) for assumptions of a \emph{a) smooth} and \emph{b) clumpy} unresolved density distribution (see Figure\,\ref{fig:toyfig}). The contours show the Gaussian KDE distribution in levels that include 99, 90, 75, 50, and 25 per cent of the points. Above the dashed line the \HII\ regions would be over-pressured (blue shaded), and below the \HII\ regions are under-pressured (orange shaded). {\em Upper panel:} We show the distributions for the maximum pressure limit (\pmax) calculated for the smallest volume. {\em Lower panel:} We show the minimum pressure limit (\pmin) calculated for the largest volume.}
    \label{fig:scat_PtotPDE}
\end{figure}

\begin{figure*}
    \centering
	\includegraphics[width=1\textwidth]{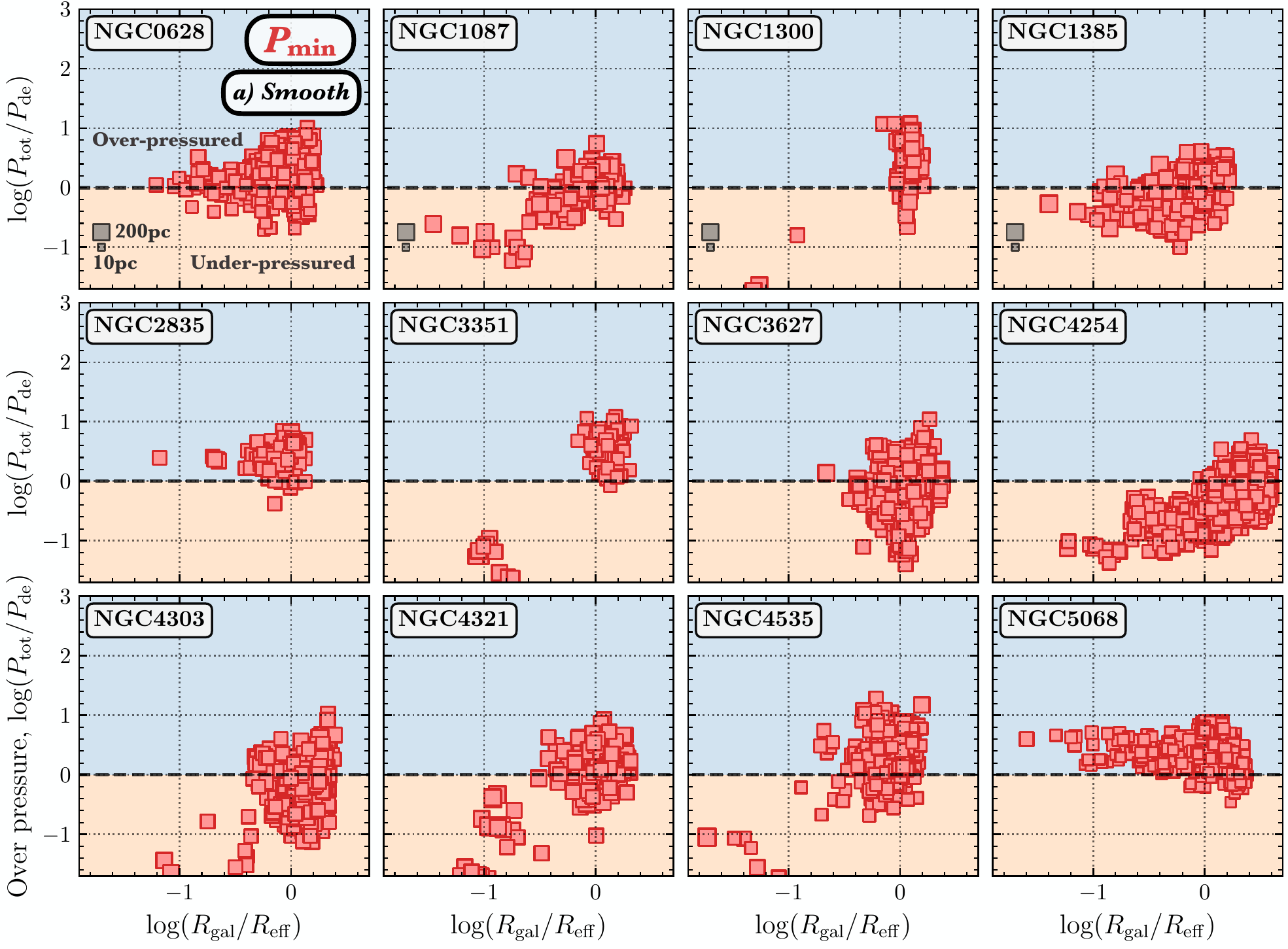}
    \caption{Over-pressure, $\ptot/\pde = (\ptherm+\pwind+\pdir)/\pde$, of each \HII\ region as a function of the galactocentric radius for each galaxy for assumptions of a \emph{a) smooth} and \emph{b) clumpy} unresolved density distribution (see Figure\,\ref{fig:toyfig}). We plot the galactocentric radius normalised to the $R_\mathrm{eff}$ radius (see Table\,\ref{tab:galprops}). The size of each point has been scaled to the radius of the \HII\ region, and corresponds to the scale shown in the lower left of each panel. The horizontal dashed black line shows where the external pressure is equal to the internal pressure. Above the dashed line the \HII\ regions would be over-pressured (blue shaded), and below the \HII\ regions are under-pressured (orange shaded).}
    \label{fig:scat_PtotPDEreff}
\end{figure*}

\begin{figure*}
    \centering
	\includegraphics[width=1\textwidth]{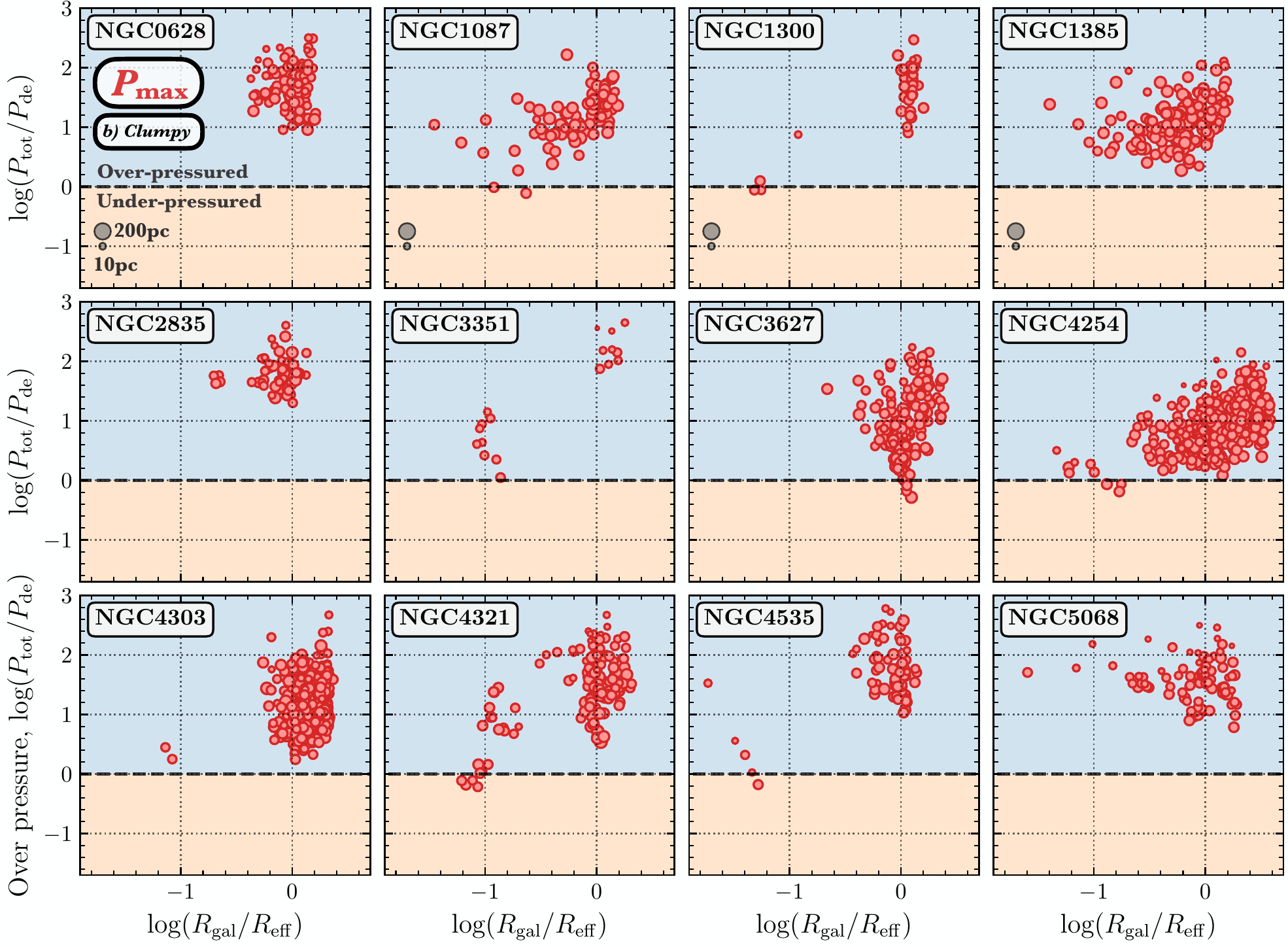}
    \contcaption{}
    \label{fig:scat_PtotPDEreffcont}
\end{figure*}

\begin{figure}
    \centering
	\includegraphics[width=0.9\columnwidth]{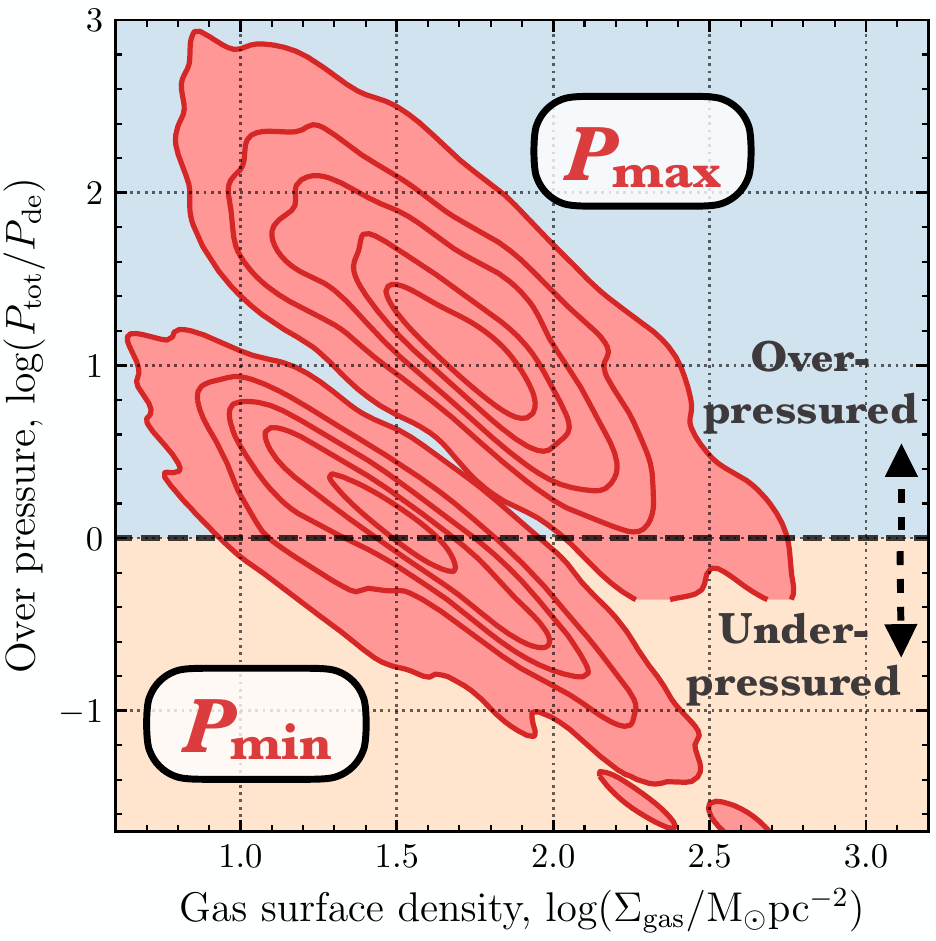}
    \caption{Over-pressure, $\ptot/\pde\ = (\ptherm+\pwind+\pdir)/\pde$, of each \HII\ region as a function of the total gas mass surface density, $\Sigma_\mathrm{gas} = \Sigma_\mathrm{H_2} + \Sigma_{\HI}$ (measured on kpc-sizescales; \citealp{sun20}). The contours show the Gaussian KDE distribution in levels that include 99, 90, 75, 50, and 25 per cent of the points. We show both the maximum pressure limits (\pmax) and minimum pressure limits (\pmin) as separate distributions (see Section\,\ref{sec:int_prescalc}). Above the dashed line the \HII\ regions would be over-pressured (blue shaded), and below the \HII\ regions are under-pressured (orange shaded).}
    \label{fig:scat_PtotSigma}
\end{figure}

In this section, we assess how the total internal pressures vary as a function of their external environment. To do so, we first determine the sum of the minimum or maximum internal pressure component limits,
\begin{equation}
\begin{gathered}
P_\mathrm{tot,min} = P_\mathrm{therm,min} + P_\mathrm{rad,min} + P_\mathrm{wind,min}~, \\
P_\mathrm{tot,max} = P_\mathrm{therm,max} + P_\mathrm{rad,max} + P_\mathrm{wind,max}~,
\end{gathered}
\end{equation}
which assumes all components act independently and combine constructively to create net positive internal pressure. Figure\,\ref{fig:scat_PtotPDE} shows the total internal pressures as a function of the external dynamical pressure. Here, we again show the Gaussian KDE distribution of the points as contours, where the contour levels include 99, 90, 75, 50, and 25 per cent of the data. The diagonal dashed line shows where \pde\ and \ptot\ are equal, and the region where $\pde < \ptot$ (over-pressured) is shaded blue and $\pde > \ptot$ (under-pressured) is shaded in orange.

We see that \pde\ spans four orders of magnitude in Figure\,\ref{fig:scat_PtotPDE}, whereas \ptot\ covers only one and two orders of magnitude for $P_\mathrm{tot,min}$ and $P_\mathrm{tot,max}$, respectively. In addition, here we see a very gradual increase of both \ptot\ limits as a function of \pde. This significantly larger range of \pde\ compared to \ptot, and the tentative correlation between the two pressures, is suggestive that the ambient environmental pressure potentially has only a minor effect in regulating the internal pressures of \HII\ regions. For example, we posit a scenario where a high ambient environmental pressure could confine an \HII\ region, and therefore cause the \HII\ region to become more highly pressured for a given size. If this is indeed the case, this effect would be relatively minor for the larger \HII\ regions we observe. There is a potential caveat to discuss here, however, that the \pde\ the \HII\ regions feel is different from what we consider in this work (see Section\,\ref{sec:ext_prescalc}). Here we use the kpc-scale average \pde\ from \citet{sun20} (i.e. $P_\mathrm{de,1kpc}$; which would be relevant if \HII\ regions are located randomly within the ISM disc), and not the cloud-scale average $\ensuremath{\langle P_\mathrm{de,\theta pc} \rangle_\mathrm{1kpc}}$ (which would be relevant if most \HII\ regions are within/near ISM overdensities). $\ensuremath{\langle P_\mathrm{de,\theta pc} \rangle_\mathrm{1kpc}}$ would have less radial variation, as at larger radii ISM overdensities are less common and, therefore, more impactful when calculating $\ensuremath{\langle P_\mathrm{de,\theta pc} \rangle_\mathrm{1kpc}}$ (i.e. because of the luminosity weighting) than at small radii where the ISM is more densely packed at a fixed measurement scale of $\theta \sim 100$\,pc (\citealp{sun20}). As the \HII\ regions studied here are large and, therefore, evolved, we expect this effect to be significant for a small sample that will be investigated further in a future work. 

Focusing on the $P_\mathrm{tot,min}$ limits in Figure\,\ref{fig:scat_PtotPDE} (lower panel), we see that just under half of the \HII\ regions ($1927$ in total) are over-pressured relative to their environment. These \HII\ regions would therefore still be expanding, despite their large measured sizes of several tens to a few $100$\,pc (\reff). These large and over-pressured \HII\ regions could then be expanding into super\-bubble-like structures, which are interesting targets to study the effect of large-scale expansion in the (ionised, atomic and molecular) gas spatial and kinematic distributions. The remaining $2206$ \HII\ regions appear to be under-pressured relative to their environment, highlighting that these have most likely stopped expanding. If they have not already, these \HII\ regions will begin to dissipate without further energy and moment injection from young stars. The low-density cavities of these large \HII\ regions present the perfect environments into which future SNe can quickly expand.     

Comparing to the $P_\mathrm{tot,max}$ limits in Figure\,\ref{fig:scat_PtotPDE} (upper panel), we see that the majority of \HII\ regions are now over-pressured. This is expected as $P_\mathrm{tot,max}$ is estimated for \HII\ region size scales of a few to a few tens of parsecs, and hence \HII\ regions that are still relatively young and expanding. Interestingly, there is a small number of \HII\ regions~(15) that have $P_\mathrm{tot,max} < \pde$ and are therefore under-pressured relative to their environment. To determine where these \HII\ regions reside within each galaxy, in Figure\,\ref{fig:scat_PtotPDEreff}, we show the ratio of the internal pressure over the external pressure, $\log(\ptot/\pde)$, as a function of the galactocentric radius (also see Table\,\ref{tab:meanpressures}). The horizontal dashed line shows where both \pde\ and \ptot\ are equal, $\log(\ptot/\pde) = 0$, and the region where $\pde < \ptot$ is shaded blue and $\pde > \ptot$ is shaded in orange. Here, the size of each point has been scaled to the effective radius (\reff) of the \HII\ region, and corresponds to the size scale shown in the lower left of each panel. Note that the galaxies IC\,5532, NGC\,1365, NGC\,1433, NGC\,1512, NGC\,1566, NGC\,1672 and NGC\,7496 have been omitted from this analysis due to their lack of available \pde\ measurements. 

Figure\,\ref{fig:scat_PtotPDEreff} shows that $\log(\ptot/\pde)$ systematically increases with increasing galactocentric radius across the sample. With the exception of NGC\,3627, we see that in the six galaxies with $P_\mathrm{tot,max} < \pde$, this occurs at a radius around $R_\mathrm{gal}/R_\mathrm{eff} < 0.1$, approximately corresponding to the central ${<}1$\,kpc (see Table\,\ref{tab:galprops}). This then highlights that centres are interesting high-pressured regions in which to assess the effects of stellar feedback \citep[see e.g.][]{barnes20b}. In the case of the strongly barred galaxy NGC\,3627, we previously mentioned the large scatter in the \pde\ measurements at a galactic radius are coincident with the prominent bar-end features \citep[see e.g.][]{Beuther2017, Beslic2021}. The build-up of gas at the bar-end regions causes an increase in the gas density, and hence an increase in the dynamical pressure similar to that within the galaxy centres. It is interesting to then assess if, more generally, the increase in gas density towards the galaxy centres and bar-end regions has a significant effect on the over- \mbox{(under-)} pressure of a \HII\ region. 

Figure\,\ref{fig:scat_PtotSigma} shows the over-pressure of each \HII\ region as a function of the total gas mass surface density ($\Sigma_\mathrm{gas} = \Sigma_\mathrm{H_2} + \Sigma_{\HI}$). Where the molecular ($\Sigma_\mathrm{H_2}$) and atomic ($\Sigma_{\HI}$) mass surface densities are taken from \cite{sun20}, and have been measured over the same $1$\,kpc hexagonal grid as the \pde\ measurements. Here, we see that both the \pmax\ and \pmin\ distributions show a decreasing $\ptot/\pde$ with increasing $\Sigma_\mathrm{gas}$ (modulo the alternative case described above that  $\ensuremath{\langle P_\mathrm{de,\theta pc} \rangle_\mathrm{1kpc}}$ is larger than \pde). This then shows that the radial trends shown in Figure\,\ref{fig:scat_PtotPDEreff} also apply between galaxies, and galaxies (or environments in general) with higher global gas surface densities are less over-pressured (more under-pressured). The simple interpretation of this is that how quickly/\linebreak[0]{}easily \HII\ regions can expand depends on the global gas surface density; i.e. more dense environments may inhibit rapid expansion (e.g. also see \citealp{Dopita2005,Dopita2006,Watkins2019}). Studying the impact of stellar feedback \citep[e.g.][]{grudic18,Kim2018,fujimoto19,li19,keller_2020} and its effect on the molecular cloud lifecycle \citep{chevance20b}, in setting the initial conditions for star formation \citep[e.g.][]{faesi18,sun18,sun20,schruba19,jeffreson20} and the subsequent star formation efficiency \citep[e.g.][]{krumholz05,blitz06,federrath12}, within dense regions is particularly important, because ISM pressures observed within starburst systems, and at the peak of the cosmic star formation history, are several orders of magnitude higher than those observed in disc galaxies today \citep[e.g.][]{genzel11,swinbank11,swinbank12,tacconi13}. 


\subsection{Pressure components within the literature}
\label{sec:litcomp}

\begin{figure*}
    \centering
	\includegraphics[width=1\textwidth]{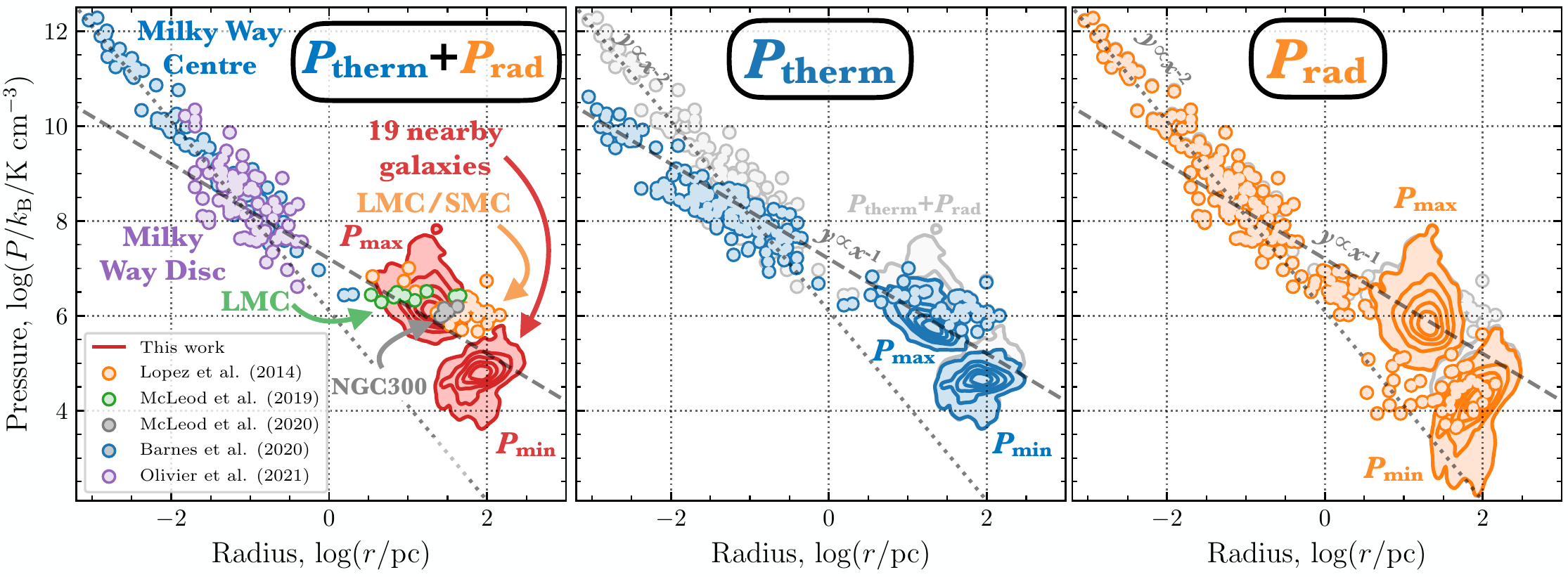}
    \caption{The relative contributions of the thermal (\ptherm) and direct radiation (\pdir) internal pressure components as a function of \HII\ region size ($r$) for a sample of galactic and extragalactic \HII\ regions taken from the literature. We show as points the Small and Large Magellanic Clouds \citep{lopez_2014}, the Large Magellanic Cloud \citep{mcleod_2019}, NGC\,300 \citep{mcleod_20}, the centre ($\rgal < 100$\,pc) of Milky Way \citep{barnes20b}, and disc of the Milky Way \citep{olivier2020}. The contours show the Gaussian KDE of the distribution of pressure terms determined in this work, where the levels are as shown in Figure\,\ref{fig:scat_Pcompreff_all}. We show the distributions for both the maximum pressure limit (\pmax) calculated for the smallest volume, and minimum pressure limit (\pmin) calculated for the largest volume as labelled. {\em Left panel:} We show the sum \ptherm\,+\,\pdir\ for each of the literature samples. {\em Centre and right panels:} We show the separate \ptherm\ and \pdir\ distributions in blue and orange, respectively. The shaded grey distributions are of the \ptherm\,+\,\pdir\ for reference. Shown as diagonal black dashed and dotted lines are power-law dependencies of $y \propto x^{-1}$ and $y \propto x^{-2}$.}
    \label{fig:plit_reff}
\end{figure*}

We now compare the internal pressure components to those determined in previously studied samples of \HII\ regions taken from the literature. Although the wind pressure, and additional internal pressure components such as that from the heated dust, have been determined within the literature, the methodologies for calculating these differ; e.g.\ \pwind\ has been inferred from the shocked $X$-ray emitting gas \citep[e.g.][]{lopez_2011, lopez_2014}. Therefore, here we focus on only the \ptherm\ and \pdir\ pressure components from the literature, as these have been determined using a methodology consistent to that used in this work.

We make use of measurements of \ptherm\ and \pdir\ for a sample of extragalactic \HII\ regions within the Small and Large Magellanic Clouds from \citet{lopez_2014}, the Large Magellanic Cloud from \citet{mcleod_2019},\footnote{\citet{mcleod_2019} used a different expression for the calculation of \pdir, which estimates the radiation force density at the rim of a shell rather than the volume-averaged radiation pressure. We then multiply their values by a factor of three to account for this difference \citep[see e.g.][]{barnes20b}.} and NGC\,300 from \citet{mcleod_20}. We also compare to Galactic measurements focusing on \HII\ regions within in the central regions ($\rgal < 100$\,pc) of the Milky Way from \citet{barnes20b},\footnote{\citet{barnes20b} used varying resolution observations to study a sample of \HII\ regions within the Galactic Centre. Here, we take only the highest resolution measurements towards the three most prominent \HII\ regions covered in that work: Sgr~B2, G0.6 and Sgr~B1 (\citealp{mehringer_1992, schmiedeke_2016}).} and the disc of the Milky Way from \citet{olivier2020}. 

In total, the literature sample comprises $293$ \HII\ regions. This represents the entire sample of \HII\ regions with consistent internal pressure estimates currently available within the literature. The addition of the ${\sim}6000$ \HII\ regions studied in this work represent a significant advancement in the number of measurements available, and for the quantitative assessment of (large-scale) \HII\ region dynamics.    

Figure\,\ref{fig:plit_reff} shows relative contributions of \ptherm\ and \pdir\ as a function of the radius for all the literature measurements mentioned above, and including the measurements determined in this work. In the left panel, we show the sum of \ptherm\ and \pdir, and colour the points by the reference. In the centre panel, we show only the distribution of \ptherm\ (in blue), whereas in the right panel we only show the distribution of the \pdir\ (in orange). For reference, we also show the distribution of the \ptherm+\pdir\ in these panels in faded grey. 

The first thing to note in the left panel of Figure\,\ref{fig:plit_reff} is that the Galactic sources have a much smaller scale and more highly pressured measurements than the extragalactic sources ($\reff \lesssim 1$\,pc). This is likely due to two reasons. Firstly, this could be a resolution effect. As it is possible to more easily achieve a higher spatial resolution with observations within the Milky Way due to its relative proximity compared to other galaxies. The observed extragalactic \HII\ regions could then fragment on smaller scales, and these \HII\ regions could be more highly pressured. Secondly, the \HII\ regions within the sample we observe could be at a later stage in their evolution compared to the Galactic samples, and hence larger and less pressured. Differentiating between these two possibilities is ultimately the aim of the future investigations discussed in the next section. 

As previously noted, radii larger than $\sim 1$~pc have somewhat larger thermal pressures compared to their direct radiation pressure. Although, this is only true when considering their \pmin\ estimates. Whereas, their \pmax\ are similar (see Figure\,\ref{fig:scat_PtotPDEreff}). 
On scales of $0.01{-}1$\,pc, however, \ptherm\ and \pdir\ are comparable, and on the smallest scale of ${<}0.01$\,pc, $\ptherm < \pdir$ \citep[also see][]{barnes20b, olivier2020}. Shown in Figure\,\ref{fig:plit_reff} are diagonal black dashed and dotted lines highlighting power-law dependencies of $y \propto x^{-1}$ and $y \propto x^{-2}$ for reference (note that these are not fits to the data). We see that \ptherm\ approximately follows a $r^{-1}$ relation. The \pdir\ follows $r^{-2}$, albeit with a significant scatter for the larger \HII\ regions (i.e. for the extragalactic observations). These are both in agreement with their expected radial dependencies (e.g. by construction from Equation\,\eqref{eq:Pdir}; see \citealp{barnes20b}). Moreover, these are in broad agreement with \citet{olivier2020}, who compared their Milky Way measurements to the LMC and SMC measurements from \citet{lopez_2014}. These authors found radial power-law dependencies for \pdir\ of $r^{-1.36}$ and \ptherm\ of $r^{-0.74}$ (also thermal dust pressure is found to scale as $r^{-1.43}$), showing that the thermal pressure is typically sub-dominant on the smallest scales, yet does not decrease as rapidly with increasing size, and hence becomes mildly dominant on the largest scales. The transition in pressure terms shows that the impact of the different feedback mechanisms evolves and that, for example, direct radiation pressure is more significant at early times \citep[e.g.][]{Arthur2006, Tremblin2014}. However, we note that we do not consider trapped radiation pressure in this work (e.g. in the form of dust heating), which could contribute significantly at later times (e.g. \citealp{lopez_2014, olivier2020}).  


\subsection{Open questions and future avenues}\label{subsec:future}

This work represents a significant milestone in observationally quantifying the feedback properties in a large sample of evolved \HII\ regions. However, there are limitations to our study that leave several questions unanswered. We end this section by outlining these questions, and noting the possible future avenues for building on our analysis. 

Firstly, we had to make several simplistic assumptions about the unresolved sub-structure of the \HII\ regions within our sample. With these current data, it is difficult to address, for example, which of the smooth (i.e. \pmin) vs clumpy (i.e. \pmax) models may be favoured? Comparison to the literature trends in Figure\,\ref{fig:plit_reff}, shows that \pmax\ is more similar to the \ptherm\ measurements, whereas \pmin\ could favour \pdir. In addition, the literature shows that Galactic centre \HII\ regions have maximum sizes of few $\sim\,1$\,pc \citep{barnes20b}, yet our \rmin\ within these central regions is still $\gtrsim$\,10pc. Could we then be overestimating the sizes, particularly in this environment? This could be due to confusion from the DIG, which is particularly extended within centres, and clustering, which could cause multiple smaller \HII\ regions to be merged in our observations. To address these questions, higher-spatial resolution data sets are required to constrain the true sizes and separations of the \HII\ regions, and hence allow us to place tighter constraints on the internal pressures. Such observations could be obtained from the {\em Hubble Space Telescope}, the {\em James Webb Space Telescope} ({\em JWST}), or integral field spectroscopy observations of more nearby targets (e.g. NGC\,300; \citealp{mcleod_20}, SIGNALS; \citealp{Rousseau-Nepton2019}; and the \mbox{SDSS-V} Local Volume Mapper survey; \citealp{Kollmeier2017}). 

Secondly, we have inferred several key properties of the stellar populations within the \HII\ regions using the \textsc{starburst99} models. When doing so, we used a representative age range for the sample of 0 to 4\,Myr, defined by the time for $L_{\Ha}$ to drop by half an order of magnitude from the zero-age main sequence (see Section\,\ref{sec:sb99_lbol}). Could our results differ when accounting for the \HII\ regions at various evolutionary stages? This could be addressed by using age estimates of the cluster associations within the \HII\ regions from the, e.g., PHANGS-HST survey (e.g. \citealp{Lee2021}). In addition, by confirming the presence of the young stellar population driving the \HII\ regions, we will be able to mitigate contamination from other sources of ionisation in our sample (e.g. shocks; e.g. see \citealp{Espinosa-Ponce2020} and references therein). Efforts to link the ionised gas properties from PHANGS-MUSE and cluster properties from PHANGS-HST are currently underway. 

Thirdly, in this work, we have focused on the internal pressure components of \ptherm, \pdir\ and \pwind, yet could the additional contribution from, e.g., trapped radiation from heated dust be important in driving the large scale expansion of the \HII\ regions (e.g. \citealp{Krumholz2009, draine_2011})? The inclusion of heated dust pressure is, however, difficult for distant extragalactic sources. As, for temperatures of ${\sim}100$\,K the blackbody function peaks within the mid-infrared (${\sim}30$\,\micron). Hence, modelling the heated dust emission requires observations within the infrared regime. Currently, available data sets are limited in resolution (e.g. from the {\em Spitzer} and {\em Herschel} space observatories), yet this may be possible in the near future using the scheduled {\em JWST} observations for this sample of galaxies (PI: Lee). 

Lastly, it would be interesting to see how the balance of internal and external pressures of the \HII\ regions varies with both local and global galactic environments. We could ask: does the pressure balance within \HII\ regions differ for arm, inter-arm and central regions? Are the \HII\ regions still embedded within molecular clouds, and how does this affect their expansion? To assess the effect of the local environment, one can compare to the measurements of the \pdess\ characterized at scales of $\theta \sim 100$\,pc, which accounts for the clumpy molecular and diffuse atomic ISM (\citealp{sun20}; also see \citealp{Barrera-Ballesteros2021a,Barrera-Ballesteros2021b}). We could then assess the pressure balance relative to \pdess\ for those \HII\ regions that are still potentially embedded, which could be identified as having a higher extinction or associated with \mbox{CO(2--1)} emission from the PHANGS-ALMA survey (\citealp{Leroy2021a}). To assess the effect of the global galactic environment on the \HII\ region properties, we can compare the pressure balance to the environmental masks produced by \citet{Querejeta2021}. These masks were produced using the {\it Spitzer} 3.6\,\micron\ images, and differentiate stellar structures that form the galaxy centres, bars, spiral arms, and inter-arms regions. In this work, we inferred that a higher fraction of \HII\ regions may be under-pressured within the galaxy centres, yet it would be interesting to assess if this could be found within these various other galaxy environments.
\section{Summary}\label{sec:summary}

In this paper, we compare the internal and external pressures acting on a sample of 5810 \HII\ regions across 19 nearby spiral galaxies. The \HII\ region sample is identified using Multi Unit Spectroscopic Explorer (MUSE) data taken as part of the Physics at High Angular resolution in Nearby GalaxieS survey (PHANGS-MUSE; \citealt{Emsellem2021}). We constrain the internal pressure components of the thermal pressure from the warm ionised gas (\ptherm; Section\,\ref{sec:therm_prescalc}), direct radiation pressure (\pdir; Section\,\ref{sec:dir_prescalc}), and mechanical wind pressure (\pwind; Section\,\ref{sec:wind_prescalc}), which we compare to the confining external pressure of their host environment, or their dynamical pressure (\pde; Section\,\ref{sec:ext_prescalc}). With the MUSE observations, we cannot constrain the unresolved density distribution within the ionised gas, and hence we place upper and lower limits on each of the internal pressure components. The lower limit (\pmin) corresponds to the assumption of a smooth density profile, where the measured radius (\reff) is assumed to be representative of the \HII\ volume over which the pressure is acting (Section\,\ref{subsubsec:reff}). The upper limit (\pmax) corresponds to the assumption of a more clumpy density profile, where the minimum radius (\rmin) is derived from the electron density measurement (Section\,\ref{sec:lowerlimits}). Of the sample of 5810 \HII\ regions studied in this work, 2238 \HII\ regions have both \pmax\ and \pmin\ measurements, whereas 3431 have only \pmax, and 141 have only \pmin\ (see Table\,\ref{tab:galobsprops}). Due to our observational selection criteria (section\,\ref{sec:lowerlimits}), these samples are biased towards the brightest and largest \HII\ regions within the galaxies. The main conclusions from the analyses of these samples are summarised below. 

We assess the relative differences of the \pmax\ or \pmin\ measurements for each pressure term. We see that the maximum internal pressures are all relatively similar, with mean values of around $\pmax/k_\mathrm{B} \sim 10^{6}$\,\Kcmcb. On the other hand, the minimum values appear relatively different, with the direct radiation and wind pressures having values around $\pmin/k_\mathrm{B} \sim 10^{4}$\,\Kcmcb\ and the thermal pressures being around a factor of~4 higher ($P_\mathrm{therm,min}/k_\mathrm{B} \sim 10^{4.6}$\,\Kcmcb). This shows that at best the pressure terms are comparable if they have a compact density distribution (i.e. at \pmax). However, it is likely that the vast majority of \HII\ regions have at least some extended structure (i.e. tending to $\pmin$) that would then cause \ptherm\ to become dominant. 

Comparison to a sample of \HII\ region pressure measurements available within the literature shows that on the scales of several tens to a couple of hundred parsecs \ptherm\ is expected to be the highest internal pressure (e.g. \citealp{lopez_2011, lopez_2014, mcleod_2019, mcleod_20}). In addition, we compare to \HII\ regions within the Milky Way and more nearby galaxies such as the LMC and SMC; combined with the presented measurements of this work, the sample covers spatial scales that span a total of six orders of magnitude ($0.001$ to $300$\,pc). Indeed, above scales of around~$0.1$ to~$1$\,pc the thermal pressure is marginally dominant, yet below $0.1$\,pc the direct radiation pressure is dominant \citep{lopez_2014, mcleod_2019, barnes20b, olivier2020}. We note that due to inconsistencies within the literature, this comparison does not include the indirect (trapped) radiation pressure from heated dust or the contribution of winds.

We compare our total internal pressures ($\ptot = \ptherm + \pdir + \pwind$) within each \HII\ region to the external pressure ($\pde$) of their host environment, which we take directly from \citet[][but also see \citealp{Barrera-Ballesteros2021a,Barrera-Ballesteros2021b} for similar calculations of $\pde$]{sun20}. We see that for the $P_\mathrm{tot,min}$ limits (see Figure\,\ref{fig:scat_PtotPDE}) just under half of the \HII\ regions ($1927$ in total) are over-pressured relative to their environment (i.e. $P_\mathrm{tot,min} > \pde$). These \HII\ regions would still be expanding despite their large measured sizes of several tens to a few $100$\,pc (\reff), and would, therefore, represent interesting targets to study the effect of large-scale expansion in the (ionised, atomic and molecular) gas spatial and kinematic distributions. The remaining $2206$ \HII\ regions appear to be under-pressured relative to their environment, highlighting that these have most likely stopped expanding.

We find that for the $P_\mathrm{tot,max}$ limits, the majority of \HII\ regions are now over-pressured. This is expected as $P_\mathrm{tot,max}$ is estimated assuming an \HII\ region size scale typically of the order a few to a few tens of parsecs. In this case, \HII\ regions would be still relatively young and expanding. Interestingly, however, there is a small number of compact \HII\ regions~(15) that are under-pressured relative to their environment. Plotting the ratio of the internal pressure over the external pressure, $\log(\ptot/\pde)$, as a function of galactocentric radius (see Figure\,\ref{fig:scat_PtotPDEreff}), we see that the majority of these compact under-pressured \HII\ regions reside within galaxy centres. This then highlights that centres are interesting high-pressured regions in which to assess the effects of stellar feedback \citep[see e.g.][]{barnes20b}. To assess the effect of environment more generally, we investigate if the increase in gas density has a significant effect on the over- \mbox{(under-)} pressured nature of an \HII\ region (see Figure\,\ref{fig:scat_PtotSigma}). We see that regions of galaxies (or environments in general) with higher gas surface densities have fewer over-pressured \HII\ regions (and more under-pressured \HII\ regions). The simple interpretation of this is that a more dense environment may inhibit rapid expansion, and thus limit the effect of stellar feedback. This is of particular importance not only for current-day star formation, but also has implications for cosmic timescales, given that ISM pressures and densities observed at the peak of the cosmic star formation history are several orders of magnitude higher than those observed in disc galaxies today \citep[e.g.][]{genzel11,swinbank11,swinbank12,tacconi13}. 
\section*{Acknowledgements}

We would like to thank the referee, Sebastian Sanchez, for their constructive feedback that helped improve the quality of this paper. This work was carried out as part of the PHANGS collaboration. ATB, IB, and FB would like to acknowledge funding from the European Research Council (ERC) under the European Union’s Horizon 2020 research and innovation programme (grant agreement No.726384/Empire). SCOG and RSK acknowledge support from the DFG via SFB 881 ``The Milky Way System'' (sub-projects B1, B2 and B8) and from the Heidelberg cluster of excellence EXC 2181-390900948 “STRUCTURES: A unifying approach to emergent phenomena in the physical world, mathematics, and complex data”, funded by the German Excellence Strategy. KK gratefully acknowledges funding from the German Research Foundation (DFG) in the form of an Emmy Noether Research Group (grant number KR4598/2-1, PI Kreckel). The work of ECO is partly supported by the US National Science Foundation under grant AST-1713949. CE acknowledges funding from the Deutsche Forschungsgemeinschaft (DFG) Sachbeihilfe, grant number BI1546/3-1. JMDK and MC gratefully acknowledge funding from the Deutsche Forschungsgemeinschaft (DFG, German Research Foundation) through an Emmy Noether Research Group (grant number KR4801/1-1) and the DFG Sachbeihilfe (grant number KR4801/2-1), as well as from the European Research Council (ERC) under the European Union's Horizon 2020 research and innovation programme via the ERC Starting Grant MUSTANG (grant agreement number 714907). ES, TS, TGW, FS, and RM acknowledge funding from the European Research Council (ERC) under the European Union’s Horizon 2020 research and innovation programme (grant agreement No. 694343).

Based on observations collected at the European Southern Observatory under ESO programmes 1100.B-0651, 095.C-0473, and 094.C-0623 (PHANGS-MUSE; PI Schinnerer), as well as 094.B-0321 (MAGNUM; PI Marconi), 099.B-0242, 0100.B-0116, 098.B-0551 (MAD; PI Carollo) and 097.B-0640 (TIMER; PI Gadotti). This paper makes use of data gathered with the 2.5 meter du Pont located at Las Campanas Observatory, Chile, and data based on observations carried out at the MPG 2.2m telescope on La Silla, Chile.

\section*{Data availability}

Science-level MUSE mosaicked datacubes and high-level analysis products (e.g. emission line fluxes) are provided via the ESO archive phase 3 interface\footnote{\url{https://archive.eso.org/scienceportal/home?data_collection=PHANGS}}. A full description of the the first PHANGS data release is presented in \citet{Emsellem2021}. The complete \HII\ region catalogue used in this work will be made available in a forthcoming publication. 




\bibliographystyle{mnras}
\bibliography{references}


\appendix

{\footnotesize \it \noindent \\
$^{1}$\aifa\\
$^{2}$\ita\\
$^{3}$\ari\\
$^{4}$\princeton\\
$^{5}$\arcetri\\
$^{6}$\carnegie\\
$^{7}$\Uchile\\
$^{8}$\wyoming\\
$^{9}$\stern\\
$^{10}$\eso\\
$^{11}$\anu\\
$^{12}$\UWaus\\
$^{13}$\iwr\\
$^{14}$\ohio\\
$^{15}$\ljmu\\
$^{16}$\mpia\\
$^{17}$\UGent\\
$^{18}$\nrao\\
$^{19}$\alberta\\
$^{20}$\mpe\\
}

\section{}\label{sec:nelimcomp}

\begin{figure}
    \centering
	\includegraphics[width=\columnwidth]{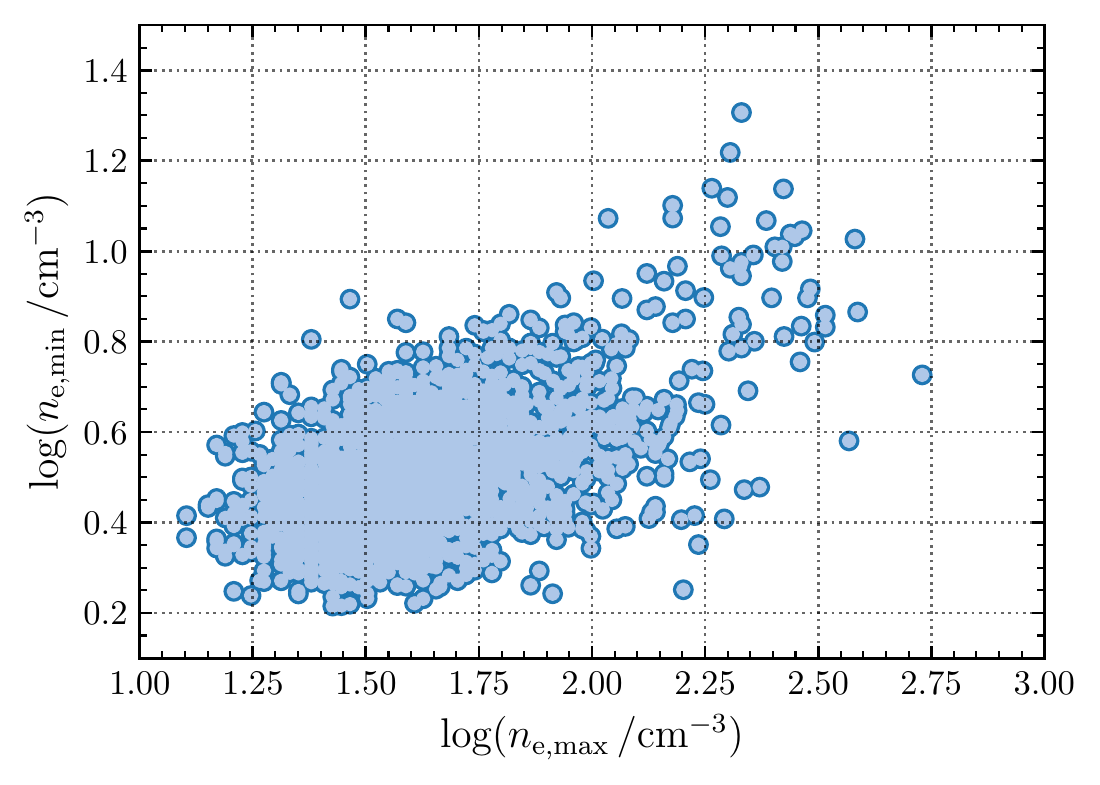}
    \caption{Comparison the electron densities assuming a smooth ($n_\mathrm{e,min}$) and clumpy ($n_\mathrm{e,max}$) unresolved density distribution (see Figure\,\ref{fig:toyfig}). We plot \HII\ regions for which we can derive both \rmin\ and \nemin\ (2238 regions in total; see section\,\ref{sec:lowerlimits}).}
    \label{fig:nelimcomp}
\end{figure}

In Figure\,\ref{fig:nelimcomp}, we show a comparison between the electron densities assuming a smooth ($n_\mathrm{e,min}$) and clumpy ($n_\mathrm{e,max}$) unresolved density distribution (see Figure\,\ref{fig:toyfig}). We plot \HII\ regions for which we can derive both \rmin\ and \nemin\ (2238 regions in total; see section\,\ref{sec:lowerlimits}). We see that the $n_\mathrm{e,max}/n_\mathrm{e,min}\sim10$, highlighting that the volume filling factor of the unresolved \HII\ regions could be of the order $\sim$1\,per cent (see equation\,\ref{equ:volume}).

\section{Physical properties of the sample}
\label{sec:appendix_physprops}


\begin{figure}
    \centering
	\includegraphics[width=0.9\columnwidth]{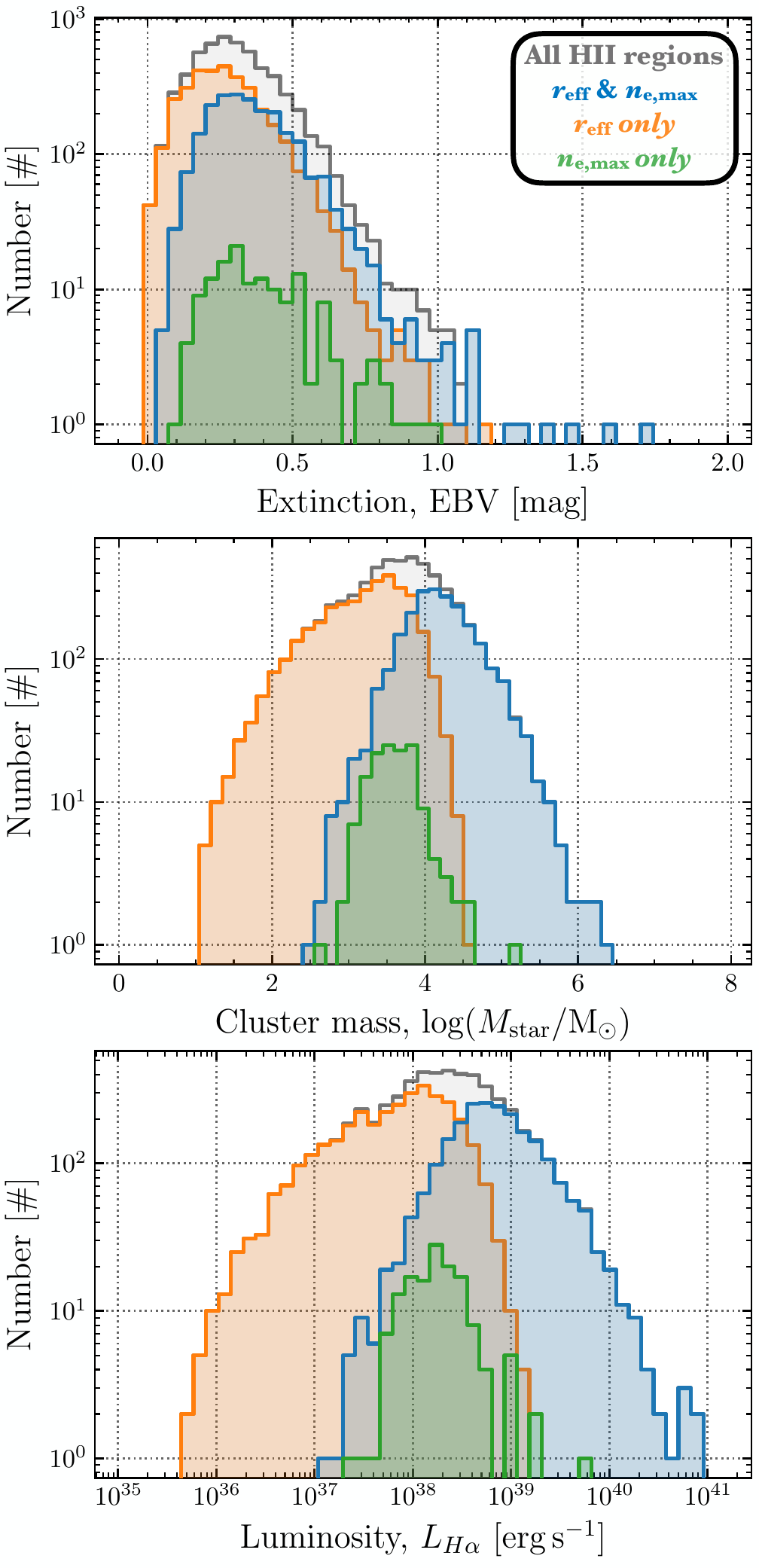}
    \caption{Distribution of the extinction ({\em top panel}), extinction-corrected \Ha\ luminosity ({\em middle panel}), and cluster mass ({\em bottom panel}; see Section\,\ref{sec:physprops}) for the various subsets of the \HII\ region sample. We show the resolved ($\reff \ge \mathrm{FWHM}_\mathrm{PSF}$; see Section\,\ref{subsubsec:reff}) sources as both blue and orange histograms. The blue and orange histograms differentiate the samples with and without electron density measurements, respectively (see Section\,\ref{subsubsec:electrondensity}). The distribution of the unresolved sources $\reff < \mathrm{FWHM}_\mathrm{PSF}$ with electron density measurements are shown in green.}
    \label{fig:histprops}
\end{figure}

In Figure\,\ref{fig:histprops} (top panel), we show the distributions for the reddening ${E(B-V)}$ across the various sub-sample of \HII\ regions. These colour excess measurements have been used to correct the \Ha\ fluxes, which are used to determine the \Ha\ luminosity for each \HII\ region (see middle panel of Figure\,\ref{fig:histprops}). Synthetic stellar population models (\textsc{starburst99}; \citealp{Leitherer1999}) are used to estimate the cluster mass, mass loss rate and mechanical luminosity ($M_\mathrm{cl}$, $\dot M$ and $L_\mathrm{mech}$; see Section\,\ref{sec:sb99}). The distribution of $M_\mathrm{cl}$ across the sample is shown in the bottom panel of Figure\,\ref{fig:histprops}. Above $M_\mathrm{cl} \sim 10^{3}$\,\sol\ the IMF is generally fully sampled, so the ratios $L_\mathrm{bol}/M_\mathrm{cl}$, $Q/M_\mathrm{cl}$, $L_\mathrm{mech}/M_\mathrm{cl}$ and $\dot M/M_\mathrm{cl}$ are relatively independent of $M_\mathrm{cl}$. Here we see that the majority of the \HII\ regions within our sample have $M_\mathrm{cl} > 10^{3}$\,\sol. However, we find that that around ${\sim}20$ per cent ($1299$) of the \HII\ regions in our sample are below this mass limit, and may be affected by increased uncertainties on their derived properties.

\section{Effect of galactic environment on feedback}\label{sec:hstcomp}

\begin{figure*}
    \centering
	\includegraphics[width=\textwidth]{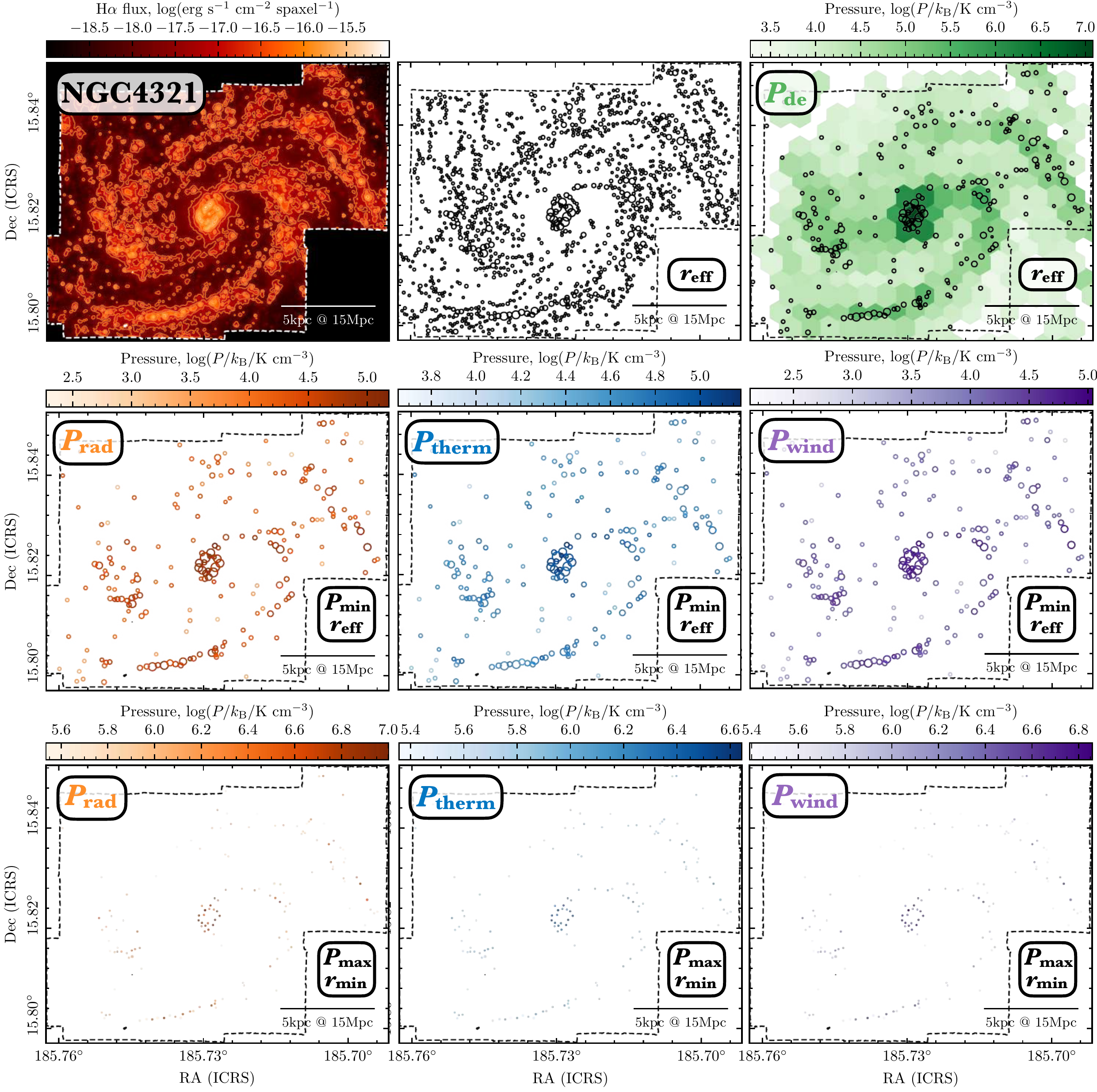}
   \caption{Summary of the various pressure components for the \HII\ regions across the galaxy NGC\,4321. {\em Upper left:} We show the MUSE \Ha\ emission map taken as part of the PHANGS-MUSE survey \citep{Emsellem2021}. A~scale bar of $5$\,kpc is shown in the lower right corner of the panel. {\em Upper centre}: We show the full \HII\ region sample identified within each galaxy (\citealt{Santoro2021}; see Section\,\ref{sec:sample}). The size of the circles represent the physical sizes (\reff\ or \rmin) of the \HII\ regions denoted in the lower right corner of the panel (see Sections\,\ref{subsubsec:reff} and~\ref{sec:lowerlimits}). {\em Upper right:} The background colour-scale shows the \pde\ measurements that have been sampled on a $1$\,kpc hexagonal grid \citep{sun20}. This is overlaid with the sample of \HII\ regions with resolved \reff\ size measurements. We show the lower ({\em centre row of panels}) and upper ({\em bottom row of panels}) limits of the direct radiation (\pdir), thermal (\ptherm) and wind pressures (\pwind).}
    \label{fig:maps_pres1}
\end{figure*}

In Figure\,\ref{fig:maps_pres1}, we show an example how the different pressures vary as a function of position across one of the galaxies in our sample, NGC\,4321. Here we show the MUSE \Ha\ emission map taken as part of the PHANGS-MUSE survey \citep{Emsellem2021}, from which the \HII\ region sample has been identified (see top centre panel). In the upper right panel, in the background colour scale we show the \pde\ measurements that have been sample on a $1$\,kpc hexagonal grid \citep{sun20}. In the central row of panels we show the lower (\pmin) limits of the direct radiation (\pdir), thermal (\ptherm) and wind pressures (\pwind). In the bottom row, we show the upper (\pmax) limits of the pressures, where the size of the points corresponds to the lower size limit (\rmin). 

\bsp	
\label{lastpage}
\end{document}